\documentclass[11pt]{article}
\usepackage{amsthm,amsfonts,amsmath}
\usepackage{latexsym}
\swapnumbers
\theoremstyle{plain}

\theoremstyle{definition}

\theoremstyle{remark}

\newcommand{\upchi}{\raise1pt\hbox{$\chi$}}

\renewcommand{\|}{{\Vert}}
\numberwithin{equation}{section}
\newcommand{\p}{\partial}
\newcommand{\og}{\omega}

\newcommand{\dt}{\delta}
\newcommand{\Dt}{\Delta}

\newcommand{\ld}{\lambda}
\newcommand{\Ld}{\Lambda}

\newcommand{\gm}{\gamma}
\newcommand{\vp}{\varphi}
\newcommand{\vep}{\varepsilon}

\newcommand{\fr}{\frac}
\newcommand{\sg}{\sigma}

\newcommand{\be}{\begin{equation}}
\newcommand{\ee}{\end{equation}}
\newcommand{\ba}{\begin{array}}
\newcommand{\ea}{\end{array}}
\newcommand{\bea}{\begin{eqnarray}}
\newcommand{\eea}{\end{eqnarray}}
\newcommand{\beas}{\begin{eqnarray*}}
\newcommand{\eeas}{\end{eqnarray*}}

\newcommand{\bR}{{\bf R}^3 }
\newcommand{\bS}{{\bf S}^2 }

\newcommand{\bRN}{{\bf R}^N}
\newcommand{\bSN}{{\bf S}^{N-1}}
\newcommand{\bRRN}{{\bRN}\times {\bRN}}
\newcommand{\bRRSN}{{\bRRN}\times{\bSN}}

\newcommand{\bRSN}{{\bRN}\times{\bf S}^{N-1}}

\begin{document}           

\title{On Strong Convergence to Equilibrium for\\ the Boltzmann Equation with
Soft Potentials}

\author{ Eric A. Carlen$^1$,  Maria  C. Carvalho$^2$ and  Xuguang Lu$^3$\\}

\maketitle

\tableofcontents
\vspace{1cm}

\footnotetext
[1]{ Department of Mathematics, Rutgers University,
Piscataway, NJ  08854, U.S.A. Work partially
supported by U.S. National Science Foundation
grant DMS 06-00037.  }

\footnotetext
[2]{Department of Mathematics and CMAF, University of Lisbon,
1649-003 Lisbon, Portugal. Work partially supported by POCI/MAT/61931/2004}
\footnotetext
[3]{Department of Mathematical Sciences,
Tsinghua University, Beijing 100084, China. Work partially supported  by 
NSF of China grant 10571101\\
\copyright\, 2009 by the authors. This paper may be reproduced, in its
entirety, for non-commercial purposes.}

\date{}      

\begin{abstract}
The paper concerns $L^1$- convergence to equilibrium for weak solutions
of the spatially homogeneous Boltzmann Equation for soft potentials
$(-4\le \gm<0$), with and without angular cutoff. We prove the time-averaged $L^1$-convergence to equilibrium for all weak solutions
whose initial data have finite entropy and finite moments up to
order greater than $2+|\gm|$. For the usual $L^1$-convergence we
prove that the convergence rate can be controlled from below by the initial energy tails, and
hence, for initial data with long energy tails,  the convergence can be arbitrarily slow. We
also show that under the integrable angular cutoff on the collision kernel with $-1\le \gm<0$,
there are algebraic upper and lower bounds on
the rate of $L^1$-convergence to equilibrium. Our
methods of proof are based on entropy inequalities and moment estimates.
\end{abstract}

\\

\section{Introduction}
\medskip

While convergence to equilibrium for solutions of the spatially
homogeneous Boltzmann equation has been extensively studied for hard
potentials and Maxwellian molecules, much less is known in the case
of soft potentials. For instance, for the hard sphere model, it has
been proven that solutions of the equation for all initial data that
have finite mass and energy always converge strongly to equilibrium
at an exponential rate. The same result holds for all hard
potentials with angular cutoff under only mild additional
assumptions on the initial data $f_0$; e.g., that $f_0$ is square
integrable. See \cite{15} and references therein.

For Maxwellian molecules with angular cutoff, one again has
exponential convergence to equilibrium if the initial data have
finite moments up to order $s>2$, although for $s= 2$ the
convergence rate can be arbitrarily slow \cite{5}.  (In the
Maxwellian case, as in the hard sphere case, there is no need to
make any assumption such as square integrability of $f_0$, or even
that the initial entropy be finite.)

For soft potentials, existing bounds \cite{10,17} on the rate of
convergence to equilibrium are only algebraic, and so far have been
obtained under certain cutoffs, not only in the angle, but also on
the singularity in small relative velocities (which is present for
soft potentials). It is commonly believed that for soft potentials,
the convergence rate {\em actually  is} generally worse than the
former cases, and that the algebraic bounds found in the references
cited above are at least qualitatively sharp.

In this paper, we show that this is indeed the case. Moreover, we
also prove convergence results for a very broad class of weak
solutions, and for a range of {\em very soft} potentials. The
convergence results that we obtain at this broadest level of
generality are exactly that: They assert convergence, in either a
time averaged sense, or in the usual sense, but with no rate at all.

However, in the case of angular cutoff, and with a potential that is
not {\em too soft}, we are able to prove much more: In particular,
we shall show that for a natural  class of weak solutions, if the
initial data has moments of all orders, then the solution converges
strongly in $L^1({\bf R}^N,(1+|v|^2)dv)$ to its Maxwellian
equilibrium at a {\em super-algebraic rate}; i.e., faster than any
inverse power of the time $t$. In proving this result we rely in
part on an
 entropy production inequality of
Villlani \cite{20}, but also introduce a new strategy to avoid the
use of pointwise lower bounds on the solution $f$ that were used in
\cite{20},

One crucial difference between hard potentials, Maxwellian
molecules, and soft potentials shows up in the different behavior
concerning energy tails: In the case of hard potentials with an
angular cutoff, even if the initial data has no moments of order
higher than $2$,  the solution at any strictly positive time will
have moments of {\em all} orders \cite{23}. That is, long energy
tails, which are an obstacle to rapid convergence, are immediately
eliminated for hard potentials.  This is not the case for Maxwellian
molecules, but at least whatever control one has on the energy tails
of the initial data is propagated {\em uniformly in time}.  For soft
potentials, the situation is much less favorable, and there is no
propagation of higher moments uniformly in time.  Instead, one has
bounds on the growth of such moments, or uniform bounds on their
time averages, as found in \cite{9}. Such moment bounds play a
crucial role in this paper, and we shall prove several new and
strengthened results of this type.

Before  proceeding with the introduction of our results, let us
first precisely specify the equation to be studied and the notation
that we shall use.

\subsection{The Boltzmann Equation for soft and very soft potentials}

The spatially homogeneous
Boltzmann equation is given by (see \cite{6,7,8})
$$\fr{\p}{\p t}f(v,t)=Q(f)(v,t)\,,\quad (v,t)\in(0,\infty)\times {\bRN}
\eqno({\rm B})$$  where  $N\ge 2$,
\begin{equation}\label{Qdef}
Q(f)(v,t)=\int_{{\bRSN}}B(v-v_*,\sg)(f'f_*'-ff_*)d\sg dv_*\,,
\end{equation}
$f=f(v,t), f'=f(v',t), f_*=f(v_*,t), f_*'=f(v_*',t)$,
$$v'=\fr{v+v_*}{2}+\fr{|v-v_*|\sg}{2}\,,\quad v_*'=\fr{v+v_*}{2}-\fr{|v-v_*|\sg}{2}
\,,\quad \sg\in{\bSN}$$ and ${\bSN}$ is the unit sphere in ${\bRN}$.
The collision kernel $B(z,\sg)$ is a nonnegative Borel function of
$(|z|,\cos\theta)$, i.e.
$$B(z,\sg)=B(|z|,\cos\theta)\,,\quad \cos\theta=\langle{z/|z|,\sg}\rangle\,,\quad
 z=v-v_*\neq 0\,.$$
In the case of main physical interest, $N =3$. Then, if the potential
energy function that governs the interaction between pairs of
molecules in the dilute gas is an inverse power of the distance
separating them,  $B$ takes the form
\begin{equation}\label{factored}
B(z,\sg)=b(\cos\theta)|z|^{\gm}
\end{equation}
with the exponent $\gamma$ depending on the power in the
interaction. The following ranges of $\gamma$ are  distinguished
\cite{6,7} by the methods required to treat them: The range
$0<\gm<1$ corresponds to {\em hard potentials}, $\gm=0$ to {\em
Maxwellian molecules}, and $\gm<0$ to {\em  soft potentials}, with
the case $\gm<-2$  corresponding to {\em very soft potentials}
\cite{20}.

These  distinctions pertain to the different strategies that must be
employed in studying solutions of  Eq.(B), or even interpreting it,
for $\gamma$ in the different ranges. When $\gamma$ is negative,
both the singularity in $B(z,\sg)$ at $z=0$, and the vanishing of
$B(z,\sg)$ at $z=\infty$ cause difficulties that partially account
for the fact that soft potentials have been less intensively
investigated than Maxwellian molecules or hard potentials. The
problems caused by the vanishing of   $B(z,\sg)$ at $z=\infty$
include the fact that with soft potentials, one does not have
uniform in time bounds on higher order moments of solutions; we
shall return to this shortly.

The problems caused by the singularity in $B(z,\sg)$ at $z=0$ are
more immediate: This singularity precludes a naive approach to
making sense of the integral in (\ref{Qdef}), and hence complicates
the interpretation of the equation itself.

$Q(f)$ is a difference of two integrals,
and if each of them is to be integrable, it would have to be the case that
$$B(v-v_*,\sg)f'f_*'\qquad{\rm and}\qquad B(v-v_*,\sg)ff_*$$
would both be integrable on ${\bf R}^3\times {\bf R}^3 \times {\bf
S}^{2}$. When $B$ takes the form $B(z,\sg)=b(\cos\theta)|z|^{\gm}$
as in (\ref{factored}), with the exponent $\gamma$ and $0< \gamma\le
2$, at least the integration over ${\bf R}^3\times {\bf R}^3$ poses
no problem: As is well known, solutions of the Boltzmann equation
({\rm B}) should conserve energy, and so an {\em a priori} bound on
$\int_{{\bf R}^3}(1+|v|^2)f(v) dv$ is natural to assume. Granted
this, the integrability over  ${\bf R}^3\times {\bf R}^3$ is
obvious.

There still remains the fact that for inverse power law potentials, the function $b(\cos\theta)$  in (\ref{factored})  is not integrable on ${\bf S}^{2}$, and so in many studies of Eq.(B)
for hard potentials, one invokes a ``Grad angular cut-off''
to truncate  $b(\cos\theta)$ so that it becomes integrable.

For soft potentials, the situations is  more delicate: Lack of
integrability of $b(\cos\theta)$ is not the only problem. When
$\gamma$ is negative, the function
\begin{equation}\label{badprod}
|v-v_*|^\gamma f(v)f(v_*)
\end{equation}
is not in general integrable on ${\bf R}^3\times {\bf R}^3$ under
any natural hypothesis on $f$. The finite energy condition does not
help, nor does the $H$-Theorem, which would justify assuming that
$f\log  f$ is integrable. By the Hardy-Littlewood-Sobolev
inequality, the function in (\ref{badprod}) would be integrable if
$f$ belonged to $L^{6/(6+\gamma)}({\bf R}^3)$, but there is no
reason to expect control on this $L^p$ norm along any general class
of solutions of Eq.(B).

To proceed, let $\varphi$ be a test function, and note that
by standard formal calculations
(see e.g. \cite{18}),  if we define
\begin{equation}\label{(1.0)}\Dt\vp(v', v_*', v, v_*):=
\vp(v')+\vp(v_*')-\vp(v)-\vp(v_*)
\end{equation}
and define
$$Q(f\,|\Dt\vp)(v)=\int_{{\bf R}^3\times{\bf S}^2}B(v-v_*,\sg)
(f'f_*'-ff_*)\Dt\vp\, d\sg dv_*\ ,$$
then we would have
$$\int_{{\bf R}^3}\vp(v)Q(f)(v)dv=-\fr{1}{4}
\int_{{\bf R}^3}Q(f\,|\Dt\vp)(v)dv\,.$$

It can be shown (see Lemma 2.1 below, and the references cited
there) that the following pointwise bound holds:
\begin{equation}\label{cancel}
|\Delta \vp| \le  C|v-v_*|^2\sin\theta\ ,
\end{equation}
where $C$ is a constant depending on the second derivatives of
$\vp$. Moreover, if one first averages $\Delta \vp$ with respect to
the angle around the axis defined by $v -v_*$, one can improve the
right hand side to
$$ C|v-v_*|^2\sin^2\theta\ .$$

For $-2 \le \gamma \le 0$, the factor of $|v-v_*|^2$ is enough to
deal with the factor $|v-v_*|^\gamma$ in (\ref{badprod}), and thus --
neglecting for the moment problems with $b(\cos\theta)$ -- the bound
(\ref{cancel}) provides what is needed to make sense of a weak form
of Eq.(B) for $\gamma$ in this range.

The analysis of this case was initiated by Arkeryd \cite{1}, who
actually considered only $-1\le \gamma < 0$, and it was carried
forward by a number of authors. See \cite{18} for a discussion of
the history.

The case $\gamma < -2$ is more subtle; there is nothing more to be
squeezed out of  $\Delta\vp$ to help with the singularity at $z=0$.
Results in this very soft range were first obtained by
 Villani   \cite{18}.  A key idea
in his work is to use an additional regularity estimate on the
solutions $f$ coming not from the entropy itself, but from the {\em
entropy production}. Later, we shall return to this point in more
detail. Hopefully now at least it is clear where the distinction
between soft and very soft potentials comes from.

There is still the problem  that for inverse power law interactions,
the function $b(\cos\theta)$ is not integrable on ${\bf S}^2$. The
problem comes from a singularity in the small $\theta$ collisions;
i.e., the grazing collisions. Whenever one wishes to consider $Q(f)$
as a difference of two separate integrals -- the {\em gain} and {\em
loss} terms -- it is necessary to impose a {\em Grad angular
cut-off}  which is the assumption that  $b(\cos\theta)$ {\em  is}
integrable on ${\bf S}^2$.

However, for many purposes, this is unnecessary, and one can takes
advantage of the weak form $Q(f|\Delta\vp)$ and the extra factors of
$\sin\theta$ in (\ref{cancel}) and the bound below it. This takes
care of the singularity in $b(\cos\theta)$ for $-3 < \gamma < 1$,
since in this case one has
$$
\int_{{\bS}}B(z,\sg)\sin^2\theta \,d\sg={\rm const.}|z|^{\gm}
<\infty\ . $$

The case $\gm=-3$ is the Coulomb potential, and is therefore of
particular interest. However, in this case
$B(z,\sg)=C_0(\sin(\theta/2))^{-4}|z|^{-3}$, so that
$$
\int_{{\bS}}B(z,\sg)\sin^{2}\theta \,d\sg=\infty\ .$$ This
difficulty with the Coulomb interaction is a genuine part of the
physics, and not a weakness of current technical tools. Without an
angular cut-off, the Boltzmann equation does not make sense for the
Coulomb interaction.  See \cite{18} for further discussion of this,
and what is done in plasma physics to study the kinetics of plasmas
nonetheless.

Here, we stay within the framework of the Boltzmann equation (B)
with $N\ge 2$, and often will simply require of $b(\cos\theta)$ the
mild cut-off hypothesis that $b(\cos\theta)\sin^2\theta$ is
integrable on ${\bf S}^{N-1}$:

In this paper we shall write that  $B(z,\sg)$ satisfies  a {\em mild
angular cut-off}, provided
\begin{equation}\label{(1.1)}
\int_{{\bSN}}B(z,\sg)\sin^2\theta\, d\sg
\le A^*|z|^{\gm}\,,\qquad -4\le \gm<0
\end{equation}
 for some constant
$0<A^*<\infty$. Again, the difference between this and the stronger
Grad angular cut-off  is the factor of $\sin^2\theta$ in the
integral, and the possibility of making such a mild cut-off
assumption in the context of weak solutions has been exploited by a
number of authors; again we refer to \cite{18} for an account of the
history.

In addition to the upper bound in (\ref{(1.1)}), we shall also
sometimes need to invoke a corresponding
 lower bound.
For instance, to prove the moment estimates mentioned above, and prove the
convergence to equilibrium, we assume in addition that $B(z,\sg)>0$
for almost every $(z,\sg)\in{\bRSN}$ and there is a constant
$0<A_*<\infty$ such that
\begin{equation}\label{(1.2)}
\int_{{\bSN}}B(z,\sg)\sin^2\theta d\sg \ge  A_*(1+|z|^2)^{\gm/2}
\,,\qquad -4\le \gm<0\,. \end{equation}

\subsection{Weak solutions of the Boltzmann equation}

Having explained the difference between soft and very soft
potentials, and the kinds of cut-off assumptions we shall consider,
we are ready to introduce the class of weak solutions of Eq.(B) that
we shall study.

Eq.(B) for soft potentials is usually investigated by entropy and
moment methods with working spaces of Lebesgue measurable functions
$f:{\bRN}\to {\bf R}$ \beas&& L^1_0({\bRN})=L^1({\bRN}),\quad
L^1_s({\bRN})=\left\{f\,\,\bigg|\,\,
\|f\|_{L^1_s}:=\int_{{\bRN}}\langle{v}\rangle^{s}|f(v)|
dv<\infty\right\}\,,\qquad s\in{\bf R}\,,\\
&&L^1_s{\rm log}L({\bRN})=\left\{
f\,\,\bigg|\,\,\int_{{\bRN}}\langle{v}\rangle^{s}|f(v)|(1+|\log|f(v)|)
dv<\infty \right\}\eeas where and throughout the paper we use the notation
$$\langle{v}\rangle=(1+|v|^2)^{1/2}\,.$$
 The entropy (Boltzmann $H$-functional)
and the entropy dissipation are given by
$$H(f)=\int_{{\bRN}}f(v)\log f(v) dv\,,\qquad 0\le f\in L^1{\rm log}L({\bRN})\,,$$
\begin{equation}\label{(1.3)}
D(f)=\fr{1}{4}\int_{{\bRRSN}}B(v-v_*,\sg)(f'f_*'-ff_*)
\log\left(\fr{f'f_*'}{ff_*}\right)d\sg dv_* dv\,.
\end{equation} Here and below
we define $(a-b)\log(a/b)=\infty$ if $a>b=0$ or $b>a=0;$
$(a-b)\log(a/b)=0$ if $a=b=0$.

As noted above, in order to establish a weak form of Eq.(B), we need
to be able to make sense of the expression
$$\int_{{\bf S}^{N-1}}B(|v-v_*|,\cos\theta)\Dt\vp\, d\sigma$$
even when, due to the singularity in $B$ at $\theta = 0$, the
integrand is not integrable. As we shall see in Section 2, this can
be done under smoothness assumptions on $\varphi$ provided we {\em
first} integrate over all the variables in ${\bf S}^{N-1}$ except
$\theta$, and {\em then} integrate over $\theta$. That, is with
${\bf k} = (v-v_*)/|v-v_*|$, we can parameterize  ${\bf S}^{N-1}$ by
$(\theta,\omega)\in [0,\pi]\times {\bf S}^{N-2}({\bf k})$ through
$\sigma = \cos(\theta){\bf k} + \sin(\theta)\omega$. Using this
parameterization, and interpreting the integral as an iterated
integral, we shall show in Section 2 that when $\varphi$ is
sufficiently smooth, integrating first in $\omega$ renders the
$\theta$ integral convergent.  On this basis (see Section 2 for
details), we define for all $\vp\in C^2({\bRN})$
\begin{equation}\label{(1.4)}
L[\Dt\vp](v,v_*)
=\int_{0}^{\pi}B(|v-v_*|,\cos\theta)\sin^{N-2}\theta\left(\int_{{\bf
S}^{N-2}({\bf k})}\Dt\vp \,d\og\right)d\theta
\end{equation}
where $\Dt\vp=\Dt\vp(v',v_*',v,v_*)$ is given by (\ref{(1.0)}).

The relevant space ${\cal T}$ of test functions $\varphi$ for which
this construction works is given by
$${\cal T} =
\bigg\{\vp\in C^{2}({\bRN})\,\bigg|\, \sup_{v\in
{\bRN}}\left(\langle{v}\rangle^{-2}|\vp(v)|+\langle{v}\rangle^{-1}|\p\vp(v)|+
|\p^2\vp(v)|\right)<\infty\bigg\}\,,$$ where
$\p\vp(v)=({\p}_{v_1}\vp(v), ...,{\p}_{v_N}\vp(v)),\,
\p^2\vp(v)=\left({\p}^2_{v_iv_j}\vp(v)\right)_{N\times N}$,
${\displaystyle |\p\vp(v)|=\left(\sum_{1\le i\le N} |{\p}_{v_i}
\vp(v)|^2\right)^{1/2}}$ and ${\displaystyle |\p^2\vp(v)|
=\left(\sum_{1\le i,j\le N} |{\p^2}_{v_i v_j}
\vp(v)|^2\right)^{1/2}}$.

As in the previous subsection, we also define, here for all
$\vp\in {\cal T}$,
$$Q(f\,|\Dt\vp)(v)=\int_{{\bRSN}}B(v-v_*,\sg)
(f'f_*'-ff_*)\Dt\vp\, d\sg dv_*\ .$$

By formal calculation we have
$$\int_{{\bRN}}\vp(v)Q(f)(v)dv=-\fr{1}{4}
\int_{{\bRN}}Q(f\,|\Dt\vp)(v)dv
=\fr{1}{2}\int_{{\bRRN}}L[\Dt\vp](v,v_*)ff_* dv_*dv\,.$$

Referring to  Arkeryd \cite{1} and Goudon \cite{11} (for $-1\le
\gm<0$ and $-2\le \gm<0$ respectively) and Villani \cite{18} ( for
$-4<\gm<0$), we introduce

\medskip

\noindent{\bf  Definition of Weak Solutions}. {\it Suppose the
kernel $B$ satisfies (\ref{(1.1)}). Let $0\le f_0\in L^1_2\cap L^1{\rm
log}L^1({\bRN})$. A nonnegative measurable function $f(v,t)$ on
${\bRN}\times[0,\infty)$ is called a weak solution of Eq.(B) with
$f(v,0)=f_0(v)$ if the following ${\rm (i), (ii)}$  hold:

{\rm (i)}\, $f\in L^{\infty}([0,\infty); L^1_2\cap L^1{\rm
log}L^1({\bRN}))$ and
\begin{equation}\label{(1.5)}
H(f(t))+\int_{0}^{t}D(f(\tau))d\tau\le H(f_0)\,,\quad t\ge 0\,.
\end{equation}

{\rm (ii)}\, For all $\vp\in {\cal T}$, if $-4\le \gm<-2$, then
\begin{equation}\label{(1.6)}
\int_{{\bRN}}\vp(v)f(v,t)dv=\int_{{\bRN}}\vp(v)f_0(v)dv -
\fr{1}{4}\int_{0}^{t}d\tau\int_{{\bRN}}Q(f\,|\Dt\vp)(v,\tau)dv\,,\quad
t\ge 0\,;
\end{equation}
and if $-2\le \gm<0$, then
\begin{equation}\label{(1.7)}
\int_{{\bRN}}\vp(v)f(v,t)dv=\int_{{\bRN}}\vp(v)f_0(v)dv +
\fr{1}{2}\int_{0}^{t}d\tau\int_{{\bRRN}}L[\Dt\vp]ff_* dv_*dv\,,\quad
t\ge 0\,.
\end{equation}
 }
\medskip

Note that the particular functions $\vp(v)=1, v_i\,(i=1,2,...,N) $
and $|v|^2$ all belong to ${\cal T}$ and satisfy $\Dt\vp\equiv 0$.
So the above definition implies that {\it every weak solution $f$ of
Eq.(B) conserves the mass, momentum and energy, i.e.}
$$\int_{{\bRN}}(1, v, |v|^2)f(v,t)dv=\int_{{\bRN}}(1, v, |v|^2)
f_0(v)dv\,,\quad t\ge 0\,.$$ It will be seen that the collision
integrals in (ii) are absolutely convergent with respect to the
total measure $d\sg dv_* dv d\tau$ and $dv_*dvd\tau$ respectively
(see Lemma 2.2 below). For very soft potentials, $-4\le \gm<-2$,
this is essentially due to the entropy inequality (\ref{(1.5)}) as first
noted in \cite{18}; the corresponding weak solutions are also called
$H$-solutions.

We shall prove in Section 3 that the integral equations (\ref{(1.6)}) and
(\ref{(1.7)}) are both equivalent to a full and common version like (\ref{(1.6)})
with $\vp\in C^{1}_b({\bRN}\times[0,\infty))\cap
L^{\infty}([0,\infty); C^2_{b}({\bRN}))$ where \beas&&
C^{2}_b({\bRN})=\bigg\{\vp\in C^{2}({\bRN})\,\bigg|\,\sup_{v\in
{\bRN}}(|\vp(v)|+|\p\vp(v)|+|\p^2\vp(v)|)<\infty \bigg\}\,.\eeas The
precise statement of this equivalence is given in the following
proposition:

\noindent{\bf Proposition 1.1}. {\it Suppose the kernel $B$
satisfies (\ref{(1.1)}). Let $0\le f_0\in L^1_2\cap L^1{\rm
log}L^1({\bRN})$, $0\le f\in L^{\infty}([0,\infty); L^1_2\cap L{\rm
log}L({\bRN}))$ satisfy the entropy inequality (\ref{(1.5)}) and
$f|_{t=0}=f_0$. Then the following are equivalent (for total range
$-4\le \gm<0$):

$(a)$ $f$ is a weak solution of Eq.(B).

$(b)$  $f$ satisfies the equation (\ref{(1.6)}) for all $\vp\in
C_b^2({\bRN})$.

$(c)$  $f$ satisfies the following equation: For all $\vp\in
C^{1}_b({\bRN}\times[0,\infty))\cap L^{\infty}([0,\infty);
C^2_{b}({\bRN}))$
$$
\int_{{\bRN}}\vp(v,t)f(v,t)dv=\int_{{\bRN}}\vp(v,0)f_0(v)dv +
\int_{0}^{t}d\tau \int_{{\bRN}}\fr{{\p \vp}(v,\tau)}{\p
\tau}f(v,\tau) dv$$
\begin{equation}\label{(1.8)}
 \qquad \qquad-
\fr{1}{4}\int_{0}^{t}d\tau\int_{{\bRN}}Q(f\,|\Dt\vp)(v,\tau)dv\,,
\qquad  t\ge 0\,.
\end{equation} }
\medskip

The existence of weak solutions has been proven respectively by
Arkeryd \cite{1} for $-1\le \gm<0$, Goudon \cite{11} for  $-2\le
\gm<0$, and Villani \cite{18} for $-4<\gm<0$.   In Proposition 1.2
below we summarize these results, with one improvement: We also treat the case
$\gamma =-4$.

\medskip

\noindent{\bf Proposition 1.2}. {\it Let $B(z,\sg)$ satisfy (\ref{(1.1)}).
Then for any $0\le f_0\in L^1_2\cap L^1{\rm log}L({\bRN})$, the
Eq.(B) has a weak solution $f$ satisfying $f|_{t=0}=f_0$. }
\medskip

We shall provide a proof of Proposition 1.2 in Section 3 below.
Despite the fact that apart from the case $\gamma =-4$, a proof can
be found in the references cited above, there are motivations for
presenting the details here.

First, the only paper covering the range $-4 < \gamma < -2$ is
Villani's \cite{18}, and he bases his analysis on the relation
between the Boltzmann equation and the Landau equation. In fact, he
gives a complete proof for the case of the Landau equation for
$-4<\gm<0$, and then simply discusses the main ideas of proof for
the Boltzmann equation. While the discussion is quite clear, and
while there are good physical reasons for making a connection with
the Landau equation, it is possible to proceed somewhat more
directly for the Boltzmann equation, as we do here: Our proof is
direct, relatively short, complete and covers the case $\gm=-4$. (In
\cite{18}, the hypothesis $-4<\gm $, was used two times for Landau
equation, and hence needed for Boltzmann equation.)

A second reason for presenting a proof here is that to go beyond
$\gamma =-2$, one must use entropy production estimates. We shall
use simple entropy production arguments systematically throughout
the paper, not only to construct weak solutions. But using them to
construct weak solutions for very soft potentials provides an
excellent topic with which to introduce them.

Finally, various approximation procedures that are used in the proof
of existence are also used in our study of convergence to
equilibrium, and for this reason it is quite useful to have them
included explicitly in this paper.

\subsection{The main results}

By changing scales one can assume without loss of generality that
initial data have unite mass, zero momentum and unit temperature,
i.e.
$$f_0\in L^1_{(1,0,1)}({\bRN}):=\bigg\{ 0\le f\in L^1_2({\bRN})\,\bigg|\,
 \int_{{\bRN}}(1,v,\fr{1}{N}|v|^2)f(v)dv=(1,0,1)\bigg\}\,.$$
 The Maxwellian in $L^1_{(1,0,1)}({\bRN})$ is given by
 \begin{equation}\label{(1.9)}
 M(v)=(2\pi)^{-N/2}\exp(-|v|^2/2) \,,\quad v\in {\bRN}\,.
 \end{equation}
To study $L^1$- convergence to equilibrium, we shall use a property
that the $L^1$-distances $\|f-M\|_{L^1}$ and
$\|f-M\|_{L^1_2}$ are almost equivalent \cite{5}:
There is an explicit constant $0<C_N<\infty$ depending only on $N$,
such that
$$\|f-M\|_{L^1_2}\le C_N\|f-M\|_{L^1}\log
\left(\fr{6}{\|f-M\|_{L^1}}\right)\,\qquad \forall\,f\in L^1_{(1,
0,1)}({\bRN})\,.$$ This implies that ( with a different $C_N<\infty$)
\begin{equation}\label{(1.10)}
\|f-M\|_{L^1}\le \|f-M\|_{L^1_2}\le C_N\sqrt{\|f-M\|_{L^1}}\qquad \forall\,f\in
L^1_{(1,0,1)}({\bRN}) \,.
\end{equation}

 Our main results are Theorems 1-3 below; their proofs will be given
 in latter sections.

\medskip

\noindent{\bf Theorem 1}. {\it Let $B(z,\sg)$ satisfy (\ref{(1.1)}) and
(\ref{(1.2)}). For any initial datum $f_0\in L^1_{(1,0,1)}\cap L^1_{s}\cap
L^1{\rm log}L({\bRN})$ with $s>2$, let $f(v,t)$ be a weak solution
of Eq.(B) with $f|_{t=0}=f_0$ . Then

{\rm (I)} {\rm (}Moment Estimates{\rm )}.
\begin{equation}\label{(1.11)}
\|f(t)\|_{L^1_s}\le C_{s}(1+t),
\qquad \fr{1}{t}\int_{0}^{t}\|f(\tau)\|_{L^1_{s+\gm}}d\tau \le C_{s},
\qquad\forall\, t> 0
\end{equation}
 where the constant
$0<C_{s}<\infty$ depends only on $N, \gm, A_*, A^*, s$ and
$\|f_0\|_{L^1_s}$,  and in case $-4\le \gm<-2$, $C_s$ depends also
on $H(f_0)$.

{\rm (II)} {\rm (}Time Averaged Convergence{\rm
)}. If $s> 2+|\gm|$, then
\begin{equation}\label{(1.12)}
\lim_{T\to \infty}\fr{1}{T}\int_{0}^{T}\|f(t)-M\|_{L^1_2} d t=0
\end{equation}
where $M\in L^1_{(1,0,1)}({\bRN})$ is the Maxwellian (\ref{(1.9)}).\quad }
\medskip

\noindent{\bf Remarks:} (1) The moment estimates in (\ref{(1.11)})
were first established by Desvillettes \cite{9} under the Grad
angular cut-off on $B$ with $-1<\gm<0$. Villani \cite{18} then
proved (\ref{(1.11)})
  under the mild cut-off  assumption
(\ref{(1.1)})-(\ref{(1.2)}) for $-4< \gm<0$ and $s\le 4$, and in
\cite{19} he concluded further that $\forall\,s>2,\,\,\exists\,
\ld_s>0$ such that $\|f(t)\|_{L^1_s}\le C_s(1+t)^{\ld_s}$. Here we
prove that $\ld_s\equiv 1$ for all $s>2$ (thanks to the
integrability $\int_{0}^{\infty}D(f(t))dt<\infty$).

(2)  {\em Theorem 2 provides the first convergence results for weak
solutions for very soft potentials without any cut-off, or with very
weak angular cutoff}. (Recall that  (\ref{(1.1)}) holds for free if
$\gamma > -3$). For the usual $L^1$ convergence, it has been proven
in \cite{10} and \cite{17} that $\|f(t)-M\|_{L^1}\le C(1+t)^{-\ld}$
($\ld>0$) only under quite strong cut-off assumptions soft
potentials; e.g., that $z\mapsto B(z,\sg)$ is bounded near $z=0$.

Our next results concern lower and upper bounds of convergence rate
to equilibrium for certain classes of solutions of Eq.(B). We show
that a general lower bound can be obtained for such initial data
$f_0\in L^1_{(1,0,1)}\cap L^1_s\cap L^1{\rm log}L({\bRN})$ that have
energy long-tails:
\begin{equation}\label{(1.13)}
\limsup_{R\to\infty}R^{\beta}\int_{|v|>R}|v|^2 f_0(v)dv=\infty
\end{equation}
where $\beta=\min\{s, \, s-2+|\gm|\}$ for $s\ge 2\,,\, -4\le \gm<0$.
Note that the condition (\ref{(1.13)}) implies that for any constant $K>0$,
the equation
\begin{equation}\label{(1.14)}
({\rm R}(t))^{\beta}\int_{|v|>{\rm R}(t)}|v|^2 f_0(v)dv=K(1+t)^{2-[2/s]}
\,,\quad t\in[0,\infty)
\end{equation}
has a minimal solution ${\rm R}(t)>0$. Here $[x]$ denotes the
largest integer not exceeding $x$.

\medskip

\noindent{\bf Theorem 2}.  {\it Let $f_0\in  L^1_{(1,0,1)}\cap
L^1_s\cap L^1{\rm log}L({\bRN})$ satisfy (\ref{(1.13)}) for some $s\ge 2$,
and let $f(v,t)$ be a weak solution of Eq.(B) with initial datum
$f|_{t=0}= f_0$. Then

{\rm (I)}  For any constant $K_0\in(0,\infty)$ there exists
$K\in[K_0,\infty)$ which depends only on $N, \gm, A^*, A_*, s,
\|f_0\|_{L^1_s}, H(f_0)$ and $K_0$   such that for the minimal
solution ${\rm R}(t)$ of (\ref{(1.14)}) we have
\begin{equation}\label{(1.15)}
\|f(t)-M\|_{L^1_2}\ge \int_{|v|> {\rm R}(t)}|v|^2 f_0(v)dv\qquad
\forall\, t\ge 0\,.
\end{equation}
As a consequence we have the
following explicit lower bounds:

{\rm (II)} Suppose $s>2,\, \beta=\min\{s\,,\,s-2+|\gm|\}$, and
there are constants $s-2<\dt<\beta$ and $0<\vep_0\,,
R_0<\infty$ such that
\begin{equation}\label{(1.16)}
f_0(v)\ge \vep_0 \langle{v}\rangle^{-(N+2+\dt)}
\end{equation}
for all $|v|\ge R_0$. Then there is a computable constant $C>0$ such that
\begin{equation}\label{(1.17)}
\|f(t)-M\|_{L^1_2}\ge C\, (1+t)^{-\ld}\qquad \forall\, t\ge 0
\end{equation}
where $\ld=2\dt/(\beta-\dt) \,.$

{\rm (III)} Suppose $s=2,\,\beta=\min\{2\,,\, |\gm|\} $. Let $A\in
C^1_{b}([0,\infty))$ satisfy
\begin{equation}\label{(1.18)}
\lim_{t\to\infty}A(t)=0\,,\quad \inf_{t\ge
0}(1+t)^{\dt}A(t)>0\,,\quad A_1(t):=-\fr{d}{dt}A(t)\ge 0 \quad {\rm
on}\quad [0,\infty)
\end{equation}
where  $0<\dt <\beta$. Suppose for
some $0<\vep_0, R_0<\infty$
\begin{equation}\label{(1.19)}
f_0(v) \ge \vep_0 |v|^{-(N+1)}A_1(|v|)\qquad \forall\,|v|\ge R_0\,.\
\end{equation}
Then there are constants $0<c, C<\infty$ such that
\begin{equation}\label{(1.20)}
\|f(t)-M\|_{L^1_2}\ge C A(c\, t^{\alpha})\qquad
\forall\, t\ge 0
\end{equation}
where $\alpha=1/(\beta-\dt)$.}

\medskip

There are many initial data $f_0$ that satisfy all
conditions in Theorem 2. For example, in the case $s=2$, one can
choose
$$A(t)=(1+t)^{-\dt}\,,\quad [1+\log(1+t)]^{-1}\,,\quad
[1+\log(1+\log (1+t))]^{-1}\,,...$$ which means that  the rate of
convergence to equilibrium can be arbitrarily slow for $s=2$. This
fact has been observed in \cite{5} for Maxwellian molecules
($\gm=0$) with angular cutoff. Note also that for any initial
datum $f_0\in L^1_{(1,0,1)}({\bRN})$, the mass tail of $f_0$ always
decays at least with algebraic order 2,  $\int_{|v|>R}f_0(v) dv\le N
R^{-2}$, but the energy tail $\int_{|v|>R}|v|^2f_0(v) dv$ may
decay very slowly.
This is why we consider the energy tail (hence
$L^1_2$ norm) rather than the mass tail.

We now turn to upper bounds on the rate of convergence. Here, we
must impose more restrictive conditions on the collision kernel: We
assume that $B(z,\sg)$ satisfies the following cutoff conditions
(with constant $K_*>0$):
\begin{equation}\label{(1.21)}
 K_*(1+|z|^2)^{\gm/2}\le B(z,\sg)\le b(\cos\theta)|z|^{\gm}\,,\quad
-1\le \gm<0\quad
\end{equation}
\begin{equation}\label{(1.22)}
A_0:=|{\bf S}^{N-2}|\int_{0}^{\pi} b(\cos\theta)
\sin^{N-2}\theta \,d\theta<\infty\,.
\end{equation}

\smallskip

\noindent{\bf Theorem 3}. {\it Let $B(z,\sg), \gm $ satisfy
(\ref{(1.21)})-(\ref{(1.22)}) and let $f_0\in L^1_{(1,0,1)}\cap
L^1_s{\rm log}L^1({\bRN})$ with $s>10$. Then there exist a finite
constant $C$ and a weak solution $f(v,t)$ of Eq.(B) with
$f|_{t=0}=f_0$ such that
\begin{equation}\label{(1.23)}
\|f(t)-M\|_{L^1_2}\le C\,(1+t)^{-\ld}\,,\qquad\,t\ge 0
\end{equation}
 where
\begin{equation}\label{(1.24)}
\ld=\fr{s-10}{12}\,>0\,.
\end{equation} }
\medskip

\noindent{\bf Remark:}  In this theorem we do not assume that $f_0$
has any strictly positive pointwise lower bounds, nor shall we make
use of any pointwise lower bounds on the weak solutions.

\medskip

In Theorem 3, if the initial datum satisfies
(\ref{(1.16)}), then the corresponding solution $f$ satisfies both (\ref{(1.17)})
and (\ref{(1.23)}), i.e., the convergence rate to equilibrium satisfies both
upper and lower bounds that are only algebraic:
$$C_1(1+t)^{-\ld_1}\le \|f(t)-M\|_{L^1_2}\le C\,(1+t)^{-\ld}\,,
\qquad\,t\ge 0\,.$$

One of the main tools we use to prove Theorem 3 is an entropy
production bound of Villani \cite{20} for {\em super hard}
potentials; i.e., $\gamma = +2$.  As Villani showed in \cite{20},
for super hard potentials, there is an especially nice inequality
relating the entropy production and the relative entropy. And
moreover, while super hard potentials are themselves non-physical,
one can use the super hard entropy production bound to obtain
entropy production bounds for physically interesting hard
potentials, using {\em moment bounds} and {\em pointwise lower
bounds} on the solutions.

Our Theorem 1 provides moment bounds for soft potentials  that are
good enough to proceed with an adaptation of this part of Villani's
argument to soft potentials, but the pointwise lower bounds are more
problematic in this setting.

The pointwise lower bounds enter Villani's argument as follows: To
estimate the entropy production $D(f)$ for $\gamma < 2$ in terms of
${\cal D}_2(f)$, the entropy production for $\gamma =2$, a simple
H\"older argument explained in Section 7 leads to the consideration
of the quatitiy
$${\cal D}_k(f)=\fr{1}{4}\int_{{\bRRSN}}(1+|v-v_*|^2)^{k/2}
(f'f_*'-f f_*)\log\left(\fr{f'f_*'}{ff_*}\right) d\sg dvdv_*$$ for
$k>2$. It is easy to see that ${\cal D}_k(f)$ can be estimated in
terms of $L^1$ bounds on $\langle v\rangle^k f(v)\log f(v)$, and
clearly the negative part of this function is integrable if $f$
satisfies a bound of the type $f(v) \ge C e^{-c|v|^p}$ and $f$ has
moments of sufficiently high orders. The details are somewhat more
complicated than indicated in this sketch, but the sketch should
nonetheless give a fair indication of the interplay between moment
bounds and pointwise lower bounds in Villani's arguments.  While
suitable pointwise lower bounds are available for the hard
potentials that Villani considers, they are not available for soft
potentials, at least not for the sort of general initial data that
we wish to consider.

A novel element in our proof of Theorem 3 is a strategy for avoiding
any pointwise bounds which is explained in Section 7. Given a
solution $f(v,t)$ of Eq.(B), we define the function $g(v,t)$ by
$$g(v,t) = (1 - e^{-t-1})f(v,t) + e^{-t-1}M(v)\ .$$
Evidently, $g(v,t)$ has good pointwise lower bounds by construction.
Although $g$ is not itself a solution to Eq.(B), it is closely
related enough to one, namely $f$, that we shall use Villani's
entropy production inequality for super hard potentials to prove and
an entropy production inequality relating $D(g)$ and $H(g|M)$ and
show that  $H(g|M)$ tends to zero at any polynomial rate provided
the initial datum $f_0$ has sufficiently many moments.

The most technically involved part of the proof is the demonstration
that the $L^1$ norm of $\langle v\rangle^k g(v,t)\log g(v,t)$ is
bounded by a constant multiple of $(1+t)^2$, again , provided the
initial data has sufficiently many moments. This approach to proving
and using such {\em entropic moment estimates} is one of the more
novel features of this paper, and may well have other applications.
The condition $\gamma \ge -1$ is used in this part of the proof of
Theorem 3.

We thank the referee of this paper who encouraged us to work harder
on the proof of Theorem 3, and relax the stronger conditions on the
initial data that we had imposed in a previous version, and we thank
this referee for his many other useful remarks and suggestions as well.

\section{Basic Lemmas Concerning Collision Integrals}

In this section we  collect some lemmas that will be used to
ensure integrability of certain collision integrals, as well as to
estimate others in terms of entropy dissipation.

There are nothing fundamentally new in this section
except some quantitative improvements and  mild generalizations of some
lemmas that can be found in
 previous works by Goudon [13] and Villani
[21], and references they cite.

The first lemma justifies the definition of
$L[\Delta\varphi](v,v_*)$ that figures in our definition of weak
solutions. We begin with a more complete explanation of the notation
used in the definition of $L[\Delta\varphi](v,v_*)$ that we have
given in the introduction.

Let
$${\bf
k}=\fr{v-v_*}{|v-v_*|}\quad {\rm if}\quad  v\neq v_*\,;\quad {\bf k}=
{\bf e}_1=(1,0,...,0)\quad {\rm if}\quad v=v_*\,.
$$
Under the spherical coordinate transform $\sg=\cos\theta\,{\bf
k}+\sin\theta \,\og\,, \theta\in[0,\pi]\,, \og\in {\bf S}^{N-2}({\bf
k})$ we have
\begin{equation}\label{(2.1)}\left\{\begin{array} {ll}v'
= \cos^2(\theta/2)v+\sin^2(\theta/2)v_*
+\fr{1}{2}|v-v_*|\sin\theta\,\og\,,\\ \\
v_*'= \sin^2(\theta/2)v+\cos^2(\theta/2)v_*-\fr{1}{2}|v-v_*|\sin\theta\,\og\,
\end{array}\right.\quad \og\in{\bf S}^{N-2}({\bf k})
\end{equation}
\begin{equation}\label{(2.2)}
|v'-v|=|v_*'-v_*|=|v-v_*|\sin(\theta/2)\,,\quad
|v'-v_*|=|v_*'-v|=|v-v_*|\cos(\theta/2)\,.
\end{equation}
Here
$${\bf S}^{N-2}({\bf k})=\{\og\in {\bSN}\,\,|\,\,
\langle{\og, {\bf k}\rangle}=0\,\}\,
\quad (\,N\ge 3)\,;\qquad {\bf S}^{0}({\bf k})=\{ -{\bf k}^{\bot}\,,\,{\bf
k}^{\bot}\}\,\quad (N=2) $$ where
${\bf k}^{\bot}\in {\bf S}^1$ satisfies
$\langle{{\bf k}^{\bot}, {\bf k}}\rangle=0$.
Also, for any  $F\in L^1({\bSN})$ or any measurable function $F\ge 0$
on ${\bSN}$, we have
$$\int_{{\bSN}}F(\sg)d\sg=\int_{0}^{\pi}\sin^{N-2}\theta
\left(\int_{{\bf S}^{N-2}({\bf
k})}F(\cos\theta {\bf k}+\sin\theta\,\og)d\og\right) d\theta$$
and in case $N=2$ we define $$\int_{{\bf S}^{0}({\bf
k})}g(\og) d\og=g(-{\bf k}^{\bot})+g({\bf k}^{\bot})\,.$$
Let $|{\bf S}^{N-2}({\bf k})|=\int_{{\bf S}^{N-2}({\bf k})}d\og$,
etc. Then $|{\bf S}^{N-2}({\bf k})|=|{\bf S}^{N-2}|$ for $N\ge 3$,
$|{\bf S}^{0}({\bf k})| =|{\bf S}^{0}|=2$ for $N=2$.

\noindent{\bf Lemma 2.1}. {\it Let $\vp\in C^2({\bRN})\,,
\Dt\vp=\vp(v')+\vp(v_*')-\vp(v)-\vp(v_*)$.  Then for all
 $\sg\in{\bSN}$, $v,v_*\in{\bRN}$
\begin{equation}\label{(2.3)}
|\Dt\vp|\le
2^{(4-3m)/2}\bigg(\sup_{|u|\le
\sqrt{|v|^2+|v_*|^2}}|\p^m\vp(u)|\bigg)|v-v_*|^m\sin\theta\,,\,\quad m=1,2\,;
\end{equation}
\begin{equation}\label{(2.4)}
\fr{1}{|{\bf S}^{N-2}|}\left|\int_{{\bf S}^{N-2}({\bf
k})}\Dt\vp \,d\og\right|\le \bigg(\sup_{|u|\le
\sqrt{|v|^2+|v_*|^2}}|\p^2\vp(u)|\bigg)|v-v_*|^{2}\sin^2\theta\,.
\end{equation}
}

\medskip

\noindent{\bf Proof}.  Observing that $-\sg=\cos(\pi-\theta)\,{\bf
k}+\sin(\pi-\theta)(-\og)$ and $\Dt\vp$ is
invariant under the reflection  $\sg \to-\sg$, we can assume without
loss of generality that $\theta\in [0,\pi/2]$. In this case we have
$\sin(\theta/2) \le (\sin\theta)/\sqrt{2}$.

By writing
$\Dt\vp=(\vp'-\vp)+(\vp_*'-\vp_*)$ one sees that (\ref{(2.3)}) for $m=1$ follows from the
first equality in (\ref{(2.2)}).
Next writing
$\Dt\vp=(\vp'-\vp)-(\vp_*-\vp_*')$ and using $v_*-v_*'=v'-v$, we compute
\beas&&\Dt\vp=\int_{0}^{1}\langle{\p\vp(v+t(v'-v))-\p\vp(v_*'+t(v'-v)),
v'-v\rangle}dt\\
&&=\int_{0}^{1}\!\!\!\int_{0}^{1}(v-v_*')\p^2\vp(\xi_{t,\tau})
(v'-v)^{T}d\tau dt\eeas with $|\xi_{t, \tau}|\le \max\{|v|, |v'|, |v_*|,
|v_*'|\}\le  \sqrt{|v|^2+|v_*|^2}\,.$ Since
$|v_*'-v||v'-v|=\fr{1}{2}|v-v_*|^2\sin\theta$, this gives (\ref{(2.3)})
for $m=2$.  To prove (\ref{(2.4)}) we write
$\Dt\vp=(\vp'-\vp)+(\vp_*'-\vp_*)$ and use $v_*'-v_*=-(v'-v)$. Then
\beas && \Dt\vp=\langle{\p\vp (v)-\p \vp (v_*), v'-v}\rangle\\
&&+\int_{0}^{1}(1-t)(v'-v)\p^2\vp(v+t(v'-v))(v'-v)^{T}dt\\
&&+\int_{0}^{1}(1-t)(v_*'-v_*)\p^2\vp(v_*+t(v_*'-v_*))(v_*'-v_*)^{T}dt\,.
\eeas Since by (\ref{(2.1)})
\beas&&\fr{1}{|{\bf S}^{N-2}|}\int_{{\bf S}^{N-2}({\bf
k})}\langle{\p\vp (v)-\p \vp (v_*), v'-v}\rangle d\og = \langle{\p\vp
(v)-\p\vp(v_*), v_*-v}\rangle\sin^2(\theta/2) \eeas where we
used $\int_{{\bf S}^{N-2}({\bf
k})}\langle{{\p }\vp(v)-{\p }\vp(v_*),\og}\rangle d\og=0$, it follows that
\beas&&\fr{1}{|{\bf S}^{N-2}|}\bigg|\int_{{\bf S}^{N-2}({\bf k})}
\Dt\vp \,d\og\bigg|\le 2\bigg(\sup_{|u|\le
\sqrt{|v|^2+|v_*|^2}}|{\p }^2\vp(u)|\bigg)\,|v-v_*|^2
\sin^2(\theta/2)\,.\eeas

\noindent$\Box$
\smallskip

 Our next lemma provides bounds on
certain collision integrals in terms of entropy dissipation. As in
the case of the local Sobolev bounds on the collision kernel first
proved by Lions \cite{Lions}, and then extended in subsequent work
\cite{V,ADVW}, the proof of our bounds depends on the pointwise
inequality (\ref{(2.10)}) below. However, as our bounds do not
involve local Sobolev norms, the proof is somewhat simpler.
\medskip

 \noindent{\bf Lemma 2.2}. {\it Let $B=B(v-v_*,\sg)$ satisfy (\ref{(1.1)}),
$0\le f\in L^1_2({\bRN})$ satisfy $D(f)<\infty$. Then:
\medskip

\noindent{\rm (I)} For any nonnegative measurable function $\Psi$
 on ${\bRRN}$ satisfying
$$\Psi(v',v_*')=\Psi(v,v_*)\quad \forall\, (v,v_*,\sg)\in {\bRRSN}$$ we have
$$\int_{{\bRRSN}}B\,\Psi(v,v_*)\sin\theta \,
|f'f_*'-ff_*|d\sg dv_*dv$$
\begin{equation}\label{(2.5)}
 \le \left(4
A^*\int_{{\bRRN}}[\Psi(v,v_*)]^2|v-v_*|^{\gm} ff_*
dv_*dv\right)^{1/2}\sqrt{D(f)}
\end{equation}
 where $A^*$ is the
constant in (\ref{(1.1)}).

\noindent({\rm II}) Let $m\in\{1,2\}$ be such that $ 0\le 2m+\gm\le 2$. Then
\begin{equation}\label{(2.6)}
\int_{{\bRRSN}}B |v-v_*|^{m}\sin\theta
|f'f_*'-ff_*|d\sg dv_*dv\le \sqrt{4A^*}\|f\|_{L^1_2}\sqrt{
D(f)}
\end{equation} and consequently for all $\vp\in C^2_b({\bRN})$
\begin{equation}\label{(2.7)}
\int_{{\bRRSN}}B|\Dt\vp|
|f'f_*'-ff_*|d\sg dv_*dv\le
\sqrt{8A^*}\|\p^m\vp\|_{L^{\infty}}\|f\|_{L^1_2}\sqrt{
D(f)}\,.
\end{equation}

\noindent({\rm III)} For all $\vp\in {\cal T}$,
if $-4\le \gm<-2$, then
\begin{equation}\label{(2.8)}
\int_{{\bRRSN}}B |\Dt\vp ||f'f_*'-ff_*|d\sg dv_*
dv\le \sqrt{A^*}\|\p^2\vp\|_{L^{\infty}}\|f\|_{L^1_2}\sqrt{
D(f)}
\end{equation}
and if $-2\le \gm<0$, then
\begin{equation}\label{(2.9)}
\int_{{\bRRN}}
\int_{0}^{\pi}B(|v-v_*|,\cos\theta)\sin^{N-2}\theta \left|\int_{{\bf
S}^{N-2}({\bf k})}\Dt\vp\, d\og\right|d\theta\,
ff_* dv_*dv \le A^*\|\p^2\vp\|_{L^{\infty}}\|f\|_{L^1_2}^2
\end{equation}

}

\noindent{\it Proof}. Applying the elementary inequality
\begin{equation}\label{(2.10)}
|a-b|\le( \sqrt{a}+\sqrt{b}\,)\sqrt{\fr{1}{4}(a-b)\log
(\fr{a}{b})}\,,\quad a,b\ge 0
\end{equation}
 to $a=f'f_*', b=ff_*$ and using
Cauchy-Schwarz inequality we have
\beas&&\int_{{\bRRSN}}B\,\Psi(v,v_*)\sin\theta\,
|f'f_*'-ff_*|d\sg dv_*dv\\
&&\le
\left(4\int_{{\bRRN}}\left(\int_{{\bSN}}B\,\sin^2\theta\,d\sg\right)
[\Psi(v,v_*)]^2ff_*\, dv_*dv\right)^{1/2}\sqrt{D(f)}\eeas which
gives (\ref{(2.5)}) by assumption (\ref{(1.1)}). The condition $0\le
2m+\gm\le 2$ implies $|v-v_*|^{2m+\gm}\le
\langle{v}\rangle^2\langle{v_*}\rangle^2$. So applying (\ref{(2.5)})
to $\Psi(v,v_*)=|v-v_*|^{m}$ gives (\ref{(2.6)}). The inequality
(\ref{(2.7)}) follows from (\ref{(2.3)}),(\ref{(2.4)}) and
(\ref{(2.6)}). The inequality (2.8) follows from (\ref{(2.3)}) and
(\ref{(2.6)}) with $m=2$. Finally from (\ref{(2.4)}) and $|{\bf
S}^{N-2}|\int_{0}^{\pi}B(|v-v_*|,\cos\theta)\sin^{N}\theta
d\theta\le A^*$
 we have
$$\int_{0}^{\pi}B(|v-v_*|,\cos\theta)\sin^{N-2}\theta
\left|\int_{{\bf S}^{N-2}({\bf k})}\Dt\vp\, d\og\right|d\theta\\
\le \|\p^2\vp\|_{L^{\infty}}A^*|v-v_*|^{2+\gm}$$
and
$|v-v_*|^{2+\gm}\le \langle{v}\rangle^2\langle{v_*}\rangle^2$ when $-2\le \gm<0$.
This gives (\ref{(2.9)}).
$\quad \Box$
\smallskip

The last lemma in this section justifies the equalities
resulting from formal calculation that are cited just above the
definition of weak solutions, at least for certain cutoff parts of
the collision integrals -- which is just what we shall need in the
next section.

\medskip

\noindent{\bf Lemma 2.3}. {\it Suppose $B(z,\sg)$ satisfies
(\ref{(1.1)}). For any $\ld>0$, let
\begin{equation}\label{(2.11)}
B^{\ld}(z,\sg)={\bf 1}_{\{|z|\le \ld\}} B(z,\sg)\,,\quad B_{\ld}(z,\sg)
={\bf 1}_{\{|z|>\ld\}} B(z,\sg)
\end{equation}
and let $Q(\cdot), Q^{\ld}(\cdot), L_{\ld}[\cdot]$ be the operators
corresponding to the kernels $B(z,\sg), B^{\ld}(z,\sg)$ and
$B_{\ld}(z,\sg)$ respectively. Then for all $0\le f\in
L^1_2({\bRN})$ satisfying $D(f)<\infty$ we have

{\rm (I)} If $-4\le \gm<-2$ then for any $\vp\in {\cal T} $ and any
$\ld>0$
\begin{equation}\label{(2.12)}
\int_{{\bRN}}Q(f\,|\Dt\vp)(v) dv
=\int_{{\bRN}}Q^{\ld}(f\,|\Dt\vp)(v) dv
-2\int_{{\bRRN}}L_{\ld}[\Dt\vp]ff_* dv_*dv\,.
\end{equation}

{\rm (II)} If $-2\le \gm<0$ then for all  $\vp\in
C^{2}_{b}({\bRN})$
\begin{equation}\label{(2.13)}
\int_{{\bRN}}Q(f\,|\Dt\vp)(v) dv =-2\int_{{\bRRN}}L[\Dt\vp]ff_*
dv_*dv\,.
\end{equation}
 }

\noindent{\it Proof}.  (I) Suppose $-4\le \gm<-2$. Given $\vp\in {\cal T}$.
By Lemma 2.1 we have
$|\Dt\vp(v', v_*', v, v_*)|\le
\|\p^2\vp\|_{L^{\infty}}|v-v_*|^{2}\sin\theta\,.$ So applying (\ref{(2.6)})
 with $m=2$ gives
\begin{equation}\label{(2.14)}
\int_{{\bRRSN}}B(v-v_*,\sg) |\Dt\vp ||f'f_*'-ff_*|d\sg dv_*
dv< \infty
\end{equation}
and thus
\begin{equation}\label{(2.15)}
\int_{{\bRN}}Q(f\,|\Dt\vp)(v) dv
=\int_{{\bRN}}Q^{\ld}(f\,|\Dt\vp)(v) dv +\int_{{\bRN}}Q_{\ld}(f\,|\Dt\vp)(v) dv
\end{equation}
with $Q^{\ld}(\cdot), Q_{\ld}(\cdot)$ corresponding to $B^{\ld}z,\sg)$
and $B_{\ld}(z,\sg)$.
Introduce further truncation
$$B_{\ld,\vep}(v-v_*,\sg)={\bf 1}_{\{\sin\theta>\vep\}}B_{\ld}(v-v_*,\sg)\,,
\qquad \vep>0
$$
and let $Q_{\ld,\vep}(\cdot), L_{\ld,\vep}[\cdot]$ correspond to
$B_{\ld,\vep}(z,\sg)$. Then using (\ref{(2.14)}) and dominated convergence
we have
\begin{equation}\label{(2.16)}
\int_{{\bRN}}Q_{\ld}(f\,|\Dt\vp)(v)dv
=\lim_{\vep\to 0+}\int_{{\bRN}}Q_{\ld,\vep}(f\,|\Dt\vp)(v)dv\,.
\end{equation}
By Lemma 2.1 and $B_{\ld,\vep}\le \fr{1}{\vep}B_{\ld}\sin\theta$ we have
$$B_{\ld,\vep}(v-v_*,\sg)|\Dt\vp |\le
\fr{\|\p^2\vp\|_{L^{\infty}}}{2\vep}|v-v_*|^2 {\bf 1}_{\{|v-v_*|>\ld\}}
B(v-v_*,\sg)\sin^2\theta\,.$$
Since $|v-v_*|^{2+\gm}{\bf 1}_{\{|v-v_*|>\ld\}}\le \ld^{2+\gm}$,
it follows that
\beas&&\int_{{\bRRSN}}B_{\ld,\vep}|\Dt\vp
|ff_*d\sg
dv_*dv\le \fr{A^*}{2\vep}\|\p^2\vp\|_{L^{\infty}}\ld^{2+\gm}\|f\|_{L^1}^2 <\infty\,.\eeas
This allows us to use the
standard derivation and obtain
\begin{equation}\label{(2.17)}
 \int_{{\bRN}}Q_{\ld,\vep}(f\,|\Dt\vp)(v) dv
=-2\int_{{\bRRN}}L_{\ld,\vep}[\Dt\vp]ff_* dv_*dv\,.
\end{equation}
Also by Lemma 2.1\beas&&
|L_{\ld}[\Dt\vp](v,v_*)|\,,\, \,\,\sup_{\vep>0}|L_{\ld,\vep}[\Dt\vp](v,v_*)|\\
&&\le \int_{0}^{\pi}B_{\ld}(|v-v_*|,\cos\theta)\sin^{N-2}\theta
\left|\int_{{\bf S}^{N-2}({\bf k})}\Dt\vp\,d\og\right|d\theta \le
A^*\|\p^2\vp\|_{L^{\infty}}\ld ^{2+\gm}\,. \eeas Therefore using
dominated convergence gives
$$\lim_{\vep\to 0+}\int_{{\bRRN}}
L_{\ld,\vep}[\Dt\vp]ff_* dv_*dv
=\int_{{\bRRN}}L_{\ld}[\Dt\vp]ff_* dv_*dv \,.$$ This together
with (\ref{(2.15)}), (\ref{(2.16)}) and (\ref{(2.17)}) proves (\ref{(2.12)}).

(II) Suppose $-2\le \gm<0$.
Consider $B_{\vep}(z,\sg)={\bf 1}_{\{\sin\theta>\vep\}}B(z,\sg)$
and let $Q_{\vep}(\cdot),\, L_{\vep}[\cdot]$ correspond to $B_{\vep}(z,\sg)$.
For any  $\vp\in C^{2}_b({\bRN})$  we have, by Lemma 2.2 (use (\ref{(2.7)})
with $m=1$) and dominated convergence, that
\begin{equation}\label{(2.18)}
\int_{{\bRN}}Q(f\,|\,\Dt\vp)(v)dv=\lim_{\vep\to
0+}\int_{{\bRN}}Q_{\vep}(f\,|\,\Dt\vp)(v)dv\,.
\end{equation}
 As shown above using $B_{\vep}\le \fr{1}{\vep}B\sin\theta$ and
 $|v-v_*|^{2+\gm}\le \langle{v}\rangle^2\langle{v_*}\rangle^2$ we have
\beas&&\int_{{\bRRSN}}B_{\vep}|\Dt\vp
|ff_* d\sg
dv_*dv<\infty\eeas
hence
\begin{equation}\label{(2.19)}
\int_{{\bRN}}Q_{\vep}(f\,|\,\Dt\vp)(v)dv
=-2\int_{{\bRRN}}L_{\vep}[\Dt\vp](v,v_*)ff_* dv_*dv\ .
\end{equation}
Since
$0\le B_{\vep}\le B$ and $B_{\vep}\to B\,\,(\vep\to 0)$ pointwise,
it follows from
(\ref{(2.9)}) and dominated convergence that
$$\lim_{\vep\to 0}\int_{{\bRRN}}L_{\vep}[\Dt\vp]ff_* dv_*dv
= \int_{{\bRRN}}L[\Dt\vp]ff_* dv_*dv
\,.$$ This together with (\ref{(2.18)}) and (\ref{(2.19)})
proves (\ref{(2.13)}). $\Box$

\section{Construction of Weak Solutions, and the Equivalence of Two Definitions of 
Weak Solutions }

In this section we prove Proposition 1.1 (equivalence) and Proposition 1.2 (existence).

\medskip
\noindent{\bf Proof of Proposition 1.1.} First, $``
(a)\Longrightarrow (b)"$ is a consequence of $C^2_b({\bRN}) \subset
{\cal T}$ and part (II) of Lemma 2.3. To prove $``
(b)\Longrightarrow (a)"$,
 we consider
approximation. Let $\chi\in C^{\infty}_c({\bRN})$ satisfy
\begin{equation}\label{(3.1)}
0\le \chi\le 1\quad {\rm on}\quad {\bRN}\,,\quad \chi(v)=1\quad
\forall\, |v|\le 1\,;\quad \chi(v)=0\quad \forall\,
|v|>2\,.
\end{equation}
 Given any $\vp\in
{\cal T}$. Let $\vp_n(v)=\vp(v)\chi(v/n)$,
$n\ge 1$. It is easily seen that $\{\vp_n\}\subset C^{2}_b({\bRN})$
and
$$\sup_{n\ge 1}\sup_{v\in{\bRN}}\left(\langle{v}\rangle^{-2}|\vp_n(v)|+
\langle{v}\rangle^{-1}|\p\vp_n(v)|+
|\p^2\vp_n(v)|\right)
<\infty.$$
Since $|\vp_n(v)|\le |\vp(v)|$ and $\vp_n(v)\to \vp(v)$\, $\forall\,
v\in {\bRN}$, it follows  that
$$
\int_{{\bRN}}\vp(v)f(v,t)dv=\int_{{\bRN}}\vp(v)f_0(v)dv -
\fr{1}{4}\lim_{n\to\infty}\int_{0}^{t}d\tau\int_{{\bRN}}
Q(f\,|\,\Dt\vp_n)(v,\tau)dv\,,
 \qquad  t\ge 0\,.$$
Assume $-4\le \gm<-2$. Since
$$
\Dt\vp_n\to \Dt\vp \quad (\,n\to\infty\,)\quad \forall\,
(v,v_*,\sg)\in {\bRRSN}$$ and $\sup\limits_{n\ge 1}|\Dt\vp_n|\le
C_{\vp}|v-v_*|^2\sin\theta$, it follows from part (III) of Lemma 2.2
and dominated convergence that
$$\lim_{n\to\infty}\int_{0}^{t}d\tau\int_{{\bRN}}Q(f\,|\,\Dt\vp_n)(v,\tau)dv
=\int_{0}^{t}d\tau\int_{{\bRN}}Q(f\,|\,\Dt\vp)(v,\tau)dv \,.$$
Next assume $-2\le \gm<0$. By part (II) of Lemma 2.3  we have
\beas&& -\int_{0}^{t}d\tau\int_{{\bRN}}Q(f\,|\,\Dt\vp_n)(v,\tau)dv =
2\int_{0}^{t}d\tau\int_{{\bRRN}} L[\Dt\vp_n]ff_* dv_* dv\eeas
Since
$\lim\limits_{n\to\infty}L[\Dt\vp_n](v,v_*)
=L[\Dt\vp](v,v_*)$ and $$
|L[\Dt\vp](v,v_*)|\,,\, \sup_{n\ge
1}L[\Dt\vp_n](v,v_*)|\le C_{\vp}|v-v_*|^{2+\gm}\le
C_{\vp}\langle{v}\rangle^{2+\gm}\langle{v_*}\rangle^{2+\gm}$$
 it follows from dominated convergence that
\beas&&-\lim_{n\to\infty}\int_{0}^{t}d\tau\int_{{\bRN}}Q(f\,|\,\Dt\vp_n)dv
=2\int_{0}^{t}d\tau\int_{{\bRRN}}L[\Dt\vp]ff_* dv_*dv\,.\eeas
Therefore $f$ is a weak
solution.

Now we are going to prove ``$(b)\Longleftrightarrow(c)"$.
``$(b)\Longleftarrow(c)"$
is trivial.  To prove ``$(b)\Longrightarrow(c)"$ we denote for
notation convenience that
$$Q(f\,|\Dt\vp(s,\cdot))(v,\tau)=Q(f(\tau)\,|\Dt\vp(s))(v)\,
\,,\quad \int_{{\bRN}}g(t)dv=\int_{{\bRN}}g(v,t)dv\,.$$
Given any $\vp\in
C^{1}_b({\bRN}\times[0,\infty))\cap L^{\infty}([0,\infty);
C^2_{b}({\bRN}))$. By Lemma 2.1
and Lemma 2.2 there is $m\in\{1,2\}$ such that
\begin{equation}\label{(3.2)}
|\Dt\vp(v',v_*'v,v_*, t)|\le C_{\vp}|v-v_*|^{m}\sin\theta\,,
\end{equation}
\begin{equation}\label{(3.3)}
\int_{t_1}^{t_2}ds\int_{{\bRRSN}}B\,|v-v_*|^{m}\sin\theta
|f'f_*'-ff_*|d\sg dv_*dv\le C_f\int_{t_1}^{t_2}\sqrt{
D(f(s))}\,ds
\end{equation}
 for all $0\le t_1<t_2<\infty$, where
$C_{\vp}=2\sup_{t\ge 0}\|\p ^m_{v}\vp(\cdot,t)\|_{L^{\infty}}$,
$C_f=\sqrt{4A^*}\sup_{t\ge 0}\|f(t)\|_{L^1_2}$. Applying (\ref{(1.6)}) to
the test function $v\mapsto \vp(v,t_2)$ we obtain
$$\int_{{\bRN}}\vp(t_2)f(t_2)dv=\int_{{\bRN}}\vp(t_2)f(t_1)dv -
\fr{1}{4}\int_{t_1}^{t_2}d\tau\int_{{\bRN}}Q(f(\tau)\,|\Dt\vp(t_2))dv\,.
$$
This gives
\beas&&\int_{{\bRN}}\vp(t_2)f(t_2)dv-\int_{{\bRN}}\vp(t_1)f(t_1)dv \\
&&=\int_{{\bRN}} \left(\vp(t_2)-\vp(t_1)\right)
f(t_1)dv-\fr{1}{4}\int_{t_1}^{t_2}d\tau\int_{{\bRN}}Q(f(\tau)\,|\Dt\vp(t_2))dv
\eeas which implies by (\ref{(3.2)}) and (\ref{(3.3)}) that $t\mapsto
\int_{{\bRN}}\vp(t)f(t)dv$ is continuous on $[0,\infty)$. Choose
$t_1=s,\, t_2=s+h, 0<h<1$. Taking integration with respect to $s\in
[0,t]$ and changing variables we compute \beas&&
\fr{1}{h}\int_{t}^{t+h}\int_{{\bRN}}\vp(s)f(s)dvds-\fr{1}{h}\int_{0}^{h}
\int_{{\bRN}}\vp(s)f(s)dv ds \\
&&=\int_{0}^{t}ds\int_{{\bRN}}\fr{1}{h}\left(\vp(s+h)-\vp(s)\right)
f(s)dv-\fr{1}{4}I(t,h)\,,
\eeas
\beas&&I(t,h):=\int_{0}^{1}d\tau\int_{0}^{t}ds\int_{{\bRN}}Q(f(s+\tau h)\,|\Dt\vp(s+h))dv\\
&&= \int_{0}^{1}d\tau\int_{0}^{t+1}ds\int_{{\bRN}}{\bf 1}_{\{\tau h\le
s\le t+\tau h\}} Q(f(s)\,|\Dt\vp(s+(1-\tau)h))dv\,.\eeas
 Since
$${\bf 1}_{\{\tau h\le
s\le t+\tau h\}}\Dt\vp(v',v_*'v,v_*, s+(1-\tau)h)\to {\bf 1}_{\{0\le
s\le t\}}\Dt\vp(v',v_*'v,v_*, s)\quad (h\to 0)$$ for almost
every  $(v,v_*,\sg,s, \tau)\in{\bRRSN}\times[0,t+1]\times[0,1]$, it
follows from (\ref{(3.2)}), (\ref{(3.3)}) and dominated convergence that
$$I(t,h)\to
\int_{0}^{t}ds\int_{{\bRN}}Q(f(s)\,|\Dt\vp(s))dv\qquad (\,h\to 0\,)\,.$$
Therefore
 \beas&&\int_{{\bRN}}\vp(t)f(t)dv-\int_{{\bRN}}\vp(0)f_0dv\\
 &&=\int_{0}^{t}ds\int_{{\bRN}}
({\p}_{s}\vp(s))
f(s)dv-\fr{1}{4}\int_{0}^{t}ds\int_{{\bRN}}Q(f(s)\,|\Dt\vp(s))dv
\eeas for all $t\in(0,\infty)$\,. Hence, $f$ satisfies (\ref{(1.8)}).
$\quad \Box$
\smallskip

\noindent{\bf Proof of Proposition 1.2}. As usual, we shall use
approximate solutions. For every $n\in{\bf N}$, let
$B_n=\min\{B\,,\,n\}$ and let $Q_n(\cdot),D_n(\cdot)$ and
$L_n[\cdot] $ correspond to the kernel $B_n$. It is well-known that
for every $n$ there is a unique strong (or mild) solution $f^n(v,t)$
of Eq.(B) with the kernel $B_n$ and the initial datum
$f^n|_{t=0}=f_0$. And $f^n(v,t)$ conserves the mass, momentum, and
energy and satisfies the entropy inequality
\begin{equation}\label{(3.4)}
H(f^n(t))+\int_{0}^{t}D_n(f^n(\tau))d\tau\le H(f_0)\,,\quad t\ge
0\,.
\end{equation}
 These imply  $\sup\limits_{n\ge 1\,,\,t\ge
0}\int_{{\bRN}}f^n(v,t)(\langle{v}\rangle^2+|\log f^n(v,t)|)dv<\infty$.

 Since $f^n$ are also weak solutions, they satisfy
equation (\ref{(1.6)}) which together with (\ref{(2.7)}) and (\ref{(3.4)})
 imply that for all
$\vp\in C^2_b({\bRN})$ and all $|t_1-t_2|\le 1$
\begin{equation}\label{(3.5)}
\sup_{n\ge 1}\left|\int_{{\bRN}}\vp(v)f^n(v,t_1)dv-\int_{{\bRN}}\vp(v)
f^n(v,t_2)dv\right|
\le C\|\p^2\vp\|_{L^{\infty}}|t_1-t_2|^{1/2}\,.
\end{equation}
 Here $C$
depends only on $A^*, \|f_0\|_{L^1_2}$ and $H(f_0)$. From this we
have for any $\psi\in L^{\infty}({\bRN})$
$$\sup_{|t_1-t_2|\le \dt}\sup_{n\ge 1}
\left|\int_{{\bRN}}\psi(v)f^n(v,
t_1)dv-\int_{{\bRN}}\psi(v)f^n(v,t_2)dv\right| \to 0\quad {\rm
as}\quad \dt\to 0+\,.$$  By a standard argument, there exist a
subsequence of $\{f^n\}$ (still denoted by $\{f^n\}$), and a
$(v,t)$-measurable function $0\le f\in L^{\infty}([0,\infty);
L^1_2\cap L^1{\rm log}L({\bRN}))$ such that
\begin{equation}\label{(3.6)}
\forall\, t\ge 0\qquad f^n(\cdot,t)\rightharpoonup f(\cdot,t)
\quad{\rm weakly
\,\,\,in}\quad L^1({\bRN})
\end{equation}
 Hence, by convexity and
Fatou's Lemma, we conclude from (\ref{(3.4)}) that $f$ satisfies the
entropy inequality (\ref{(1.5)}). (The details of such an argument
may be found in \cite{DL}.) Thus, to prove that  $f$ is a weak
solution, it only needs to show that for any $\vp \in
C^{2}_{b}({\bRN})$, $t\in [0,\infty)$,
\begin{equation}\label{(3.7)}
\lim_{n\to\infty}\int_{0}^{t}d\tau\int_{{\bRN}}
Q_n(f^n\,|\,\Dt\vp)dv =\int_{0}^{t}d\tau\int_{{\bRN}}
Q(f\,|\,\Dt\vp)dv\,.
\end{equation}
 To do this, we use the following
property (which is a consequence of weak convergence (\ref{(3.6)}) and
$\sup_{n\ge 1,\,t\ge 0}\|f^n(t)\|_{L^1_2}<\infty$):  If for some
$\alpha<2$, $\Phi(v,v_*)$ and $\Phi_n(v,v_*)$ satisfy
$$|\Phi(v,v_*)|\,,\,\sup_{n\ge 1}|\Phi_n(v,v_*)|\le
C\langle{v}\rangle^{\alpha}\langle{v_*}\rangle^{\alpha}\,,$$
$$\Phi_n(v,v_*)\to \Phi(v,v_*)\quad (\,n\to\infty\,) \quad {\rm
a.e.}\,\,(v,v_*)\in {\bRRN}$$
 then
\begin{equation}\label{(3.8)}
\lim_{n\to\infty}\int_{0}^{t}d\tau\int_{{\bRRN}}\Phi_n f^n f^n_* dv_*dv
=\int_{0}^{t}d\tau\int_{{\bRRN}}\Phi f f_* dv_*dv \,.
\end{equation}

Suppose $-2\le \gm<0$. Then
$$|L[\Dt\vp](v,v_*)|\,,\, \sup_{n\ge 1}|L_n[\Dt\vp](v,v_*)|\le
C_{\vp}\langle{v}\rangle^{2+\gm}\langle{v_*}\rangle^{2+\gm}\ ,$$
so applying
the relation (\ref{(2.13)}) in Lemma 2.3, we see that the convergence
(\ref{(3.7)})
follows from (\ref{(3.8)}) with $\Phi(v,v_*)=L[\Dt\vp](v,v_*)$
and $\Phi_n(v,v_*)
=L_n[\Dt\vp](v,v_*)$.

Next, for $-4\le \gm<-2$, we truncate: For any
$\ld>0$, let
$B_{n,\ld}={\bf 1}_{\{|v-v_*|>\ld\}}B_n
\,,\,B_{\ld}={\bf 1}_{\{|v-v_*|>\ld\}}B
$
and let $Q_{n,\ld}(\cdot), L_{n,\ld}[\cdot]$ and $Q_{\ld}(\cdot),
L_{\ld}[\cdot]$ correspond to $B_{n,\ld}$ and $B_{\ld}$
respectively. Then
\beas&&\int_{0}^{t}d\tau\int_{{\bRN}}Q_n(f^n\,|\,\Dt\vp)dv-\int_{0}^{t}d\tau\int_{{\bRN}}Q(f\,|\,\Dt\vp)dv
\\
&&=\int_{0}^{t}d\tau\int_{{\bRN}}Q_n(f^n\,|\,\Dt\vp)dv-\int_{0}^{t}d\tau\int_{{\bRN}}
Q_{n,\ld}(f^n\,|\,\Dt\vp)dv\\
&&+\int_{0}^{t}d\tau\int_{{\bRN}}
Q_{n,\ld}(f^n\,|\,\Dt\vp)dv
-\int_{0}^{t}d\tau\int_{{\bRN}}Q_{\ld}(f\,|\,\Dt\vp)dv\\
&&+
\int_{0}^{t}d\tau\int_{{\bRN}}Q_{\ld}(f\,|\,\Dt\vp)dv
-\int_{0}^{t}d\tau\int_{{\bRN}}Q(f\,|\,\Dt\vp)dv
\\
&&:=I_{n,\ld}(t)+J_{n,\ld}(t)+I_{\ld}(t)\,.\eeas
Using part (I) of Lemma 2.2 we have   \beas&&
|I_{n,\ld}(t)|
\le C_{\vp}\int_{0}^{t}d\tau\int_{{\bRRSN}}B {\bf 1}_{\{|v-v_*|\le
\ld\}}|v-v_*|^2\sin\theta |{f^n}'{f^n_*}'-f^nf^n_*|d\sg dv_* dv\\
&&\le C_{\vp}\left(\int_{0}^{t}d\tau\int_{{\bRRN}}{\bf 1}_{\{|v-v_*|\le
\ld\}}|v-v_*|^{4+\gm}f^nf^n_*dv_*
dv\right)^{1/2}\left(\int_{0}^{t}D_n(f^n(\tau))d\tau\right)^{1/2}\\
&&\le C_{\vp}\left(\int_{0}^{t}d\tau\int_{{\bRRN}}{\bf
1}_{\{|v-v_*|\le \ld\}}f^nf^n_*dv_* dv\right)^{1/2} \,.\eeas By
$\sup\limits_{n\ge 1, \tau\ge 0}\int_{{\bRN}}f^n(v,\tau)|\log
f^n(v,\tau)| dv<\infty$ we obtain for all $0<\ld <1<R$
 \beas&&\sup_{n\ge 1,\tau\ge 0}\int_{{\bRN}}f^n(v,\tau)\left(\int_{|v-v_*|\le
\ld}f^n(\tau,v_*)dv_*\right)dv \le C\left(R\,\ld^N+\fr{1}{\log
R}\right)\|f_0\|_{L^1}\,.\eeas This implies that $\sup_{n\ge
1}|I_{n,\ld}(t)|\to 0 $ as $\ld\to 0\,.$ Similarly $I_{\ld}(t)\to
0\,(\ld\to 0)\,.$ To estimate $J_{n,\ld}(t)$ we use (\ref{(2.12)})
in Lemma 2.3 to get
 \beas&& |J_{n,\ld}(t)|=
2 \left|\int_{0}^{t}d\tau\int_{{\bRRN}}
L_{n,\ld}[\Dt\vp] f^n {f^n_*} dv_* dv- \int_{0}^{t}d\tau\int_{{\bRRN}}
L_{\ld}[\Dt\vp] f f_* dv_* dv\right|\,.\eeas Since
$L_{n,\ld}[\Dt\vp](v,v_*)\to L_{\ld}[\Dt\vp](v,v_*)\,(n\to\infty)$ for all $(v,v_*)\in{\bRRN}$ and
$$|L_{\ld}[\Dt\vp](v,v_*)|\,,\, \sup_{n\ge 1}|L_{n,\ld}[\Dt\vp](v,v_*)|
\le C_{\vp}{\bf 1}_{\{|v-v_*|>\ld\}} |v-v_*|^{2+\gm}\le C_{\vp,\ld}$$ it
follows from (\ref{(3.8)}) that
$J_{n,\ld}(t)\to 0\,(n\to\infty)\,\,\forall\, \ld>0\,.$
These imply (\ref{(3.7)}) for $-4\le \gm<-2$. Therefore, $f$ is a
weak solution. $\quad \Box$

\section {Moment Estimates for  Weak Solutions }

In this section we prove the first part of Theorem 1; i.e., the moment
estimates. We need two lemmas that provide  estimates of $\Dt\vp$ for $\vp(v)=\langle{v}\rangle^s$.
These are so-called Povzner type estimates, but including averaging that allows them to be applied in our weak solution setting.

For any  $v,v_*\in{\bRN}$ let ${\bf h} = \fr{v+v_*}{|v+v_*|}$ for
$v+v_*\neq 0$ and ${\bf h}={\bf e}_1=(1,0,...,0)$ for $v+v_*=0$;  ${\bf
k}=\fr{v-v_*}{|v-v_*|} $ for $v-v_*\neq 0$ and ${\bf k}={\bf e}_1$ for $v-v_*=0$.
Then using representation (\ref{(2.2)}) we have with $\og\in{\bf S}^{N-2}({\bf
k})$ \beas &&
|v'|^2=\fr{|v|^2+|v_*|^2}{2}+\fr{|v+v_*||v-v_*|}{2}\left(\langle{{\bf
h},{\bf k}}\rangle\cos\theta
+\sqrt{1-\langle{{\bf h},{\bf k}}\rangle^2}\,\sin\theta\,\langle{{\bf j}, \og\rangle}\right),\\ \\
&&
|v_*'|^2=\fr{|v|^2+|v_*|^2}{2}-\fr{|v+v_*||v-v_*|}{2}\left(\langle{{\bf
h},{\bf k}}\rangle\cos\theta +\sqrt{1-\langle{{\bf h},{\bf
k}}\rangle^2}\,\sin\theta\,\langle{{\bf j}, \og\rangle} \right)
\end{eqnarray*}
where ${\bf j}=\fr{{\bf h}-\langle{{\bf h},{\bf k}\rangle}{\bf
k}}{\sqrt{1-\langle{{\bf h},{\bf k}\rangle}^2}}$ for $|\langle{{\bf
h},{\bf k}\rangle}|<1$ and  ${\bf j}={\bf e}_1$ for $|\langle{{\bf
h},{\bf k}\rangle}|=1$.

To prove the moment estimates  we need the following lemmas:

\noindent{\bf Lemma 4.1.} {\it For all $\,s>2$ and  $\,v,v_*\in {\bRN}$
\beas&&
\fr{1}{|{\bf
S}^{N-2}|}\int_{{\bf
S}^{N-2}({\bf
k})}\left(\langle{v'\rangle}^{s}+\langle{v_*'\rangle}^{s}-\langle{v\rangle}
^{s}-\langle{v_*\rangle}^{s}\right)d\og\\
&&\le s(s-2)\left(\langle{v\rangle}^2+\langle{v_*\rangle}^2\right)^{(s-4)/2}
|v+v_*|^2|v-v_*|^2
\left[(1-\langle{ {\bf h},{\bf k} }\rangle^2)^{{\bar s}/2}
-2^{-s/2-3}\right]\sin^2\theta\eeas where $
 {\bar s}=\min\{s-2,\,2\}.$
}

\noindent{\it Proof}. Let
$$\rho=\fr{1}{2}\left(\langle{v\rangle}^2+\langle{v_*\rangle}^2\right)\,,\quad
r=\fr{|v+v_*||v-v_*|}{\langle{v\rangle}^2+\langle{v_*\rangle}^2}
\,,\quad X=r\langle{{\bf h},{\bf k}}\rangle,\quad
Y=r\sqrt{1-\langle{{\bf
h},{\bf k}}\rangle^2}\,.$$
Then from the representation of $|v'|^2, |v_*'|^2$  we have
\beas
&&
\langle{v'\rangle}^2=\rho\left(1+X\cos\theta\,
+Y\sin\theta\,\,\langle{{\bf j}, \og\rangle}\,\right)\,,\quad \langle{v\rangle}^2=\rho(1+X)\,,\\
&&\langle{v_*'\rangle}^2=\rho\left(1-X\cos\theta\,
-Y\sin\theta\,\,\langle{{\bf j}, \og\rangle}\,\right)\,,\quad \langle{v_*\rangle}^2=
\rho(1-X)\eeas
and so
$$W(v,v_*,\theta):=
\rho^{-k}\int_{{\bf S}^{N-2}({\bf
k})}\left(\langle{v'\rangle}^{2k}+\langle{v_*'\rangle}^{2k}-\langle{v\rangle}
^{2k}-\langle{v_*\rangle}^{2k}\right)d\og$$
\begin{equation}\label{(4.1)}
= \int_{{\bf S}^{N-2}({\bf k})}\left\{
\sum_{i=1,-1}\bigg(1+i X\cos\theta +i Y\sin\theta\,\langle{{\bf j},
\og\rangle}\bigg)^{k} -\sum_{i=1,-1}\left(1+i
X\right)^{k}\right\}d\og
\end{equation}
where $k=s/2>1$. To estimate the integrand $\{\cdots\}$ we shall
use the following inequality:
For all $a\in [-1,1]$ and $t\in [-1,1]$
\begin{equation}\label{(4.2)}
(1+at)^{k}+(1-at)^{k}-(1+a)^{k}-(1-a)^{k}+\fr{k(k-1)}{2} a^2
(1-t^2)\le 0\,.
\end{equation}
This inequality is easily proven by
checking that the left hand side is a convex
function in $t\in[-1,1]$.
Applying (\ref{(4.2)}) to $a=X$ we have
\begin{equation}\label{(4.3)}
\sum_{i=1,-1}\left(1+i X\cos\theta\right)^{k}-
\sum_{i=1,-1}\left(1+i X\right)^{k}\le
-\fr{k(k-1)}{2}X^2\sin^2\theta
\end{equation}
 from which we see that
if $Y=0$ then the lemma holds true. Suppose $Y\neq 0$. Then
$|\langle{{\bf h},{\bf k}}\rangle|<1$. By the Cauchy-Schwarz
inequality and
$r=\fr{|v+v_*||v-v_*|}{\langle{v\rangle}^2+\langle{v_*\rangle}^2}<1$,
we have that for all $t\in[0,1]$,
\begin{equation}\label{(4.4)}
1-(|X|\cos\theta\,+
t|Y|\sin\theta) >1- \sqrt{\langle{{\bf
h},{\bf k}\rangle}^2+t^2(1-\langle{{\bf h},{\bf k}}\rangle^2)}\ge
\fr{1}{2}(1-\langle{{\bf h},{\bf k}}\rangle^2)(1-t^2)\,.
\end{equation}
Applying Taylor's formula to  the
function \beas&& t\mapsto \sum_{i=1,-1}\left(1+i X\cos\theta+ ti\,Y
\sin\theta\,\langle{{\bf j}, \og\rangle})\right)^{k}-
\sum_{i=1,-1}\left(1+i X\right)^{k}\,,\quad t\in
[0,1]\ ,\eeas we compute \beas&& \bigg\{\cdots\bigg\}
=\sum_{i=1,-1}\left(1+i X\cos\theta\right)^{k}-
\sum_{i=1,-1}\left(1+i X\right)^{k}
\\&& +k\sum_{i=1,-1}\left(1+i X\cos\theta\right)^{k-1}
 i Y\sin\theta\,\langle{{\bf j}, \og\rangle}  \\
&&+k(k-1)\left(Y\sin\theta\,\langle{{\bf j}, \og\rangle}\right)^2
\int_{0}^{1}(1-t)\sum_{i=1,-1}\bigg(1+i X\cos\theta+t i
Y\sin\theta \,\langle{{\bf j}, \og\rangle}\bigg)^{k-2}dt\,. \eeas
 Since $\int_{{\bf S}^{N-2}({\bf
k})}\,\langle{{\bf j}, \og\rangle} d\og=0$, it follows from (\ref{(4.1)}) and (\ref{(4.3)})
that
\begin{equation}\label{(4.5)} W(v,v_*,\theta)
\le -\fr{k(k-1)}{2}|{\bf S}^{N-2}|X^2\sin^2\theta +k(k-1)Y^2\sin^2\theta\,\int_{{\bf S}^{N-2}({\bf k})}
Z_k(\og)d\og
\end{equation}
where
$$Z_k(\og)=\int_{0}^{1}(1-t)\sum_{i=1,-1}\bigg(1+ i X\cos\theta+ t i
Y\sin\theta\,\langle{{\bf j}, \og\rangle}\,\bigg)^{k-2}dt\,.$$
By considering $1<k<2$ (for which we use (\ref{(4.4)})) and $k\ge 2$
respectively,
we compute for all $k>1$
\begin{equation}\label{(4.6)}
Y^2 Z_k(\og) \le
2^{k+1}r^2(1-\langle{{\bf
h},{\bf k}}\rangle^2)^{{\bar k}}\qquad \forall\, \og
\in{\bf S}^{N-2}({\bf k})
\end{equation}
 where ${\bar k}=\min\{k-1,1\}$.
Since  $-X^2\le r^2\left((1-\langle{{\bf h},{\bf
k}}\rangle^2)^{{\bar k}}-1\right)$, it
follows from (\ref{(4.5)}) and (\ref{(4.6)}) that
$$W(v,v_*,\theta)
\le \fr{k(k-1)}{2}|{\bf S}^{N-2}|
r^2\sin^2\theta\bigg\{
2^{k+3}(1-\langle{{\bf h},{\bf k}}\rangle^2)^{{\bar k}}-1\bigg\}
\,.$$  This proves the lemma. $\quad \Box$
\smallskip

\noindent{\bf Lemma 4.2.} {\it Let $B(v-v_*,\sg)$ satisfy (\ref{(1.1)}) and
(\ref{(1.2)}) with the constants $A^*, A_*$. Let
$L[\Dt\langle{\cdot\rangle}^{s}](v,v_*)$ be defined in (\ref{(1.4)}) for
$\vp(v)=\langle{v\rangle}^{s}$. Then for any $s>2$ we have

{\rm (I)} If $-2 \le \gm<0$, then for any $\vep>0$
\beas&&L[\Dt\langle{\cdot\rangle}^{s}](v,v_*)\\
&&\le
-c_s\left(\langle{v\rangle}^{s+\gm}+\langle{v_*\rangle}^{s+\gm}
\right) +\vep
C_s\left(\langle{v\rangle}^{s+\gm}\langle{v_*\rangle}^{2}
+\langle{v_*\rangle}^{s+\gm}\langle{v\rangle}^{2}\right)
+C_{s,\vep}\langle{v\rangle}^{2}\langle{v_*\rangle}^{2}\,.\eeas

{\rm (II)} If $-4\le \gm <-2$, then for any $\ld\ge 1$
\beas&&{\bf 1}_{\{|v-v_*|>\ld\}}L[\Dt\langle{\cdot\rangle}^{s}](v,v_*) \\
&&\le
-c_s\left(\langle{v\rangle}^{s+\gm}+\langle{v_*\rangle}^{s+\gm}
\right)+C_s\ld^{2+\gm}\left(\langle{v\rangle}^{s+\gm}\langle{v_*\rangle}^{2}+
\langle{v_*\rangle}^{s+\gm}\langle{v\rangle}^{2}
\right)+C_{s,\ld}\langle{v\rangle}^{2}\langle{v_*\rangle}^{2}\,.\eeas
Here the constants $0<c_s, C_s<\infty$  depend only on $N, A^*, A_*$
and $s$, while $0<C_{s,\vep}, C_{s,\ld}<\infty$ depend also on
$\vep$ and $\ld$ respectively. }
\smallskip

\noindent{\it Proof}. By symmetry
$L[\Dt\langle{\cdot\rangle}^{s}](v,v_*)=
L[\Dt\langle{\cdot\rangle}^{s}](v_*,v)$, we may  assume that $|v|\ge
|v_*|$. We note also that
\begin{equation}\label{(4.7)}
\langle{v\rangle}^{s+\gm}\ge \fr{1}{2}\left(\langle{v\rangle}^{s+\gm}
+\langle{v_*\rangle}^{s+\gm}\right)\qquad
{\rm if}\quad |v|\ge |v_*|\quad {\rm and}\quad s+\gm\ge
0\,.
\end{equation}
 By Lemma 4.1 and
$\Dt\langle{\cdot\rangle}^{s}=
\langle{v'\rangle}^{s}+\langle{v_*'\rangle}^{s}-\langle{v\rangle}
^{s}-\langle{v_*\rangle}^{s}$, we have
\beas&&L[\Dt\langle{\cdot\rangle}^{s}](v,v_*)
 \\
 &&\le s(s-2)\left(\langle{v\rangle}^2+\langle{v_*\rangle}^2\right)^{(s-4)/2}|v+v_*|^2|v-v_*|^2
\left[(1-\langle{ {\bf h},{\bf k} }\rangle^2)^{{\bar s}/2}
-2^{-s/2-3}\right]\\
&&\times \int_{{\bSN}}B(v-v_*,\sg)\sin^2\theta\,d\sg \,.\eeas Let
$R_s=2^{({\bar s}+4+s/2)/{\bar s}}$, and consider $L[\Dt\langle{\cdot\rangle}^{s}](v,v_*)=
L_1(v,v_*)+L_2(v,v_*)$ where
$$L_1(v,v_*):
=L[\Dt\langle{\cdot\rangle}^{s}](v,v_*){\bf 1}_{\{|v|\le R_s|v_*|\}}
\,,\quad L_2(v,v_*):=L[\Dt\langle{\cdot\rangle}^{s}]
(v,v_*){\bf 1}_{\{|v|>R_s|v_*|\}}\,.$$ By assumption (\ref{(1.1)}) we have
\begin{equation}\label{(4.8)}
L_1(v,v_*)
\le C_s
\left(\langle{v\rangle}^2+\langle{v_*\rangle}^2\right)^{(s-4)/2}
|v+v_*|^2|v-v_*|^{2+\gm}{\bf 1}_{\{|v|\le R_{s}|v_*|\}}\,.
\end{equation}
To estimate $L_2(v,v_*)$, observing that
$$|v|>R_{s}|v_*|\,
\Longrightarrow\, 1-\langle{ {\bf h},{\bf k} }\rangle^2\le
\fr{4|v|^2|v_*|^2}{(|v|^2+|v_*|^2)^2} <\fr{4}{R_s^2}\ ,$$ we have by the
choice of $R_s$ that
$$
\left((1-\langle{ {\bf h},{\bf k} }\rangle^2)^{{\bar s}/2}
-2^{-3-s/2}\right) {\bf 1}_{\{|v|>R_{s}|v_*|\}} \le
-2^{-4-s/2}{\bf 1}_{\{|v|>R_{s}|v_*|\}}\,.$$ Thus, using the assumption
(\ref{(1.2)}), we obtain
\begin{equation}\label{(4.9)}
L_2(v,v_*)\le-c_s\left(\langle{v\rangle}^2+\langle{v_*\rangle}^2
\right)^{(s-4)/2}|v+v_*|^2|v-v_*|^2(1+|v-v_*|)^{\gm}
{\bf 1}_{\{|v|>R_{s}|v_*|\}}\,.
\end{equation}

(I) Assume $-2\le \gm<0$. Using  $|v\pm v_*|\le
2^{1/2}\left(\langle{v\rangle}^2+\langle{v_*\rangle}^2
\right)^{1/2}$  and recalling $s>2$ and $|v|\ge |v_*|$, we have by
(\ref{(4.8)}) that \beas&& L_{1}(v,v_*) \le C_s
\langle{v\rangle}^{s+\gm}{\bf 1}_{\{|v|\le R_{s}|v_*|\}} \le
C_s\langle{v\rangle}^{s+\gm-2}\langle{v_*\rangle}^2
 \,.\eeas
Applying the elementary inequality
\begin{equation}\label{(4.10)}
a^{k-2}b^2\le \vep a^{k}b^2+(1+\vep^{-(k-2)/2})a^2b^2\,,\qquad a, b\ge 1\,,
\,\,k\in{\bf R}^1\,,\,\, \vep>0\ ,
\end{equation}
we get for any $\vep>0$
$$L_{1}(v,v_*)\le \vep C_s
\langle{v\rangle}^{s+\gm}\langle{v_*\rangle}^{2}+C_{s,\vep}
\langle{v\rangle}^{2}\langle{v_*\rangle}^{2}\,.$$
To estimate
$L_2(v,v_*)$, we observe by $R_s>2$ and $\gm<0$ that
$$|v|>R_{s}|v_*|\,\,\,{\rm and}\,\,\,|v|> 1\,\Longrightarrow\,\,
|v\pm v_*|\ge \fr{1}{4}\langle{v}\rangle\,\,\,{\rm and}\,\,\,
(1+|v-v_*|)^{\gm}\ge 4^{\gm}\langle{v}\rangle^{\gm}\,.$$ By (\ref{(4.9)})
(considering $s\le 4$ and $s>4$ respectively) we have \beas&&
L_{2}(v,v_*) \le L_{2}(v,v_*){\bf 1}_{\{|v|>1\}}\le -c_s
\langle{v\rangle}^{s+\gm}{\bf 1}_{\{|v|>R_{s}|v_*|\}}{\bf 1}_{\{|v|>1\}}\\
&&=c_s\langle{v\rangle}^{s+\gm}
\bigg(-1+{\bf 1}_{\{|v|>R_{s}|v_*|\}}{\bf 1}_{\{|v|\le 1\}}+{\bf 1}_{\{|v|\le
R_{s}|v_*|\}}\bigg)\,. \eeas Since
$\langle{v\rangle}^{s+\gm}{\bf 1}_{\{|v|>R_{s}|v_*|\}}{\bf 1}_{\{|v|\le 1\}}\le
C_s$, and by (\ref{(4.8)}) for any $\vep>0$,
\beas&&\langle{v\rangle}^{s+\gm}{\bf 1}_{\{|v|\le R_{s}|v_*|\}}\le
C_s\langle{v\rangle}^{s+\gm-2}\langle{v_*\rangle}^{2} \le \vep
C_s\langle{v\rangle}^{s+\gm}\langle{v_*\rangle}^{2}+C_{s,\vep}
\langle{v\rangle}^{2}\langle{v_*\rangle}^{2}\ ,\eeas it follows that
$$L_{2}(v,v_*)\le -c_s\langle{v\rangle}^{s+\gm}+\vep
C_s\langle{v\rangle}^{s+\gm}\langle{v_*\rangle}^{2}
+C_{s,\vep}\langle{v\rangle}^{2}\langle{v_*\rangle}^{2} \,.$$
Therefore,
$$L[\Dt\langle{\cdot\rangle}^{s}](v,v_*)=L_1(v,v_*)+L_2(v,v_*)\le
-c_s\langle{v\rangle}^{s+\gm}+\vep
C_s\langle{v\rangle}^{s+\gm}\langle{v_*\rangle}^{2}
+C_{s,\vep}\langle{v\rangle}^{2}\langle{v_*\rangle}^{2}\,.
$$
This together with (\ref{(4.7)}) (because $s+\gm>0$) gives the inequality in
part (I) of the lemma.
\smallskip

(II) Assume $-4\le \gm<-2$. Given any $\ld\ge 1$. By (\ref{(4.8)}) and
$2+\gm<0$ we have
\begin{equation}\label{(4.11)}{\bf 1}_{\{|v-v_*|>
\ld\}} L_{1}(v,v_*)\le C_{s}\ld^{2+\gm}{\bf 1}_{\{
|v|\le R_{s}|v_*|\}}\langle{v\rangle}^{s-2}\\
\le
C_{s}\ld^{2+\gm}\langle{v\rangle}^{s+\gm}\langle{v_*\rangle}^2\,.
\end{equation}
Note that if $s\le 4$,  then $s+\gm\le 2$ so that
$\langle{v\rangle}^{s+\gm}\le \langle{v\rangle}^{2}$ and
$\langle{v\rangle}^{s+\gm}, \langle{v_*\rangle}^{s+\gm}\le
\langle{v\rangle}^{2}\langle{v_*\rangle}^{2}$ and thus by
(\ref{(4.8)}) and neglecting the non-positive term ${\bf
1}_{\{|v-v_*|> \ld\}}L_{2}(v,v_*)$ we get  \beas&&{\bf
1}_{\{|v-v_*|> \ld\}}L[\Dt\langle{\cdot\rangle}^{s}](v,v_*) \le
C_{s}\ld^{2+\gm}\langle{v\rangle}^{2}\langle{v_*\rangle}^{2} \le
-(\langle{v\rangle}^{s+\gm}+\langle{v_*\rangle}^{s+\gm})
+C_{s,\ld}\langle{v\rangle}^{2}\langle{v_*\rangle}^{2}\ , \eeas
which is a special case of the inequality in part (II) of the lemma.
Next, assume $s>4$. To estimate ${\bf 1}_{\{|v-v_*|>
\ld\}}L_{2}(v,v_*)$ we see from $R_s>2$ and $\ld \ge 1$ that
$$|v-v_*|> \ld\,\,\, {\rm and}\,\,\, |v|>R_{s}|v_*| \,\, \Longrightarrow\,\,
|v\pm v_*|\ge \fr{1}{2}|v|\,,\,\,1<|v-v_*|\le 2|v|\,,\,\, {\rm
and}\,\, |v|\ge \fr{1}{4}\langle{v\rangle}\,.$$ This implies by
(\ref{(4.9)}) and $\gm<0$ that \beas&& {\bf 1}_{\{|v-v_*|>
\ld\}}L_{2}(v,v_*)\le
 -c_s\,{\bf 1}_{\{|v-v_*|>
\ld\}}{\bf 1}_{\{|v|>R_{s}|v_*|\}} \langle{v\rangle}^{s+\gm}\\
&&=c_s\left (-1 +{\bf 1}_{\{|v|\le R_{s}|v_*|\}}+{\bf 1}_{\{|v-v_*|\le
\ld\}}{\bf 1}_{\{|v|>R_{s}|v_*|\}}\right
)\langle{v\rangle}^{s+\gm}\,.\eeas Since $R_s>2$ and $|v|\le
R_{s}|v_*|$ imply $\langle{v}\rangle\le R_s\langle{v_*}\rangle$,  it
follows from the inequality (\ref{(4.10)}) that for any $\vep>0$,
$${\bf 1}_{\{|v|\le
R_{s}|v_*|\}}\langle{v\rangle}^{s+\gm}\le  \vep
C_s\langle{v\rangle}^{s+\gm}\langle{v_*\rangle}^{2}
+C_{s,\vep}\langle{v\rangle}^{2}\langle{v_*\rangle}^{2}\,.$$ Also, we
see that $|v-v_*|\le \ld$ and $|v|>R_{s}|v_*| \Longrightarrow |v|\le
2\ld$.  Thus,
$${\bf 1}_{\{|v-v_*|\le
\ld\}}{\bf 1}_{\{|v|>R_{s}|v_*|\}}\langle{v\rangle}^{s+\gm}\le
C_{s,\ld}\le C_{s,\ld}\langle{v\rangle}^{2}\langle{v_*\rangle}^{2}
\,.$$ Let us now choose $\vep=\ld^{2+\gm}$. Then
\begin{equation}\label{(4.12)}
{\bf 1}_{\{|v-v_*|> \ld\}}L_{2}(v,v_*)
 \le -c_s\langle{v\rangle}^{s+\gm}+C_s\ld^{2+\gm}
\langle{v\rangle}^{s+\gm}\langle{v_*\rangle}^{2}
+C_{s,\ld}\langle{v\rangle}^{2}\langle{v_*\rangle}^{2}\,.
\end{equation}
Summarizing (\ref{(4.11)}) and (\ref{(4.12)}) gives
\beas&&{\bf 1}_{\{|v-v_*|>\ld\}}L[\Dt\langle{\cdot\rangle}^{s}](v,v_*) \le
-c_s\langle{v\rangle}^{s+\gm}+C_s\ld^{2+\gm}
\langle{v\rangle}^{s+\gm}\langle{v_*\rangle}^{2}
+C_{s,\ld}\langle{v\rangle}^{2}\langle{v_*\rangle}^{2}\eeas which
together with $s+\gm>0$ and (\ref{(4.7)}) gives the inequality in part
(II) of the lemma. $\quad \Box$
\smallskip

\noindent{\bf Proof of Theorem 1: Moment Estimates}. We use a short
notation
$$\|f\|_{s}:=\|f\|_{L^1_s}\,.$$
Let $f(v,t)$ be a weak solution of Eq.(B) with $f|_{t=0}=f_0\in
L^1_{(1,0,1)}\cap L^1_s\cap L^1{\rm log}L({\bRN})$ and $s> 2$.
Recall that $f$ conserves the mass and energy: $\|f(t)\|_{0}\equiv
1\,,\,\|f(t)\|_{2}\equiv1+N$.

\noindent{\bf Step 1}. We shall prove that $f\in
L^{\infty}_{loc}([0,\infty), L^1_s({\bRN}))$, i.e.
\begin{equation}\label{(4.13)}
\sup_{t\in[0,T]} \|f(t)\|_{s}<\infty\quad
\forall\, T<\infty\ .
\end{equation}
For any $\ld\ge
1$, we split $B=B_{\ld}+B^{\ld}$ as in (\ref{(2.11)}) and
let $L_{\ld}[\cdot]$, $Q^{\ld}(\cdot)$ correspond to $B_{\ld}(z,\sg),
B^{\ld}(z,\sg)$ respectively.
Let $\chi\in C^{\infty}_{c}({\bRN})$ be the function used above
(see (\ref{(3.1)})). For any $k\ge 2$, $ n\ge 1$, let
$$\vp_k(v)=\langle{v}\rangle^{k}\,,\quad
\vp_{k,n}(v)=\vp_k(v)\chi(v/n)\,.$$ Then $\vp_{k,n}\in C^2_{b}({\bRN})$
and so for $-2\le \gm<0$
\begin{equation}\label{(4.14)}
\int_{{\bRN}}\vp_{k,n}(v)f(v,t)dv=\int_{{\bRN}}\vp_{k,n}(v)f_0(v)dv+
\fr{1}{2}\int_{0}^{t}d\tau\int_{{\bRRN}}L[\Dt\vp_{k,n}]ff_*
dv_*dv
\end{equation}
and for $-4\le \gm<-2$
$$
\int_{{\bRN}}\vp_{k,n}(v)f(v,t)dv=\int_{{\bRN}}\vp_{k,n}(v)f_0(v)dv\qquad \quad
\qquad
\qquad
\qquad \qquad\qquad \qquad\qquad \qquad
$$
\begin{equation}\label{(4.15)}-\fr{1}{4}\int_{0}^{t}d\tau\int_{{\bRN}}Q^{\ld}(f\,|\Dt\vp_{k,n})dv
 +
\fr{1}{2}\int_{0}^{t}d\tau\int_{{\bRRN}}L_{\ld}[\Dt\vp_{k,n}]ff_*
dv_*dv
\end{equation}
where we used (\ref{(2.12)}).
Moreover
$$0\le \vp_{k,n}(v)\le \vp_k(v)\,,\quad \lim_{n\to\infty}\vp_{k,n}(v)=\vp_k(v)\,,$$
$$
\lim_{n\to\infty}\Dt\vp_{k,n}(v',v_*',v,v_*)=\Dt\vp_{k}(v',v_*',v,v_*)
\,,$$
$$\lim_{n\to\infty}L[\Dt\vp_{k,n}](v,v_*)= L[\Dt\vp_k](v,v_*)
\,,\quad \lim_{n\to\infty}L_{\ld}[\Dt\vp_{k,n}](v,v_*)= L_{\ld}
[\Dt\vp_k](v,v_*)\,,$$
$$|\p^2\vp_k(v)|\,,
\,\,\sup_{n\ge 1}|\p^2\vp_{k,n}(v)|\le
C_{k}\langle{v}\rangle^{k-2}\,,$$
and by introducing
$$\Psi_{k,\alpha}(v,v_*):=(1+|v|^2+|v_*|^2)^{(k-2)/2}|v-v_*|^{\alpha}$$
and using Lemma 2.1, and (\ref{(1.4)}), we have
\begin{equation}\label{(4.16)}|\Dt\vp_{k}|\,,\quad \sup_{n\ge 1}|\Dt\vp_{k,n}|\le
C_k\Psi_{k,2}(v,v_*)\sin\theta
\end{equation}
\begin{equation}\label{(4.17)}
\bigg|\int_{{\bf S}^{N-2}({\bf k})}\Dt\vp_{k}d\og\bigg|\,,\,\sup_{n\ge 1}
\bigg|\int_{{\bf S}^{N-2}({\bf k})}\Dt\vp_{k,n}d\og\bigg|\le
C_k\Psi_{k,2}(v,v_*)\sin^2\theta
\end{equation}
\begin{equation}\label{(4.18)}
|L[\Dt\vp_{k}](v,v_*)|\,,\,\sup_{n\ge 1}|L[\Dt\vp_{k,n}](v,v_*)|\le
C_k \Psi_{k,2+\gm}(v,v_*)\,.
\end{equation}
The constants $C_{k}$  depend only on $k, N, \gm, A^*$.
In the following we assume $2\le k\le s$.

Suppose $-2\le \gm<0$. In (\ref{(4.14)})
letting $n\to\infty$
 we obtain by  Fatou's Lemma and (\ref{(4.18)}) that
$$\|f(t)\|_{L^1_k}
\le \|f_0\|_{s}+C_k\int_{0}^{t}d\tau\int_{{\bRRN}}
\Psi_{k,2+\gm} ff_* dv_* dv\,.$$ Since
$-2\le \gm<0$ implies
$$\Psi_{k,2+\gm}(v,v_*)\le
C_k\left(\langle{v\rangle}^{k+\gm}\langle{v_*\rangle}^{2}+\langle{v_*\rangle}^{k+\gm}
\langle{v\rangle}^{2}\right)$$ it
follows form the conservation of mass and energy that
$$\|f(t)\|_{k}\le\|f_0\|_{s}+C_k
\int_{0}^{t}\|f(\tau)\|_{k+\gm} d\tau \,,\quad t\ge 0\,.$$ Now take
$k=k_{m}=\min\{s, 2+m|\gm|\}, m=1,2,...,.$ By $\gm<0$ and induction on $m$
we then obtain
$$\|f(t)\|_{k_m}\le C_{k_m}(1+t)^{m}\,,\quad t\ge
0\,,\quad m=1,2,...\,.$$ This proves (\ref{(4.13)}) for $-2\le \gm<0$.
Next suppose $-4\le \gm <-2$.
 Observing that
$|v-v_*|\le \ld  \Longrightarrow 1+|v|^2+|v_*|^2\le 4\ld
\langle{v\rangle}\langle{v_*\rangle}$ and $|v-v_*|^{4+\gm}\le \ld^{4+\gm}$,
we have
$${\bf 1}_{\{|v-v_*|\le \ld
\}}[\Psi_{k,2}(v,v_*)]^2|v-v_*|^{\gm}
\le C_k\ld^{k+2+\gm}
\langle{v\rangle}^{k-2}\langle{v_*\rangle}^{k-2}\,,$$
$$\int_{{\bRRN}}{\bf 1}_{\{|v-v_*|\le \ld
\}}[\Psi_{k,2}(v,v_*)]^2 |v-v_*|^{\gm}ff_* dv_*dv\le
C_k \ld ^{k+2+\gm}\|f(\tau)\|_{k-2}^2 \,.$$ Therefore
using (\ref{(4.16)}) and Lemma 2.2
we obtain
\begin{equation}\label{(4.19)}
\left|\int_{0}^{t}d\tau\int_{{\bRN}}Q^{\ld}
(f\,|\,\Dt\vp_{k,n})dv\right| \le C_{k}\ld^{(k+2+\gm)/2}
\int_{0}^{t}\|f(\tau)\|_{k-2} \sqrt{D(f(\tau))}\,d\tau\,.
\end{equation}
Next,
by $2+\gm<0$ and $\ld\ge 1$  we have
$${\bf 1}_{\{|v-v_*|>
\ld \}}\Psi_{k,2+\gm}(v,v_*) \le C_k
\left(\langle{v}\rangle^{k-2}+\langle{v_*}\rangle^{k-2}
 \right)$$
 which gives by (\ref{(4.18)})  and $L_{\ld}[\cdot]={\bf 1}_{\{|v-v_*|>\ld\}}L[\cdot]$
 that
\begin{equation}\label{(4.20)}
\fr{1}{2}\int_{0}^{t}d\tau\int_{{\bRRN}}
L_{\ld}[\Dt\vp_{k,n}]ff_* dv_*dv\le C_{k} \int_{0}^{t}\|f(\tau)\|_{k-2}\,d\tau\,.
\end{equation}
In the equation (\ref{(4.15)}), letting $\ld=1$ and $n\to\infty$, we
obtain from (\ref{(4.19)}), (\ref{(4.20)}) and Fatou's Lemma that
$$\|f(t)\|_{k}\le
\|f_0\|_{s}+
 C_{k}\int_{0}^{t}\|f(\tau)\|_{k-2}\left(1+\sqrt{D(f(\tau))}\,\right)\,d\tau\,.$$ Now
we choose $k=k_m=\min\{s,2m\}$.  Then by
induction on $m$ it is easy to show that there are constants
$0<C_{k_m}<\infty$ such that
$$\|f(t)\|_{k_m}\le C_{k_m}(1+t)^{m-1}\,,\quad t\ge 0\,,\quad m=1,2,...\,.$$
This proves (\ref{(4.13)}) for $-4\le \gm<-2$.

\noindent{\bf Step 2}. We shall prove the following inequality
\begin{equation}\label{(4.21)}
\|f(t)\|_{s}+c_{s}\int_{0}^{t}\|f(\tau)\|_{s+\gm}d\tau\le
C_{s}(1+t)\,,\quad t\ge 0
\end{equation}
 which implies (\ref{(1.11)}) because
$\gm<0$.  Here and below the constants $0<c_{s}\,,\,C_s<\infty $
depend only on $N, \gm, A_*, A^*, s$ and $\|f_0\|_{L^1_s}$,  and  in
case $-4\le \gm<-2$, they depend also on $H(f_0)$.

>From the pointwise bounds (\ref{(4.16)})-(\ref{(4.18)})
and integrability shown in Step 1 we see that the dominated
convergence theorem can be used and we get from (\ref{(4.14)}) and (\ref{(4.15)})
with $k=s$ that for $-2\le \gm<0$
\begin{equation}\label{(4.22)}
\|f(t)\|_{s}=\|f_0\|_{s} +
\fr{1}{2}
\int_{0}^{t}d\tau\int_{{\bRRN}}L[\Dt\vp_s] ff_* dv_*
dv\,,
\end{equation}
 and for $-4\le\gm<-2$
\begin{equation}\label{(4.23)}
\|f(t)\|_{s}=\|f_0\|_{s}-\fr{1}{4}
\int_{0}^{t}d\tau\int_{{\bRN}}Q^{\ld}(f\,|\,\Dt\vp_s)dv + \fr{1}{2}
\int_{0}^{t}d\tau\int_{{\bRRN}}L_{\ld} [\Dt\vp_s]ff_*dv_*
dv\,.
\end{equation} To prove (\ref{(4.21)}), we first consider $-2\le \gm<0$.
By Lemma 4.2 (recalling $\vp_s(v)=\langle{v}\rangle^s$) and the
conservation of mass and energy we have from (\ref{(4.22)}) that for any
$\vep>0$ \beas&& \|f(t)\|_{s}\le \|f_0\|_{s}-\left(c_s-\vep
C_s\right) \int_{0}^{t}\|f(\tau)\|_{s+\gm} d\tau+C_{s,\vep}
t\,,\quad t\ge 0\,.\eeas Therefore choosing $\vep=\fr{c_s}{2C_{s}}$
leads to (\ref{(4.21)}) (with different constants).

Next assume $-4\le \gm<-2$. Using inequality (\ref{(4.19)}) for $k=s$ and letting $n\to\infty$
we have
\beas&&\left|\int_{0}^{t}d\tau \int_{{\bRN}}Q^{\ld}(f\,|\Dt\vp_s)dv\right| \le
C_{s}\ld^{(s+2+\gm)/2}\int_{0}^{t}\|f(\tau)\|_{s-2}\sqrt{D(f(\tau))}d\tau\,.\eeas By the
Cauchy-Schwarz inequality and using $s-4\le s+\gm$,
$$\|f(\tau)\|_{s-2}
\le \sqrt{\|f(\tau)\|_{s}}\sqrt{\|f(\tau)\|_{s-4}}\le
\sqrt{\|f(\tau)\|_{s}}\sqrt{\|f(\tau)\|_{s+\gm}}\,.$$  Therefore, for
any $\vep>0$
\beas&& \ld^{(s+2+\gm)/2}\|f(\tau)\|_{s-2}\sqrt{D(f(\tau))}
\le
\fr{1}{2}\vep\|f(\tau)\|_{s+\gm}+\fr{1}{2\vep}\ld^{s+2+\gm}\|f(\tau)\|_{s}
D(f(\tau))\ ,\eeas
and so \beas&&\fr{1}{4}\left|\int_{0}^{t}d\tau
\int_{{\bRN}}Q^{\ld}(f\,|\,\Dt\vp_s)dv\right|\le \vep
C_{s}\int_{0}^{t}\|f(\tau)\|_{s+\gm}d\tau+
C_{s,\vep,\ld}\int_{0}^{t}\|f(\tau)\|_{s} D(f(\tau))d\tau\,.\eeas
Also, using Lemma 4.2 we have as shown above that
$$\fr{1}{2}\int_{0}^{t}d\tau\int_{{\bRRN}}L_{\ld}
[\Dt\vp_s]ff_*dv_* dv \le
-\left(c_s-C_s\ld^{2+\gm}\right)\int_{0}^{t}\|f(\tau)\|_{s+\gm}d\tau+C_{s,\ld}\,t
\,.$$ Choose $\vep=\ld^{2+\gm}$. Then from (\ref{(4.23)}) and the above
estimates we obtain
$$\|f(t)\|_s\le
\|f_0\|_{s}-\left(c_s-C_{s}\ld^{2+\gm}\right)\int_{0}^{t}\|f(\tau)\|_{s+\gm}
d\tau+C_{s,\ld}\int_{0}^{t}\|f(\tau)\|_{s} D(f(\tau))d\tau+C_{s,\ld}
t\,.$$ We can assume that $C_s\ge c_s$. By $2+\gm<0$ we can choose
$$\ld=\left(\fr{c_s}{2C_{s}}\right)^{1/(2+\gm)}\quad (\,>1\,).$$
Then with different constants we obtain
$$\|f(t)\|_s+c_s\int_{0}^{t}\|f(\tau)\|_{s+\gm}
d\tau\le C_{s}(1+t)+C_{s}\int_{0}^{t}\|f(\tau)\|_{s}
D(f(\tau))d\tau\,,\quad t\ge 0\,.$$ By Gronwall's Lemma, this gives
$$\|f(t)\|_s+c_s\int_{0}^{t}\|f(\tau)\|_{s+\gm}
d\tau\le  C_s(1+t) \exp(C_s\int_{0}^{t}D(f(\tau))d\tau) \,,\quad
t\ge 0$$ which implies (\ref{(4.21)}) because $\int_{0}^{\infty}D(f(\tau))d\tau\le
C_{N,H(f_0)}<\infty$\,.$\quad \Box$

\section{Convergence to Equilibrium in Broad Generality}

In this section we give a unified treatment of strong convergence to
equilibrium which includes all cases (soft potentials, hard
potentials, etc., with and without cutoff).  In the following
theorem, the functions $f(v,t)$ are not even assumed to be solutions
of the Boltzmann equation: Their only connection to the Boltzmann
equation is the requirement (\ref{(5.3)}) below in the integrated entropy
dissipation.
\medskip

\noindent{\bf Theorem 4.} {\it  Let $B(z,\sg)$ be a collision kernel
satisfying $B(z,\sg)>0$  a.e. $(z,\sg)\in{\bRSN},$ and let $0\le
f\in L^{\infty}([0,\infty); L^1_2({\bRN}))$ satisfy
\begin{equation}\label{(5.1)}
f(\cdot,t)\in L^1_{(1,0,1)}({\bRN})\quad \forall\, t\ge 0\,;\quad
\sup_{t\ge 0}\int_{{\bRN}}f(v,t)\Phi(
f(v,t))dv<\infty\,\,
\end{equation}
\begin{equation}\label{(5.2)}
\lim_{|t_1-t_2|\to 0}
\int_{{\bRN}}\vp(v)\bigg( f(v,t_1)-f(v,t_2)\bigg)dv=0\quad \forall\,
\vp\in C_c^{\infty}({\bRN})\,,
\end{equation}
\begin{equation}\label{(5.3)}
\int_{0}^{\infty}D(f(t))dt<\infty\,,
\end{equation}
\begin{equation}\label{(5.4)}
\|f(t)\|_{L^1_{2+}}:=\int_{{\bRN}}\langle{v}\rangle^2\Psi(v)f(v,t)dv<\infty \quad \forall\,
t\ge t_0
\end{equation} for some $t_0\ge 0$, where $\Phi(r)\ge 0,
\Psi(v)\ge 1$ satisfy $\lim\limits_{r\to\infty}\Phi(r)=\infty,\,
\lim\limits_{|v|\to\infty}\Psi(v)=\infty\,,$ and $D(f)$ is defined
in (\ref{(1.3)}) with the present $B(z,\sg)$. Then for the Maxwellian $M\in
L^1_{(1,0,1)}({\bRN})$ given by (\ref{(1.9)}) we have

(I) If

\begin{equation}\label{(5.5)}
\sup_{T\ge t_0}\fr{1}{T}\int_{t_0}^{T}\|f(t)\|_{L^1_{2+}}dt
<\infty\,,
\end{equation}
 then
\begin{equation}\label{(5.6)}
\lim_{T\to \infty}\fr{1}{T}\int_{0}^{T}\|f(t)-M\|_{L^1_2} d t=0\,.
\end{equation}

(II) If
\begin{equation}\label{(5.7)}
\lim_{|t_1-t_2|\to 0} \|f(t_1)-f(t_2)\|_{L^1}=0\,,\quad
\sup_{t\ge t_0}\|f(t)\|_{L^1_{2+}} <\infty\,,
\end{equation}
then
\begin{equation}\label{(5.8)}
\lim_{t\to \infty}\|f(t)-M\|_{L^1_2} =0\,.
\end{equation}
}

\noindent{\it Proof}. First of all, we use
 $\|f(t)-M\|_{L^1_2}\le
C_N\sqrt{\|f(t)-M\|_{L^1}}$ (see (\ref{(1.10)})).  Hence, to prove the
theorem, it suffices to prove the apparently weaker result in which
$ \|f(t)-M\|_{L^1_2}$ is replaced by $\|f(t)-M\|_{L^1}$. Also, if we
let $D_{\rm min}(f)$ correspond to $B_{{\rm
min}}(z,\sg):=\min\{B(z,\sg)\,, 1\}$, then $D_{\rm min}(f(t))\le
D(f(t))$, so that by replacing $B(z,\sg)$ with $B_{{\rm
min}}(z,\sg)$, we can assume for notational convenience that $B$ is
bounded: $B(z,\sg)\le 1$.

>From (\ref{(5.1)}) we have
\begin{equation}\label{(5.9)}
\sup_{t\ge
0}\int_{{\bRN}}f(v,t)\left(\langle{v}\rangle^2+\Phi(f(v,t))\right)dv<\infty
\end{equation}
which implies that $\{f(\cdot,t)\}_{t\ge 0}$ is weakly compact in
$L^1({\bRN})$. Here and below  ``compact in $L^1$'' always means
``relatively compact in $L^1$".  We now split $Q(f)$ as usual as
\begin{equation}\label{(5.10)}
 Q(f)(v,t)=Q^{+}(f)(v,t)-Q^{-}(f)(v,t)
 \end{equation} where
\begin{equation}\label{(5.11)}
Q^{+}(f)(v,t)=\int_{{\bRSN}}B(v-v_*,\sg) f'f_*'d\sg
dv_*\,,
\end{equation}
\begin{equation}\label{(5.12)}
 Q^{-}(f)(v,t)=\int_{{\bRSN}}B(v-v_*,\sg)
ff_*d\sg dv_*=f(v,t)L(f)(v,t)
\end{equation}
\begin{equation}\label{(5.13)}
L(f)(v,t)=\int_{{\bRN}}\|B(z,\cdot)\|_{L^1({\bSN})}f(v-z,t)dz\,.
\end{equation}
Then using the special identity $Q^{+}(M)(v)=M(v)L(M)(v)$ we have
\begin{equation}\label{(5.14)}
(f-M)L(M)=-Q(f)
 +Q^{+}(f)-Q^{+}(M)-fL(f-M)\,.
 \end{equation}
As shown in the proof of Lemma 2.2, applying the inequality (\ref{(2.10)})
and the Cauchy-Schwarz inequality, we have $$\|Q(f)(t)\|_{L^1} \le
\int_{{\bRRSN}}B(v-v_*,\sg)|f'f_*'-ff_*|d\sg dv_*dv $$
\begin{equation}\label{(5.15)}\le \sqrt{4\int_{{\bRN}}f(v,t)L(f)(v,t)dv}
\, \sqrt{ D(f(t))}\le
\sqrt{4|{\bSN}|}\sqrt{ D(f(t))}\,.
\end{equation}
Let us write for any
$0<R<\infty$
$$f-M= {\bf 1}_{\{|v|\le R\}}\fr{1}{L(M)}(f-M)L(M)+{\bf 1}_{\{|v|> R\}}(f-M)\,.$$
Then using (\ref{(5.14)}), (\ref{(5.15)}) we have
\begin{equation}\label{(5.16)}
 \|f(t)-M\|_{L^1}\le \fr{1}{L_{R}}\bigg(C_N\sqrt{ D(f(t))}
+E(t)
 \bigg)+\fr{2N}{R^2}\,,\quad t\ge 0
 \end{equation}
 where $L_{R}=\min\limits_{|v|\le R}L(M)(v)>0$ and
 $$E(t)=\|Q^{+}(f)(t)-Q^{+}(M)\|_{L^1}+\|f(t) g(t)\|_{L^1}\,,
 \quad g(v,t)=L(f-M)(v,t)\,.$$
Here the positivity $L_R>0$ is obvious because the function
$v\mapsto L(M)(v)$ is continuous on ${\bRN}$ and, by the assumption
on $B$, $\|B(z,\cdot)\|_{L^1({\bSN})}>0$ a.e. $z\in {\bRN}$.

We next prove that for any sequence $\{t_n\}_{n\ge 1}\subset
[t_0,\infty)$ satisfying $t_n\to\infty \,(n\to\infty)$ and
\begin{equation}\label{(5.17)}
\sup_{n\ge 1}\|f(t_n)\|_{L^1_{2+}}<\infty\ ,
\end{equation}
 there exists a subsequence, still denoted by
$\{t_n\}_{n\ge 1}$, and a sequence $\{{\bar t}_n\}_{n\ge 1}$
satisfying $0\le {\bar t_n}-t_n\to 0\,(n\to\infty)$, such that
\begin{equation}\label{(5.18)}
\lim_{n\to\infty}E(t_n)=\lim_{n\to\infty}E({\bar t_n})=0\,,
\quad\lim_{n\to\infty}D(f({\bar
t_n}))=0\,.
\end{equation}
 To do this we use the fact that
$$\dt_{n}:=\sqrt{
\int_{t_{n}}^{\infty}D(f(t)) dt +\fr{1}{n}}\quad
\Longrightarrow\quad
\fr{1}{\dt_{n}}\int_{t_{n}}^{t_{n}+\dt_n}D(f(t)) dt< \dt_n$$ so that
there exist $\bar{t}_{n}\in [t_{n}\,,\, t_{n}+\dt_n]$ such that
$D(f(\bar {t}_{n}))\le \dt_n\to 0\,(n\to\infty)$ by the assumption
(\ref{(5.3)}).
 By $L^1$-weak compactness
of $\{f(\cdot,t)\}_{t\ge 0}$ there exists a subsequence of $\{(t_n,
{\bar t_n})\}_{n\ge 1}$, still denoted by $\{(t_n, {\bar
t_n})\}_{n\ge 1}$, and functions $0\le f_{\infty}, {\bar f}_{\infty}
\in L^1({\bRN})$ such that $f(\cdot, t_n) \to f_{\infty},\,f(\cdot,
{\bar t_n}) \to {\bar f}_{\infty}\,(n\to\infty)$ weakly in
$L^1({\bRN}).$ Since $0\le \bar {t}_n-t_{n}\le \dt_n\to 0$, it
follows from (\ref{(5.2)}) that $f_{\infty}={\bar f}_{\infty}$ a.e. on
${\bRN}$. On the other hand, by $f(\cdot,t_n)\in
L^1_{(1,0,1)}({\bRN})\,(\forall\, n)$ and (\ref{(5.17)}) we see that
$f_{\infty}\in L^1_{(1,0,1)}({\bRN})$. And by convexity and
nonnegativity of $(x,y)\mapsto (x-y)\log(x/y)$ and Fatou's Lemma, we
obtain
$$D(f_{\infty})\le \lim_{n\to\infty}D(f(\bar{t}_{n}))=0\,.$$
 Since  $B(z,\sg)>0$ for a.e.
$(z,\sg)\in{\bRSN}$, it follows from the well-known result that
$f_{\infty}(v)=a e^{-b|v-u_0|^2}$ a.e. on ${\bRN}$ for some
constants $a>0\,,b>0$ and $u_0\in {\bRN}$. Since $f_{\infty}, M\in
L^1_{(1,0,1)}({\bRN})$, this implies $f_{\infty}=M$ a.e. on
${\bRN}$.  We therefore conclude that $f(\cdot, t_n) \rightharpoonup
M\,, f(\cdot, {\bar t_n}) \rightharpoonup M\,\,(n\to\infty) $ weakly
in $L^1({\bRN})$.

Next, since $B(z,\sg)$ is bounded and $f(v,t)$ satisfies
(\ref{(5.9)}), it follows from Lions' compactness result (\cite{12},
see also e.g. \cite{2,13,20,21}) that the set
$\{Q^{+}(f)(\cdot,t)\}_{t\ge 0}$ is strongly compact in
$L^1({\bRN})$. For convenience of the reader, we give here a short
proof. By the criterion of $L^1$-strong compactness and the $L^1_2$-
bound $\sup\limits_{t\ge 0}\|Q^{+}(f)(\cdot,t)\|_{L^1_2}\le C_N $ we
need only to show that
\begin{equation}\label{(5.19)}
\sup_{t\ge 0}\|Q^{+}(f)(\cdot+h, t)-Q^{+}(f)(\cdot,t)\|_{L^1}
\to 0\quad {\rm as}\quad h\to 0\,.\
\end{equation}
 To do this, we use
truncation $$1<R<\infty\,,\quad f_{R}(v,t):=f(v,t) {\bf 1}_{\{|v|\le
R\}\cap \{f(v,t)\le R\}}\,.$$ Then $v\mapsto Q^{+}(f_R)(v,t)$ is
bounded and compactly supported uniformly in $t$:
\begin{equation}\label{(5.20)}
Q^{+}(f_R)(v,t)\le R^2\int_{{\bRSN}}
{\bf 1}_{\{|v'|\le R\}\cap \{|v_*'|\le R\}} d\sg dv_*\le C_R {\bf
1}_{\{|v|\le \sqrt{2}\, R\}}\,.
\end{equation}
 From e.g. \cite{13}
there is a strictly positive measurable function $\Psi_B(\xi)$
constructed from $B(z,\sg)$ satisfying $\Psi_B(\xi)\to\infty$
($|\xi|\to\infty)$ such that
$$
\int_{{\bRN}}\Psi_B(\xi)|{\cal F}(Q^{+}(f_R)(\cdot,t))(\xi)|^2d\xi
$$
\begin{equation}\label{(5.21)}
 \le C_N\int_{{\bRRN}}f_R^2(v,t)f_{R}^2(v_*,t)(1+|v-v_*|^2)^{N}dvdv_* \le
C_{R}\,.
\end{equation}
 Here and below $C_{R}<\infty$ depends only on
$ N$
 and $R$, and ${\cal F}(g)(\xi)$ is the Fourier transform:
$${\cal F}(g)(\xi)=\int_{{\bRN}} g(v) e^{-{\rm i}
\xi\cdot v} dv\,.$$
 From (\ref{(5.20)}) we have
$$Q^{+}(f_R)(v+h,t)-Q^{+}(f_R)(v,t)=
\left(Q^{+}(f_R)(v+h,t)-Q^{+}(f_R)(v,t)\right){\bf 1}_{\{|v|\le 1+
\sqrt{2}\, R\}}$$ for all $ v, h\in{\bRN}$ with $|h|\le 1\,.$
Therefore by Cauchy-Schwarz inequality, Parseval identity, and
(\ref{(5.21)}) (considering $|\xi|\le |h|^{-1/2}$ and $|\xi|> |h|^{-1/2}$ )
we obtain
 \beas&& \|Q^{+}(f_R)(\cdot+h,t)-Q^{+}(f_R)(\cdot,t)\|_{L^1}\\
&&\le C_R\left( \int_{{\bRN}} |1-e^{-{\rm i} \xi\cdot h}|^2|{\cal
F}(Q^{+}(f_R)(\cdot,t))(\xi)|^2d\xi\right)^{1/2} \\
&&\le
C_R\left(|h|+\sup_{|\xi|>|h|^{-1/2}}\fr{1}{\Psi_B(\xi)}\right)^{1/2}
=:C_R\Ld(h)\qquad \forall\, 0<|h|\le 1\,.\eeas On the other hand,
there is $R_0>0$ such that for all $R>R_0$ we have $\Phi(R)>0$ and
\beas&& \|Q^{+}(f_R)(\cdot,t)-Q^{+}(f)(\cdot,t)\|_{L^1}\le
C_N\int_{\{f(v,t)>R\}\cup\{|v|>R\}} f(v,t)dv\le
C_N\left(\fr{1}{\Phi(R)}+\fr{1}{R^2}\right)\,.\eeas Combining these
we obtain for all $0<|h|\le 1$ and all $R>R_0$ \beas&&\sup_{t\ge
0}\|Q^{+}(f)(\cdot+h, t)-Q^{+}(f)(\cdot,t)\|_{L^1}\le
C_{R}\Ld(h)+C_N\left(\fr{1}{\Phi(R)}+\fr{1}{R^2}\right)\eeas which
implies (\ref{(5.19)}) by first letting $h\to 0$ and then letting
$R\to\infty$.

For any $\psi\in L^{\infty}({\bRN})$, the function
$$\Psi(v,v_*):=\int_{{\bSN}}B(v-v_*, \sg)\psi(v')d\sg$$ belongs to
$L^{\infty}({\bRRN})$.
By $f(\cdot,t_n)-M \rightharpoonup 0$ weakly in $L^1({\bRN})$ and
(\ref{(5.9)}) we have  \beas&& \int_{{\bRN}}\psi(v)\bigg(
Q^{+}(f)(v,t_n)-Q^{+}(M)(v)\bigg)dv \\
&&=\int_{{\bRRN}}\Psi(v,v_*)
\bigg(f(v,t_n)f(v_*,t_n)-M(v)M(v_*)\bigg)dv_*dv\to 0 \quad
(n\to\infty)\,.\eeas  This implies $\lim\limits_{n\to\infty}\|
Q^{+}(f)(t_{n})-Q^{+}(M)\|_{L^1}= 0$ because
$\{Q^{+}(f)(t_n)\}_{n\ge 1}$ is strongly compact in $L^1({\bRN})$.
Also by weak convergence and $\|B(z,\cdot)\|_{L^1({\bSN})}\le
|{\bSN}|$ we have
$$g(v,t_n)=\int_{{\bRN}}\|B(v-v_*,\cdot)\|_{L^1({\bSN})}
\bigg(f(v_*,t_n)-M(v_*)\bigg)dv_*\to
0\quad (n\to\infty)$$ for all $v\in {\bRN}$. Since $|g(v,t_n)|\le
|{\bSN}|$, it follows from (\ref{(5.9)}) that
$\lim\limits_{n\to\infty}\|f(t_n) g(t_n)\|_{L^1}=0$. Thus
$\lim\limits_{n\to\infty}E(t_n)=0.$ The same argument also applies
to $f(v,{\bar t_n})$ and gives $\lim\limits_{n\to\infty}E({\bar
t_n})=0$.

Having proven (\ref{(5.18)}) (under the condition (\ref{(5.17)})),
we can now prove the timed averaging convergence (\ref{(5.6)}) and
the strong convergence (\ref{(5.8)}) for $L^1$-norm
$\|f(t)-M\|_{L^1}$. Suppose the assumptions in part (I) are
satisfied. By the assumption (\ref{(5.3)}) we have
$$\fr{1}{T}\int_{0}^{T}\sqrt{D(f(t))}\,dt\le
\sqrt{\fr{1}{T}\int_{0}^{T}D(f(t))\,dt}\to 0\quad (T\to\infty)\,.$$
Thus, from (\ref{(5.16)}) we see that to prove (\ref{(5.6)}) we need
only to prove that
\begin{equation}\label{(5.22)}
\fr{1}{T}\int_{0}^{T}E(t) dt\to 0\quad (T\to\infty)\,.
\end{equation}
We consider the following strategy:  For any given $\vep>0$,  choose
a sequence of $T_n=T_{n,\vep}\in[(2+t_0)^2,\infty)$ satisfying
$T_n\to\infty$ $(n\to\infty)$ such that
\begin{equation}\label{(5.23)}
\limsup_{T\to\infty}\fr{1}{T}\int_{t_0}^{T}\left(E(t)-\vep
\|f(t)\|_{L^1_{2+}}\right)dt=\lim_{n\to\infty}\fr{1}{|I_n|}
\int_{I_n}\left(E(t)-\vep\|f(t)\|_{L^1_{2+}} \right)dt
\end{equation}
where $I_n=[\sqrt{T_n}\,,\, T_n]$ and we have used the boundedness
$$\sup_{t\ge 0}E(t)\le 4|{\bSN}|\,,\quad C:=\sup_{T\ge
t_0}\fr{1}{T}\int_{t_0}^{T}\|f(t)\|_{L^1_{2+}}dt<\infty\,.$$ For
every $n\in{\bf N}$ there exists $t_n\in I_n$ such that
\begin{equation}\label{(5.24)}
\fr{1}{|I_n|}\int_{I_n}\left(E(t)-\vep \|f(t)\|_{
L^1_{2+}}\right)dt\le E(t_n)-\vep \|f(t_n)\|_{
L^1_{2+}}\,.
\end{equation}
 Since $T_n\ge (2+t_0)^2$, this gives an
$L^1_{2+}$-bound (\ref{(5.17)}):
 \beas&&
\|f(t_n)\|_{L^1_{2+}}\le
\fr{1}{\vep}E(t_n)+\fr{2}{T_n}\int_{t_0}^{T_n} \|f(t)\|_{
L^1_{2+}}dt \le \fr{4}{\vep}|{\bSN}|+2C\,.\eeas Therefore there is a
subsequence, still denote it as $\{t_n\}_{n\ge 1}$, such that
$E(t_n)\to 0$ as $n\to\infty$. By (\ref{(5.23)}), (\ref{(5.24)}) we thus obtain
$$\limsup_{T\to\infty}\fr{1}{T}\int_{t_0}^{T}E(t)dt\le
\limsup_{T\to\infty}\fr{1}{T}\int_{t_0}^{T}\left(E(t)-\vep\|f(t)\|_{
L^1_{2+}}\right)dt +\vep C\le \vep C\,.$$  Letting $\vep\to 0$ leads
to (\ref{(5.22)}).

Finally suppose the assumptions in part (II) are satisfied. Choose a
sequence $\{t_n\}_{n\ge 1}\subset [t_0,\infty)$ satisfying
$t_n\to\infty$ such that
$$\limsup_{t\to\infty}\|f(t)-M\|_{L^1}=\lim_{n\to\infty}\|f(t_n)-M\|_{L^1}
\,.$$ By the assumption in (\ref{(5.7)}), $\sup_{n\ge
1}\|f(t_n)\|_{L^1_{2+}} <\infty$. So there exist a subsequence,
still denote it as $\{t_n\}_{n\ge 1}$, and a sequence $\{{\bar
t}_n\}_{n\ge 1}$ satisfying $0\le {\bar t_n}-t_n\to
0\,(n\to\infty)$, such that (\ref{(5.18)}) holds true. Therefore by
assumption (\ref{(5.7)}) and applying (\ref{(5.16)}) we have
$\lim\limits_{n\to\infty}\|f(t_n)-f({\bar t_n})\|_{L^1}=0$ and
\beas&&\lim_{n\to\infty}\|f(t_n)-M\|_{L^1}\le
\limsup_{n\to\infty}\|f({\bar t_n})-M\|_{L^1}
\\
&&\le \fr{1}{L_{R}}\bigg(C_N \lim_{n\to\infty}\sqrt{ D(f({\bar
t_n}))} +\lim_{n\to\infty}E({\bar t_n})
 \bigg)+\fr{2N}{R^2}=\fr{2N}{R^2}\qquad \forall\, 0<R<\infty\,.\eeas
This proves $\lim\limits_{n\to\infty}\|f(t_n)-M\|_{L^1}=0$  by
letting $R\to \infty$. $\quad \Box$
\smallskip

\noindent{\bf Proof of Theorem 1: Time Averaged Convergence}. This
is a consequence of part (I) of Theorem 4 because the weak solution
$f$ in Theorem 1 satisfies all conditions (\ref{(5.1)})-(\ref{(5.5)}) with
$\Phi(f)=|\log f|$ and $\Psi(v)=\langle{v}\rangle^{s+\gm-2}$, where
the uniform continuity (\ref{(5.2)}) is indeed satisfied for any weak
solution (see the proof of (\ref{(3.5)})).$\quad \Box$

\section{Lower Bounds on the Convergence Rate}

We first consider in the following two special cases (a) and (b) which motivate
our work for general cases on the lower bounds of convergence rate.

{\bf Case (a)}.  $f|_{t=0}=f_0\in L^1_{(1,0,1)}({\bRN})$ and $f\in
L^{\infty}([0,\infty); L^1_2({\bRN}))$ is a mild solution of
Eq.(B) with $B(z,\sg)$ satisfying
$$\|B(z,\cdot)\|_{L^1({\bSN})}\le
K(1+|z|^2)^{\gm/2}\,,\quad -\infty<\gm<0\,.$$

{\bf Case (b)}. $f|_{t=0}=f_0\in L^1_{(1,0,1)}\cap L^1{\rm
log}L({\bRN})$ is an isotropic function and  $f$ is an isotropic
weak solution of Eq.(B) with $B(z,\sg)$ satisfying
$$
\|B(z,\cdot)\|_{L^1({\bSN})}\le K |z|^{\gm}\,,\quad -4\le
\gm\,,\,\,1-N<\gm<0\,.$$

Note that under the assumption in Case (a), the existence and uniqueness
of mild solution $f$ is well-known and $f$ conserves the mass,
momentum and energy. For Case (b), the existence of
isotopic weak solution $f$ is also obvious.
\smallskip

\noindent{\bf Theorem 5}. {\it For  each Case (a) and Case (b) there are
(explicit) constants $0<C_1,C_2<\infty$ depending only on $N, K$ and
$\gm$ such that
\begin{equation}\label{(6.1)}
\|f(t)-M\|_{L^1_2}\ge C_1\int_{|v|>t^{\alpha}}|v|^2 f_0(v)dv-C_2
\exp(-t^{2\alpha}/4)\qquad \forall\, t\ge 0
\end{equation}
where $\alpha=1/\min\{|\gm|\,,\, 2\}$ for Case (a) and $\alpha=1/|\gm|$ for Case (b).}
\medskip

\noindent{\it Proof}. Our proof relies on pointwise inequalities for mild
solutions of Eq.(B). Recall that a nonnegative measurable function
$f(v,t)$ on ${\bRN}\times[0,\infty)$ is called a mild solution of
Eq.(B) with initial datum $f_0$ if there is a null set $Z\subset
{\bRN}$ which is independent of $t$ such that
$$\int_{0}^{t}Q^{\pm}(f)(v,\tau)d\tau<\infty\qquad \forall\, v\in {\bRN}\setminus Z\,,
\quad \forall \, t\ge 0$$ and
$$f(v,t)=f_0(v)+\int_{0}^{t}Q(f)(v,\tau)d\tau \qquad \forall\, v\in {\bRN}\setminus Z\,,
\quad \forall \, t\ge 0\,.$$ Here $Q^{+}(f)$ and $Q^{-}(f)=fL(f)$
are defined in (\ref{(5.10)})-(\ref{(5.13)}).

In the following, we use the  letter $Z$ to denote a null set
of ${\bRN}$ which is independent of $t$ -- the particular null set may
change from line to line. Recall also that if $f$ is
a mild solution and satisfies that for almost every $v\in{\bRN}$ the
function $t\mapsto L(f)(v,t)$ is locally integrable on
$[0,\infty)$, then the the following integral formula
holds:
$$f(v,t)=f_0(v)\exp(-\int_{0}^{t}L(f)(v,\tau)d\tau)+
\int_{0}^{t}Q^{+}(f)(v,\tau)\exp(-\int_{\tau}^{t}L(f)(v,\tau_1)d\tau_1)d\tau
$$
for all $t\ge 0$ and all $v\in{\bRN}\setminus Z$.  This gives
\begin{equation}\label{(6.2)}
f(v,t)\ge f_0(v)\exp\left(-\int_{0}^{t}L(f)(v,\tau)d\tau\right)\qquad \forall\,
t\ge 0\,,\quad\forall\,v\in{\bRN}\setminus Z\,.
\end{equation}

\noindent{Case (a)}.  By the assumption on $B$, we have
$$L(f)(v,t)\le K\int_{{\bRN}}
\fr{f(v_*,t)}{(1+|v-v_*|)^{|\gm|}}dv_*\,.$$
Let
$\beta=\min\{|\gm|\,,\, 2\}$. Then for all $v,v_*\in{\bRN}$,
$$\fr{1}{(1+|v-v_*|)^{|\gm|}}
\le
\fr{C_{\beta}}{|v|^{\beta}}\left(1+|v_*|^2\right)\  .$$
Hence, using conservation
of mass and energy yields the bound
$$L(f)(v,t)\le
\fr{C_{K,\beta}\|f_0\|_{L^1_2}}{|v|^{\beta}}=\fr{c}{|v|^{\beta}}\,.$$
Thus,
$$f(v,t)\ge f_0(v)\exp(-\fr{c\, t}{|v|^{\beta}})\qquad \forall\,
t\ge 0\,,\quad\forall\,v\in{\bRN}\setminus Z\,,$$ and in
particular,
\begin{equation}\label{(6.3)}
f(v,t)\ge f_0(v)e^{-c}\qquad \forall\,
t\ge 0\,,\quad\forall\,v\in{\bRN}\setminus Z\quad {\rm s.t.}\quad
|v|>t^{\alpha}
\end{equation}
 with $\alpha=1/\beta$. Consequently,
$$\int_{|v|>t^{\alpha}}|v|^2 f(v,t)dv\ge e^{-c}\int_{|v|>t^{\alpha}}|v|^2 f_0(v)dv\qquad \forall\,
t\ge 0\,.$$ Since $M(v)=(2\pi)^{-N/2} e^{-|v|^2/2}$, it follows that (\ref{(6.1)}) holds true:
\beas&&\|f(t)-M\|_{L^1_2}\ge \int_{|v|>t^{\alpha}}|v|^2
f(v,t)dv-\int_{|v|>t^{\alpha}}|v|^2 M(v)dv\\
&&\ge  e^{-c}\int_{|v|>t^{\alpha}}|v|^2 f_0(v)dv-C_N
\exp(-t^{2\alpha}/4)\,\qquad \forall\, t\ge 0\,.\eeas

\noindent{Case (b)}.   We need to prove that under the assumptions
in Case (b), $f$ is also a mild solution after a modification on a
null set of ${\bRN}\times[0,+\infty)$. By $-4\le \gm<0$ there is
$m\in \{1,2\}$ such that $0\le m-|\gm|/2\le 1$. We show that
\begin{equation}\label{(6.4)}
\int_{{\bRN}}|v|^{2m}Q^{\pm}(f)(v,t)dv\le C\|f_0\|_{L^1_2}^2\qquad
\forall\, t\ge 0\,.
\end{equation}
 where $C<\infty$ depends only on
$K,\gm$ and $N$. In fact using $\|B(v-v_*,\cdot)\|_{L^1({\bSN})} \le
K |v-v_*|^{\gm}$ and $|v'|^2\le |v|^2+|v_*|^2$ we have \beas&&
\int_{{\bRN}}|v|^{2m}Q^{\pm}(f)(v,t)dv\le
K\int_{{\bRRN}}\fr{f(v,t)f(v_*,t)(|v|^2+|v_*|^2)^{m}}{|v-v_*|^{|\gm|}}dv_*
dv\,.\eeas Since $f$ is isotropic, i.e., $f(v,t)=f(|v|,t)$, and
$|\gm|<N-1$, it follows from Lemma 6.1 (see below) that
\beas&&\int_{{\bRN}}\fr{f(v_*,t)(|v|^2+|v_*|^2)^{m}}{|v-v_*|^{|\gm|}}
dv_*\le C_{N,\gm} \int_{{\bRN}}f(v_*,t)(|v|^2+|v_*|^2)^{m-|\gm|/2}
dv_*\,.\eeas  Since $(|v|^2+|v_*|^2)^{m-|\gm|/2}\le
\langle{v}\rangle^2\langle{v_*}\rangle^2$, (\ref{(6.4)}) follows
from conservation of mass and energy. By (\ref{(6.4)}) we have
$$\int_{0}^{T}Q^{\pm}(f)(v,t)dt<\infty\qquad \forall\, v\in
{\bRN}\setminus Z\,,\quad \forall\, T\in [0,\infty)\,.$$
Take any $\psi\in C^{\infty}_c({\bRN})$ and let
$$\vp(v)=\rho(v)\psi(v)\,,\quad \rho(v)=\left(\fr{|v|^2}{1+|v|^2}\right)^{m}\,.$$
Then using (\ref{(6.4)}) we have
$$\int_{{\bRN}}\vp(v)
Q^{\pm}(f)(v,t)dv\le C_{\psi} \|f_0\|_{L^1_2}^2<\infty\qquad
\forall\, t\ge 0\,.$$ Therefore there is no integrability problem
and we have with this $\vp(v)=\psi(v)\rho(v)$
$$-\fr{1}{4}\int_{0}^{t}d\tau\int_{{\bRN}}Q(f\,|\Dt\vp)(v,\tau)dv=
\int_{0}^{t}d\tau\int_{{\bRN}}\vp(v) Q(f)(v,t) dv\,.$$ Since $f$ is
a weak solution, this gives for all $\psi\in C^{\infty}_c({\bRN})$ \beas&&
\int_{{\bRN}}\psi(v)\rho(v)\left\{f(v,t)-f_0(v)-\int_{0}^{t}Q(f)(v,\tau)d\tau
\right\}dv=0\quad \forall\, t\ge 0\,.\eeas By $L^1$ -integrability
of $\rho(v)\{\cdots\}$ and the strict positivity of $\rho(v)$ on
$v\neq 0$, this implies
$$f(v,t)=f_0(v)+\int_{0}^{t}Q(f)(v,\tau)d\tau
\qquad \forall\, t\ge 0\,,\quad \forall\, v\in {\bRN}\setminus
Z_t$$ where $Z_t$ are null sets that may depend on $t$.
Let
$${\bar f}(v,t):=\left|f_0(v)+\int_{0}^{t}Q(f)(v,\tau)d\tau\right|\,,
\quad (v,t)\in{\bRN}\times[0,\infty)\,.$$ Then it is not difficult
to prove that ${\bar f}(v,t)$ is a mild solution of Eq.(B) and for any $t\in
[0,\infty)$, $f(v,t)={\bar f}(v,t)$ a.e. $v\in{\bRN}$.
Thus we can replace ${\bar f}$ with $f$.

Now for all $v\in{\bRN}\setminus\{0\}$ we compute using Lemma 6.1 below
and $\|f(t)\|_{L^1}=1$ that \beas&&L(f)(v,t)\le
K\int_{{\bRN}}\fr{f(v_*,t)}{|v-v_*|^{|\gm|}}dv_*\le
\fr{c\|f(t)\|_{L^1}}{|v|^{|\gm|}}=\fr{c}{|v|^{|\gm|}}\qquad \forall\,
t\ge 0\,.\eeas Therefore (\ref{(6.3)}) holds for
$\alpha=1/|\gm|$. This gives
$$\int_{|v|>t^{\alpha}}|v|^2f(v,t)dv
\ge e^{-c}\int_{|v|>t^{\alpha}}|v|^2f_0(v)dv\qquad \forall\, t\ge
0$$ and thus, (as shown above) $f$ satisfies the inequality (\ref{(6.1)}).
$\qquad \Box$
\smallskip

\noindent{\bf Lemma 6.1 }  {\it Let $f(v)=f(|v|)$ be a nonnegative
isotropic measurable  function on ${\bRN}$ with $N\ge 2$. Then
For all $0\le \alpha, \beta <N-1$ and $v\in {\bRN}$
$$ \int_{{\bRN}}\fr{f(v_*)dv_*}{|v+v_*|^{\alpha}|v-v_*|^{\beta}}\le C_{N,\alpha,\beta}
\int_{{\bRN}} \fr{f(v_*)dv_*}{(|v|^2+|v_*|^2)^{(\alpha+\beta)/2 }}
$$ where $C_{N,\alpha,\beta}=2^{(N+1)/2}\fr{|{\bf
S}^{N-2}|}{|{\bSN}|}\left(\fr{1}{N-1-\alpha}+\fr{1}{N-1-\beta}\right)
$\,.}
\smallskip

\noindent{\it Proof}. For any $v\in{\bRN}$, let $ v=\rho\og, \rho\ge 0,
\og\in{\bSN}$. Then \beas
&&\int_{{\bRN}}\fr{f(|v_*|)dv_*}{|v+v_*|^{\alpha}|v-v_*|^{\beta}}=\int_{0}^{\infty}r^{N-1}f(r)
\left(\int_{{\bSN}} \fr{d\sg}{|\rho\og
+r\sg|^{\alpha}|\rho\og-r\sg|^{\beta}}\right) dr\,. \eeas Since
$\rho^2+r^2\pm 2\rho r\,t\ge \fr{1}{2}(\rho^2+r^2)(1-t^2)$ for $\pm t\le
0$, it follows
that
 \beas&&\int_{{\bSN}} \fr{d\sg}{|\rho\og
+r\sg|^{\alpha}|\rho\og-r\sg|^{\beta}}\\
&&\le\fr{|{\bf S}^{N-2}|}{(\rho^2+r^2)^{(\alpha+\beta)/2}}
\left(2^{\alpha/2}\int_{0}^{1}(1-t^2)^{(N-3-\alpha)/2} dt+
2^{\beta/2}\int_{0}^{1}(1-t^2)^{(N-3-\beta)/2} dt\right)
\\
&&\le
C_{N,\alpha,\beta}|{\bSN}|\fr{1}{(\rho^2+r^2)^{(\alpha+\beta)/2}}\,.\eeas
This implies the inequality in the lemma.$\quad \Box$

It is not clear whether the lower bound estimate (\ref{(6.1)}) can be extended
to all weak solutions. Our proof for Theorem 2 is very
different from the above argument.

\noindent{\bf Proof of Theorem 2}.

Part (I). By identity $|a-b|=b-a+2(a-b)^{+}$ ( with
$(y)^{+}=\max\{y,0\}$) and conservation of mass and energy we have,
for all $t\ge 0,\,R>0$,
\begin{equation}\label{(6.5)}
\|f(t)-M\|_{L^1_2}\ge 2\int_{|v|>R}|v|^2f(v,t)dv-C_N
e^{-R^2/4}
\end{equation}
 where we used the
exponential decay of  $M(v)=(2\pi)^{-N/2} e^{-|v|^2/2}$. We now
prove that there is finite constant $C_{\gm}>0$ depending only on
$N, \gm, A^*, A_*, s,
\|f_0\|_{L^1_s}, H(f_0)$ and $K_0$  such that for all $R\ge 3$ and all $t\ge 0$
\begin{equation}\label{(6.6)}
\int_{|v|>R}|v|^2f(v,t)dv
\ge
\int_{|v|>2R}|v|^2f_0(v)dv-\fr{C_{\gm}(1+t)^{2-[2/s]}}{R^{\beta}}\,.
\end{equation}
 where $\beta=\min\{s,\, s-2+|\gm|\}$. To prove (\ref{(6.6)}),
we use truncation. Let $\chi\in C^{\infty}({\bRN})$ be the function
given in (\ref{(3.1)}). For any $R>1$, let $\psi_{R}(v)=|v|^2\chi(v/R).$
Then  $\psi_R\in {\cal T}$,
\begin{equation}\label{(6.7)}
|v|^2{\bf 1}_{\{|v|> 2R\}}\le \psi_{R}(v)\le |v|^2{\bf 1}_{\{|v|>R\}}\qquad
\forall\, v\in{\bRN}\,,
\end{equation}  and
$\sup\limits_{R>1}
\|\p^2\psi_R\|_{L^{\infty}}\le C_N\,.$ Also using
$|v'|^2+|v_*'|^2=|v|^2+|v_*|^2$ we have
\begin{equation}\label{(6.8)}
\Dt\psi_{R}(v', v_*', v, v_*)=\Dt\psi_{R}(v', v_*', v,
v_*){\bf 1}_{\{|v|^2+|v_*|^2>R^2\}}\,.
\end{equation}

Suppose $-4\le \gm<-2$.  Using the equation (\ref{(1.6)}) to  $\vp=\psi_{R}$
we obtain from (\ref{(6.7)}) \beas&& \int_{|v|>R}|v|^2f(v,t)dv\ge
\int_{{\bRN}}\psi_{R}(v)f(v,t)dv
\\
&&\ge \int_{|v|>2R}|v|^2f_0(v)dv-\fr{1}{4}\int_{0}^{t}d\tau
\int_{{\bRN}}Q(f\,|\,\Dt\psi_R)(v,\tau)dv\,.\eeas Let $Q^{1}(\cdot,
\cdot)$ and $L_{1}[\cdot] $ be the operators defined in Lemma 2.3
for $\ld=1$ corresponding to the kernels $B^1(z,\sg)={\bf 1}_{\{|v-v_*|\le
1\}}B(z,\sg)$ and $B_1(z,\sg)={\bf 1}_{\{|z|>
1\}}B(z,\sg)$ respectively.  Then by Lemma 2.3 we have \beas&&
\int_{0}^{t}d\tau
\int_{{\bRN}}Q(f\,|\,\Dt\psi_R)(v,\tau)dv
\\
&&=
\int_{0}^{t}d\tau\int_{{\bRN}}Q^{1}(f\,|\,\Dt\psi_R)(v,\tau)dv
-2\int_{0}^{t}d\tau\int_{{\bRRN}}L_{1}[\Dt\psi_R]ff_*
dv_*dv\,. \eeas We compute as  in the proof of Lemma 2.2 that
\beas&&
\left|\int_{0}^{t}d\tau\int_{{\bRN}}Q^{1}(f\,|\,\Dt\psi_R)(v,\tau)dv\right|
\\
&&\le C\int_{0}^{t}d\tau \left(\int_{|v|^2+|v_*|^2\ge
R^2}|v-v_*|^{4+\gm}{\bf 1}_{\{|v-v_*|\le 1\}} ff_* dv_*
dv\right)^{1/2}\sqrt{D(f(\tau))}\\
&&\le C\left(\int_{0}^{t}d\tau
\int_{|v|^2+|v_*|^2\ge R^2}{\bf 1}_{\{|v-v_*|\le 1\}}
ff_* dv_*
dv\right)^{1/2}
\eeas
By $R\ge 3$, we see that $|v|>R/\sqrt{2}$ and $|v-v_*|\le 1$ imply $|v_*|>R/3$
and so \beas&&
\int_{0}^{t}d\tau\int_{|v|^2+|v_*|^2\ge R^2}{\bf 1}_{\{|v-v_*|\le
1\}} ff_* dv_* dv \le 2\int_{0}^{t}d\tau\left(\int_{|v|\ge R/3}
f(v,\tau)dv\right)^2 \\
&&\le \fr{C}{R^{2s}}\int_{0}^{t}(1+\tau)^{2-2[2/s]}
d\tau\le \fr{C(1+t)^{3-2[2/s]}}{R^{2s}} \eeas where we have used the
moment estimates (for $s>2$) and the conservation of mass and energy (for $s=2$).
 Thus
\beas&&
\left|\int_{0}^{t}d\tau\int_{{\bRN}}Q^{1}(f\,|\,\Dt\psi_R)(v,\tau)dv\right|
\le \fr{C(1+t)^{3/2-[2/s]}}{R^{s}}\,. \eeas Also by $2+\gm<0$ we
have \beas&&
\left|\int_{0}^{t}d\tau\int_{{\bRN}}L_{1}[\Dt\psi_R]ff_*
dv\right|\le
C\int_{0}^{t}d\tau\int_{|v|^2+|v_*|^2\ge
R^2}|v-v_*|^{2+\gm}{\bf 1}_{\{|v-v_*|>1\}}ff_* dv_*dv
\\
&&\le C\int_{0}^{t}d\tau\int_{|v|\ge R/\sqrt{2}}
f(v,\tau)dv\le \fr{C}{R^s}\int_{0}^{t}(1+\tau)^{1-[2/s]}d\tau\le \fr{C(1+t)^{2-[2/s]}}{R^{s}}\,.\eeas
Therefore
\begin{equation}\label{(6.9)}
\int_{|v|>R}|v|^2f(v,t)dv\ge \int_{|v|>2R}|v|^2f_0(v)dv
-\fr{C_{\gm}(1+t)^{2-[2/s]}}{R^{s}}\,.
\end{equation}

Next suppose $-2\le \gm<0$. In this case we use
(\ref{(6.8)}) and moment estimates to get
 \beas&&
\left|\int_{0}^{t}d\tau
\int_{{\bRRN}}L[\Dt\psi_R]ff_*dv_*v\right|
\le C\int_{0}^{t}d\tau\int_{|v|^2+|v_*|^2>R^2}
|v-v_*|^{2+\gm}ff_*dv_* dv\\
&&\le \fr{C}{R^{s-2+|\gm|}}\int_{0}^{t}d\tau\int_{{\bRRN}}
(|v|^2+|v_*|^2)^{s/2}ff_*dv_* dv\le
C\fr{(1+t)^{2-[2/s]}}{R^{s-2+|\gm|}} \,,\quad t\ge 0\,.\eeas
Therefore using equation (\ref{(1.7)}) with  $\vp=\psi_R$ we obtain
\begin{equation}\label{(6.10)}
\int_{|v|>R}|v|^2f(v,t)dv\ge \int_{|v|>2R}|v|^2f_0(v)dv
-\fr{C_{\gm}(1+t)^{2-[2/s]}}{R^{s-2+|\gm|}}\,.
\end{equation}
The inequality (\ref{(6.6)}) follows from (\ref{(6.9)}) and (\ref{(6.10)}).

 Combining (\ref{(6.5)}) and (\ref{(6.6)}) and using $e^{-R^2/4}\le \fr{C}{R^{\beta}}\le\fr{C(1+t)^{2-[2/s]}}{R^{\beta}}$
we get with a larger constant $0<K<\infty$
such  that for all $R\ge 3$
\begin{equation}\label{(6.11)}
 \|f(t)-M\|_{L^1_2}
\ge 2\int_{|v|>2R}|v|^2f_0(v)dv-\fr{K(1+t)^{2-[2/s]}}{(2R)^{\beta}}
\qquad t\ge 0\,.
\end{equation}
  We can assume $K\ge \max\{ K_0,\,
6^{\beta} N\}$ so that the solution ${\rm R}(t)$ of the equation
(\ref{(1.14)}) satisfies ${\rm R}(t)\ge 6$ for all $t\ge 0$. Therefore
inserting $R=\fr{1}{2}{\rm R}(t)$ into (\ref{(6.11)}) and using equation
(\ref{(1.14)}) gives (\ref{(1.15)}).

Part (II). Recall the assumptions in the theorem. We have for all
$R\ge R_0$
\begin{equation}\label{(6.12)}
 \int_{|v|>R}f_0(v)|v|^2dv\ge C\int_{R}^{\infty}
r^{-1-\dt}dr=C R^{-\dt}.
\end{equation}
  Since $\dt<\beta$, $f_0$
satisfies (\ref{(1.13)}). By Part(I) of Theorem 2, the solution $f(v,t)$
satisfies (\ref{(1.15)}) with the function ${\rm R}(t)$ defined through
(\ref{(1.14)}) for some constant $K\ge N(R_0)^{\beta}$ which implies that
${\rm R}(t)\ge R_0$ for all $t\ge 0$. By (\ref{(1.14)}) and (\ref{(6.12)}) with
$R={\rm R}(t)$ we have
$$ K(1+t)^{2}\ge C({\rm R}(t))^{\beta-\dt}\,.$$
So ${\rm R}(t)\le C(1+t)^{2/(\beta-\dt)}$ and hence using (\ref{(6.12)})
with $R={\rm R}(t)$ again we obtain
$$\int_{|v|>{\rm R}(t)}f_0(v)|v|^2dv\ge C({\rm R}(t))^{-\dt}
\ge C(1+t)^{-2\dt/(\beta-\dt)}\,,\quad t\ge 0\,.$$ This proves
(\ref{(1.17)}).

Part (III). Let $s=2$.  By the assumption on $f_0$, and recalling that
$A_1(t)=-\fr{d}{dt}A(t)\ge 0$, we have
\begin{equation}\label{(6.13)}
\int_{|v|>R}f_0(v)|v|^2dv\ge C\int_{R}^{\infty} A_1(r)
dr=CA(R)\qquad \forall\,R\ge R_0.
\end{equation}
 On the other hand, by
assumptions on $A(t)$, there is a constant $C>0$ such that  $A(R)\ge
C R^{-\dt}$ for all $R\ge R_0$. Since $\dt<\beta$, this implies that
$f_0$ satisfies (\ref{(1.13)}) and thus from the equation
(\ref{(1.14)}) (with some constant $K\ge N(R_0)^{\beta}$) we get as
shown above that ${\rm R}(t)\ge R_0$ and
$$K(1+t)\ge C({\rm R}(t))^{\beta-\dt}\,, \qquad t\ge 0\,.$$
So ${\rm R}(t)\le c (1+t)^{\alpha}$ with $\alpha=1/(\beta-\dt)$.
Applying (\ref{(6.13)}) with $R={\rm R}(t)$,
and recalling that $A(t)$ is non-increasing,
we obtain
$$ \int_{|v|>{\rm R}(t)}f_0(v)|v|^2dv
\ge C A(c(1+t)^{\alpha})\,,\quad t\ge 0\,.$$ Finally by assumption
(\ref{(1.18)}) on $A(t)$ it is easily proved that there are constants
$0<c_1, C_1<\infty$ such that $A(c(1+t)^{\alpha})\ge
C_1A(c_1t^{\alpha})$ for all $t\ge 0$. This proves (\ref{(1.20)}). $\quad
\Box$

\section{Upper Bounds on the Convergence Rate for $\gamma \ge -1$ and Grad Angular Cutoff  }

This section is devoted to the proof of Theorem 3.   There are several
somewhat long and technical arguments, as well as some that are shorter
and more plainly motivated. We begin with one of the more technical lemmas
which shall later in this section be used to bound the {\em entropic moments}
as explained in the introduction.  It is through this lemma that the condition
$\gamma \ge -1$ enters Theorem 3. Next, in Lemma 7.2, we prove the Gronwall
type lemma that we shall use to get power law convergence out of the  entropy
production estimate that we shall eventually prove here, and then,
with the crucial preliminaries behinds us, in Lemma 7.3, we recall Villani's
entropy production bound for super hard potentials, and explain how we shall
use it here.  On a first reading of this section, the reader may wish to turn
to Lemma 7.3, and start from there, though in our exposition we now turn to
Lemma 7.1.

Instead of considering directly the growth of entropic moments defined in
terms of
$\langle v\rangle ^k$, we shall instead
use a more symmetric function $S$:
\begin{equation}\label{(7.1)}
S(v,v_*, v', v_*')=\min\{\max\{\Phi(v)\,, \Phi(v_*)\}\,,\,
\max\{\Phi(v')\,, \Phi(v_*')\}\}
\end{equation}
where
\begin{equation}\label{(7.2)}
\Phi(v)=\min\{\langle{v}\rangle^k\,, R\}\,,\quad k>0\,, R>0\,.
\end{equation}
The idea is that the difference $\Phi-S$ contributes a factor
$|v-v_*|$ that kills the singularity $|v-v_*|^{\gm}$ of
$B(v-v_*,\sg)$ and this is why we  assume that $-1\le \gm<0$, while
the integral involved $S$ can be treated as an entropy dissipation.
Also to overcome the problem of $f_0$ having no pointwise lower
bounds we consider a suitable convex combination of the solution and
the Maxwellian.

 We need several lemmas.
\smallskip

{\bf Lemma 7.1.} {\it Let $b(\cdot)\ge 0$ satisfy (\ref{(1.22)}),
$0<\beta\le 1$, and let $\Phi$, $S$ be introduced above with $k\ge
1$. Then for all $0\le f\in L^1_{k}\cap L^1{\rm log }L({\bRN})$ we
have (with $\cos\theta=\langle{(v-v_*)/|v-v_*|,\sg}\rangle$)
$$\int_{{\bRRSN}}\fr{b(\cos\theta)}{|v-v_*|^{\beta}}
\bigg(S(v,v_*,v',v_*')-\Phi(v)\bigg)^{+}ff_*\log^{+}f\, d\sg dv_* dv$$
\begin{equation}\label{(7.3)}
\le 2k A_0\left(\|f\log ^{+}f\|_{L^1}+
\|f\|_{L^1}\right)\|f\|_{L^1_{k-\beta}}\,.
\end{equation}
$$\int_{{\bRRSN}}\fr{b(\cos\theta)}{|v-v_*|^{\beta}}
\bigg(\Phi(v')-S(v,v_*,v',v_*')\bigg)^{+}ff_*\log^{+}f'\, d\sg dv_* dv$$
\begin{equation}\label{(7.4)}
\le 4^{k}NA_0\left(\|f\log ^{+}f\|_{L^1}+
\|f\|_{L^1}\right)\|f\|_{L^1_{k-\beta}}\,.
\end{equation}
 where
$(y)^{+}=\max \{y, 0\}$ and $A_0$ is given in (\ref{(1.22)}). }

Note that the right hand side of the inequalities do not depend on $R$,
so letting $R\to\infty$ one sees that
(\ref{(7.3)})-(\ref{(7.4)}) hold also for $\Phi(v)=\langle{v\rangle}^k$
by Fatou's Lemma.

To prove the lemma we will use the following formula which are
results of changing variables: Let $W(t)\ge 0, f(v)\ge 0 $ be
measurable on $[-1,1]$ and ${\bRN}$ respectively. Then for all
$v\in{\bRN}$
\begin{equation}\label{(7.5)}\int_{{\bRSN}}W(\cos\theta) f(v')d\sg dv_*=
\left(|{\bf
S}^{N-2}|\int_{0}^{\pi}\fr{W(\cos\theta)\sin^{N-2}\theta}{\sin^{N}(\theta/2)}
d\theta\right) \int_{{\bRN}}f(v_*)dv_*\,,
\end{equation}
\begin{equation}\label{(7.6)}
\int_{{\bRSN}}W(\cos\theta) f(v_*')d\sg dv_*=
\left(|{\bf
S}^{N-2}|\int_{0}^{\pi}\fr{W(\cos\theta)\sin^{N-2}\theta}{\cos^{N}(\theta/2)}
d\theta\right) \int_{{\bRN}}f(v_*)dv_*\,.
\end{equation}
\smallskip

\noindent{\bf Proof of Lemma 7.1.}   We first prove the following
inequalities:
\begin{equation}\label{(7.7)}
\bigg(S(v,v_*,v',v_*')-\Phi(v)\bigg)^{+}
\le 2k \langle{v_*}\rangle^{k-\beta}
|v-v_*|^{\beta}\,,
\end{equation}
\begin{equation}\label{(7.8)}
\bigg(\Phi(v')-S(v,v_*,v',v_*')\bigg)^{+}
\le
 k 2^{k}\bigg(\langle{v}\rangle^{k-\beta}+
\langle{v_*}\rangle^{k-\beta}\bigg)|v-v_*|^{\beta}m(\theta)\,,
\end{equation}
\begin{equation}\label{(7.9)}
m(\theta)  f_*\left(\log^{+}f'+\log^{+}f_*'
\right)
\le  2 f_*
\log^{+}f_* +2Nf_*+(m(\theta))^N\left(  f'\log^{+}f'+f_*'\log^{+}f_*'\right)\,,
\end{equation}
where
$$m(\theta)=\min\{\sin(\theta/2)\,,\, \cos(\theta/2)\}\,.$$
Given any $(v,v_*,\sg)\in{\bRRSN}$.
Suppose $\Phi(v)<S(v,v_*,v',v_*')$. In this case we have by definition of $S$
that
$S(v,v_*,v',v_*')\le \max\{\Phi(v),\,\Phi(v_*)\}$. Consequently,
 $\langle{v}\rangle<\langle{v_*}\rangle$ and so, $|v- v_*| \le |v|+|v_*|
 \le 2\langle{v_*}\rangle$.
 Therefore,
\beas&&0<S(v,v_*, v', v_*')-
\Phi(v)\le  k\langle{v_*}\rangle^{k-1}|v-v_*|
\le k 2\langle{v_*}\rangle^{k-\beta}|v-v_*|^{\beta}\,.
\eeas
The other case to consider is slightly more involved.
Suppose $\Phi(v')>S(v,v_*,v',v_*')$. Then,
$$S(v,v_*,v',v_*')=\max\{\Phi(v),\,\Phi(v_*)\}\,.$$
Since
$|v'-v|=|v-v_*|\sin(\theta/2),\, |v'-v_*|=|v-v_*|\cos(\theta/2)$,
it follows that
\beas&& 0<\Phi(v')-S(v,v_*,v',v_*')=\min\{\Phi(v')-\Phi(v)\,,
\Phi(v')-\Phi(v_*)\}\\
&&
\le k2^{k}\left(\langle{v}\rangle^{k-\beta}+
 \langle{v_*}\rangle^{k-\beta}\right)|v-v_*|^{\beta}
m(\theta)\,.
\eeas
Next, if $f(v')\le f(v_*)(m(\theta))^{-N}$, then
$$\log^{+}f(v')\le \log^{+}f(v_*) +N|\log m(\theta)|$$
so that (using $x|\log x|\le 1$ for $0\le x\le 1$)
\beas&& m(\theta)f(v_*)\log^{+}f(v')\\
&&
\le m(\theta)f(v_*)
\log^{+}f(v_*) +N m(\theta)|\log m(\theta)|f(v_*)\\
 &&\le f(v_*)
\log^{+}f(v_*) +Nf(v_*)\,.
\eeas
If $f(v')\ge f(v_*)(m(\theta))^{-N}$, then $f(v_*)\le (m(\theta))^N
f(v')$ so that
\beas&& m(\theta)f(v_*)\log^{+}f(v')\le
 (m(\theta))^Nf(v')\log^{+}f(v')\,.\eeas
Thus
\beas&&
m(\theta)f_*\log^{+}f'
\le f_*
\log^{+}f_* +Nf_*+(m(\theta))^N
 f'\log^{+}f'\,.\eeas
With the same argument one has
\beas&&
m(\theta)f_*\log^{+}f_*'
\le f_*
\log^{+}f_* +Nf_*+(m(\theta))^N
 f_*'\log^{+}f_*'\,.\eeas
So the inequalities (\ref{(7.7)})-(\ref{(7.9)}) hold.

We now prove (\ref{(7.3)}). From (\ref{(7.7)})-(\ref{(7.9)}) we get
\beas&&\int_{{\bRRSN}}\fr{b(\cos\theta)}{|v-v_*|^{\beta}}
\bigg(S(v,v_*,v',v_*')-\Phi(v)\bigg)^{+}ff_*\log^{+}f\, d\sg dv_* dv\\
&&\le 2k  A_0\int_{{\bRRN}}\langle{v_*}\rangle^{k-\beta}ff_*\log^{+}f\, dv_* dv
=2k A_0\|f\log ^{+}f\|_{L^1}\|f\|_{L^1_{k-\beta}}\,,
\eeas

Next, let $I(f)$ denote the integral on the left side of (\ref{(7.4)}). Then
\beas&&I(f):=\int_{{\bRRSN}}\fr{b(\cos\theta)}{|v-v_*|^{\beta}}
\bigg(\Phi(v')-S(v,v_*,v',v_*')\bigg)^{+}ff_*\log^{+}f'\, d\sg dv_* dv\\
&&
\le k2^{k} \int_{{\bRRSN}}b(\cos\theta)m(\theta)
\left(\langle{v}\rangle^{k-\beta}+
 \langle{v_*}\rangle^{k-\beta}\right)
ff_*\log^{+}f'\, d\sg dv_* dv\\
&&=k2^{k} \int_{{\bRN}}\left(
\int_{{\bRSN}}b(\cos\theta)m(\theta) f_*(\log^{+}f' +
\log^{+}f_*') d\sg dv_* \right)\langle{v}\rangle^{k-\beta} f\,dv
\\
&&\le
k2^{k} \int_{{\bRN}}\left(
\int_{{\bRSN}}b(\cos\theta)(2f_*
\log^{+}f_* +2Nf_*) d\sg dv_*\right) \langle{v}\rangle^{k-\beta} f\,dv\\
&&+k2^{k} \int_{{\bRN}}\left(
\int_{{\bRSN}}b(\cos\theta)(m(\theta))^N( f'\log^{+}f'+f_*'\log^{+}f_*')
d\sg dv_*\right)\langle{v}\rangle^{k-\beta} f\,dv
\\
&&:=I_1+I_2\,.\eeas

Evidently,
\beas&& I_1\le k2^{k} A_0 \left(2\|f\log ^{+}f\|_{L^1}+2N\|f\|_{L^1}\right)
\|f\|_{L^1_{k-\beta}}\,.\eeas
To estimate $I_2$, we use the formulas (\ref{(7.5)})-(\ref{(7.6)})
to compute the inner integral
\beas&&
\int_{{\bRSN}}b(\cos\theta)(m(\theta))^N( f'\log^{+}f'+f_*'\log^{+}f_*')
d\sg dv_*\\
&&=|{\bf S}^{N-2}|
\int_{0}^{\pi}b(\cos\theta)\sin^{N-2}\theta\left(
\fr{(m(\theta))^N}{\sin^{N}(\theta/2)}+\fr{(m(\theta))^N}
{\cos^{N}(\theta/2)}
\right)d\theta
\|f\log ^{+}f\|_{L^1}
\\
&&
\le 2|{\bf S}^{N-2}|
\int_{0}^{\pi}b(\cos\theta)\sin^{N-1}\theta d\theta
\|f\log ^{+}f\|_{L^1}=2 A_0 \|f\log ^{+}f\|_{L^1}\,.\eeas
Consequently,
\beas&&
I_2\le 2A_0k2^{k}\|f\log ^{+}f\|_{L^1}
\int_{{\bRN}}\langle{v}\rangle^{k-\beta} f\,dv=
 2A_0k2^{k}\|f\log ^{+}f\|_{L^1}\|f\|_{L^1_{k-\beta}}\,.\eeas
Combining the above estimates,
\beas&&
I(f)\le
k2^{k} A_0 \left(4\|f\log ^{+}f\|_{L^1}+2N\|f\|_{L^1}\right)
\|f\|_{L^1_{k-\beta}}\\
&&\le 2^{2k}NA_0\left(\|f\log ^{+}f\|_{L^1}+\|f\|_{L^1}\right)
\|f\|_{L^1_{k-\beta}}\,.\eeas
$\Box$

{\bf Lemma 7.2.}  {\it Let $u(t)\ge 0$ defined on $[0,\infty)$ be
absolutely continuous on $[0,T]$ for all $0<T<\infty$ and satisfy
for some constants $C_1>0,\,C_2\ge 0,\, k\ge 0\,,\vep>0,$ $\eta<1$,
$$\fr{d}{dt}u(t)\le -C_1(1+t)^{-\eta}[u(t)]^{1+\vep}+C_2(1+t)^{k}e^{-t}\,,\qquad
{\rm a.e.}\quad  t\ge 0\,.$$  Then there is a constant $0<C<\infty$
depending only
on $C_1,\,C_2,\, k\,,\vep,\eta$,
and $u(0)$, such that
$$u(t)\le C(1+t)^{-\alpha}\,\,\qquad \forall\,t\ge 0$$
where ${\displaystyle \alpha=\fr{1-\eta}{\vep}}$. }

\noindent{\bf Proof}. Choose a constant $C\ge \max\{u(0), 1\}$ large
enough such that
$$C^{\vep}\ge \fr{1}{C_1}(\alpha+C_2(1+t)^{k+\alpha+1}e^{-t})\qquad
\forall\, t\ge 0\,. $$
Let
$$U(t)=C(1+t)^{-\alpha}\,,\quad t\ge 0\,.$$
Then using $\alpha+1=\alpha(\vep+1)+\eta$ and $C\ge 1$ we compute
\beas&& \fr{d U(t)}{dt}+ C_1(1+t)^{-\eta}[U(t)]^{1+\vep}-C_2(1+t)^{k}e^{-t}\\
&&=(1+t)^{-\alpha-1}\left(C_1C^{1+\vep}-C\alpha\right)-C_2(1+t)^{k}e^{-t}\\
&&\ge
(1+t)^{-\alpha-1}C\bigg(C_1C^{\vep}-\alpha-C_2(1+t)^{k+\alpha+1}e^{-t}\bigg)\ge
0 \,.\eeas
By the absolute continuity of $u(t)$, and $u(0)\le U(0)$,
we have for any $t>0$, \beas&&\bigg(u(t)-U(t)\bigg)^{+}
=\int_{0}^{t}
\bigg(\fr{d}{d\tau}u(\tau)-\fr{d}{d\tau}U(\tau)\bigg){\bf
1}_{\{u(\tau)>U(\tau)\}} d\tau\\
&&\le \int_{0}^{t}
\bigg(-C_1(1+\tau)^{-\eta}[u(\tau)]^{1+\vep}+C_2(1+\tau)^{k}
e^{-\tau}-\fr{d}{d\tau}U(\tau)\bigg){\bf
1}_{\{u(\tau)>U(\tau)\}} d\tau\\
&&\le \int_{0}^{t}
\bigg(-C_1(1+\tau)^{-\eta}[U(\tau)]^{1+\vep}+C_2(1+\tau)^{k}
e^{-\tau}-\fr{d}{d\tau}U(\tau)\bigg){\bf
1}_{\{u(\tau)>U(\tau)\}} d\tau\le 0\,.\eeas
So $u(t)\le U(t)\,\,\,\,\forall\, t\ge 0$\,.
$\qquad \Box$
\smallskip
\smallskip
\smallskip

{\bf Lemma 7.3} (Villani's Inequality \cite{20}). {\it Let
$${\cal D}_2(f)=\fr{1}{4}
\int_{{\bRRSN}}(1+|v-v_*|^2)
(ff_*-f'f_*')\log\left(\fr{ff_*}{f'f_*'}\right) d\sg dvdv_*$$
and let $M\in L^1_{(1,0,1)}({\bRN})$ be the Maxwellian (1.9).
Then for all $f\in L^1_{(1,0,1)}\cap L^1{\rm log}L({\bRN})$,
$$\fr{|{\bSN}|}{4(2N+1)}(N-T^*_{f})H(f|M)\le
{\cal D}_2(f)
$$ where $H(f|M)$ is the relative entropy:
$$H(f|M)=\int_{{\bRN}}f\log(f/M)dv=H(f)-H(M)\ge 0$$
and
$$T^*_{f}=\max_{{\bf e}\in
{\bSN}}\int_{{\bRN}}\langle{v,{\bf e}}\rangle^2 f(v) dv\,.$$
Moreover for any $H_0\in [0, +\infty)$
\begin{equation}\label{(7.11)}
\inf_{f\in {\cal H}_{0}}(N-T^*_{f})>0
\end{equation}
where ${\cal H}_{0}= \{f\in L^1_{(1,0,1)}\cap L^1{\rm
log}L({\bRN})\,|\, H(f|M)\le H_0\}$. }

\smallskip

This result is Theorem 2.1 in \cite{20}, except for the
lower bound (\ref{(7.11)}), which is an elaboration
of the remark following Theorem 2.1 in \cite{20}.
There it is observed that for $T^*_{f} = N$ to hold,
$f$ would have to be concentrated on a line, which is
inconsistent with the finite entropy hypothesis.
That a uniform bound of the type (\ref{(7.11)}) holds
was communicated to us
by Villani, along with a sketched proof, and so we list
it here as part of his theorem, and provide a detailed proof of his bound:

First of all, one has the
relation
$$\inf_{f\in {\cal H}_{0}}(N-T^*_{f})=
\inf_{f\in {\cal H}_{0}}\int_{{\bRN}}|\hat{v}|^2 f(v)dv\,,\quad
\hat{v}=(v_2, ..., v_{N})\,.$$ Take $\vep>0, R>0, K>1$ and consider
$$\int_{{\bRN}}|\hat{v}|^2 f(v)dv\ge
\vep^2\int_{|\hat{v}|\ge \vep}f(v)dv =\vep^2\left(1-\int_{|\hat{v}|<
\vep}f(v)dv\right)\,,$$ and \beas&& \int_{|\hat{v}|< \vep}f(v)dv\le
\int_{|\hat{v}|< \vep, |v_1|\le R, f(v)\le K}f(v)dv+
\int_{|v_1|> R}f(v)dv+\int_{f(v)> K}f(v)dv\\
&& \qquad \qquad \qquad \le K2R(2\vep)^{N-1}+\fr{N}{R^2}
+\int_{f(v)> K}f(v)dv\,.\eeas

To estimate the remaining integral, we use Young's inequality in the form
$$ab \le ca \log a + ce^{b/c-1} $$
valid for all $a\ge 0,b\ge 0,c>0$. (The $c=1$ case is standard. To
generalize this to the present case, replace $b$ by $b/c$, and
multiply through by $c$.) Taking $a= f/M$ and $b = 1_{\{f(v)>K\}}$,
and keeping $c$ arbitrary for the moment, integrating this
inequallity against $M$ this leads to
\begin{eqnarray}
\int_{f(v)> K}f(v)dv &\le& c\int_{{\bf R}^N}\left(\frac{f}{M}\right)
\log\left(\frac{f}{M}\right)M(v) dv dv +
c \int_{{\bR}}e^{(1/c)1_{\{f(v)>K\}} -1}M(v) dv\nonumber\\
&=& cH(f|M) + \frac{c}{e}\int_{f(v)\le  K}M(v)dv + c\int_{f(v)> K}
e^{(1/c) -1}M(v) dv\nonumber\\
&\le&
c H_0 + \frac{c}{e} +  \frac{c}{K}e^{(1/c) -1} := C_{c,K,H_0}\nonumber\\
 \end{eqnarray}
 where in the last line, we have used the $H$-theorem, the fact that $M< 1$,
 and Chebychev's inequality to estimate the Lebesgue measure of
 $\{f(v)> K\}$.

Thus
\begin{equation}\label{(7.*)}
\int_{|\hat{v}|< \vep}f(v)dv\le  2^{N}K
R\vep^{N-1}+\fr{N}{R^2}+C_{c,K,H_0}\,.
\end{equation}
 With suitable choice of $c>0$, $K>1, R>1$ and $\vep>0$, the right hand side
of (\ref{(7.*)}) is less than $1/2$.  This gives (\ref{(7.11)}).

\smallskip

\smallskip

 \noindent{\bf Proof of Theorem 3.}  First of all, using inequality
 (\ref{(1.10)}) and the Csiszar-Kullback inequality
$\|f-M\|_{L^1}\le \sqrt{2H(f|M)}\,$, we have
\begin{equation}\label{(7.12)}
\|f-M\|_{L^1_2}\le C_N [H(f|M)]^{1/4}\qquad \forall\,
f\in L^1_{(1,0,1)}\cap L^1{\rm log}L({\bRN})\,.
\end{equation}
>From (\ref{(7.12)}), we see that to prove the theorem, it suffices to prove
that there is a weak solution $f$ of Eq.(B) with $f|_{t=0}=f_0$ such
that for some constant $0<C<\infty$
\begin{equation}\label{(7.13)}
H(f(t)|M)\le C(1+t)^{-4\ld}\qquad \forall\, t\ge 0
\end{equation}
where $\ld$ is given in (\ref{(1.24)}).

The proof consists of three steps.

\noindent{\bf Step 1}.  In the first two steps we assume in addition that
$B(z,\sg)$ is bounded: $B(z,\sg)\le {\rm const}.$ In this case it is
well-known that Eq.(B) has a unique mild solution $f\in C^1([0,
\infty); L^1({\bRN}))
 \cap L^{\infty}([0,\infty); L^1_2\cap L^1{\rm log}L({\bRN}))$ satisfying
 $f|_{t=0}=f_0$, and moreover, $f$ conserves the mass, moment and energy,
 and satisfies the entropy identity (see e.g.
 \cite{14} ).
 Hence by Proposition 1.1 (b), the boundedness of $B(z,\sg)$  implies  that
the mild solution $f$ is also a weak solution, and thus by Theorem 1,
it satisfies the moment
 estimate:
 $$\|f(t)\|_{L^1_s}\le C(1+t)\,,\quad
 \int_{0}^{t}\|f(\tau)\|_{L^1_{s-|\gm|}}d\tau\le C(1+t)\,,\quad t\ge 0\,.$$
where the constant $C$ is given in Theorem 1, in particular it does not
depend on the
$L^{\infty}$ bound of $B(z,\sg)$.

 To overcome the trouble that $f$ may not have a lower bound, we consider
 a suitable
 convex combination of  $f$ and the Maxwellian $M$:
 \begin{equation}\label{(7.14)}
 g(v,t)=(1-e^{-t-1})f(v,t)+e^{-t-1}M(v)\,.
 \end{equation}
 It is obvious that the flow $t\mapsto g(t)$ has the same mass, momentum,
 and energy as $f(t)$
and holds the
following properties that will be proven in this step:
\begin{equation}\label{(7.15)}
 \log^{+}g\le \log^{+}f\,,\quad g\log^{+}g\le f\log^{+}f\,,
 \end{equation}
\begin{equation}\label{(7.16)}
\log^{+}(1/g(v,t))\le C(1+t)\langle{v}\rangle^2\,,
\end{equation}
\begin{equation}\label{(7.17)}
H(f(t)|M)\le H(g(t)|M)+C(1+t)e^{-t}\,,
\end{equation}
\begin{equation}\label{(7.18)}
\fr{d}{d t}H(g(t)|M)\le -D(g(t))+C(1+t)e^{-t}\,\quad {\rm a.e.}\quad t\ge 0\,,
\end{equation}
\begin{equation}\label{(7.19)}
 \|g(t)\log^{+}g(t)\|_{L^1_k}\le C(1+t)^2\,.\qquad ( \,k=s-2\,)
 \end{equation}
Here and below all constants $0<C<\infty$ depend only $N,  K_*, A_0, \gm,, s, \|f_0\|_{L^1_s}$ and
$\|f_0\log f_0\|_{L^1_s}$.

Proof of (\ref{(7.15)})-(\ref{(7.17)}):  By $M(v)=(2\pi)^{-N/2} e^{-|v|^2/2}<1$
 we have
$$ g(v,t)\ge 1
\,\, \Longrightarrow\,\,   g(v,t)\le f(v,t)\,\, \Longrightarrow\,\, 0\le
\log g(v,t)\le \log f(v,t)\,.$$
So (\ref{(7.15)}) is true. (\ref{(7.16)}) is obvious. To prove (\ref{(7.17)})
 we denote $\dt=e^{-t-1}$. Then
\beas&& H(g(t))\ge \int_{{\bRN}}(1-\dt)f\log((1-\dt)f )dv+\int_{{\bRN}}\dt M
\log(\dt M )dv
\\
&&\qquad \quad\,\,\, =(1-\dt)\log (1-\dt)+(1-\dt)H(f(t))+\dt\log\dt+\dt H(M)
\,.\eeas
So using $(1-\dt)\log (\fr{1}{1-\dt})\le \dt$ and $0\le H(f(t)|M)\le H(f_0|M)$
we obtain (\ref{(7.17)}):
\beas&& H(f(t)|M)\le H(g(t)|M)+(1-\dt)\log (\fr{1}{1-\dt})+\dt H(f(t)|M)
+\dt\log(1/\dt)\\
&&\qquad \qquad \quad\, \le H(g(t)|M)+[H(f_0|M)+2+t]e^{-t-1}
\,.\eeas

Now we are going to prove (\ref{(7.18)}). Since $f(v,t)$ is a mild solution
and $g(v,t)\ge e^{-t-1}M(v)$,
the function $t\mapsto g(v,t)\log g(v,t)$ is also absolutely continuous on
any finite
interval for almost every $v\in{\bRN}$. So we have for almost every
$v\in{\bRN}$ and for all $t\ge 0$
\begin{multline}\label{(7.20)}
g(v,t)\log g(v,t)=g_0(v)\log g_0(v)\\
+\int_{0}^{t}(1+\log g(v,\tau))
\bigg( e^{-\tau-1}(f(v,\tau)-M(v))+(1-e^{-\tau-1})Q(f)(v,\tau)\bigg)d\tau\,.
\end{multline}
We need to show that there are no problems of integrability in
$v\in{\bRN}$. In fact, it is easily seen that the functions \beas&&
g(v,t)|\log g(v,t)|\,,\quad |g_0(v)|\log g_0(v)|\,,\\
&& \int_{0}^{t}(1+|\log g(v,\tau)|)
(f(v,\tau)+M(v))d\tau\,,\\
&&
\int_{0}^{t}(1+\log^{+}(1/g(v,\tau))[Q^{+}(f)(v,\tau)+Q^{-}(f)(v,\tau)]d\tau\,,
\\&& \int_{0}^{t}Q^{-}(f)(v,\tau)\log^{+}g(v,\tau)d\tau
\eeas
are all integrable on ${\bRN}$, while
from the equation (\ref{(7.20)}) we see that the rest term
$$\int_{0}^{t}(1-e^{-\tau-1})Q^{+}(f)(v,\tau)\log^{+}g(v,\tau) d\tau$$
is bounded by the summation of the above functions and thus it is also
 integrable on ${\bRN}$.
Since $1-e^{-\tau-1}>1/2$, it follows that
$$\int_{0}^{t}\int_{{\bRN}}Q^{+}(f)(v,\tau)\log^{+}g(v,\tau) dv d\tau<\infty\,.$$
Therefore there are no problems of integrability and we obtain for any
$0\le \vp\in L^{\infty}({\bRN})$ and for all $t\ge 0$
\begin{multline}\label{(7.21)}
\int_{{\bRN}}\vp(v)g(v,t)\log g(v,t)dv
=\int_{{\bRN}}\vp(v)g_0(v)\log g_0(v)dv\\
+\int_{0}^{t} d\tau\int_{{\bRN}}
\vp(v)
\bigg( e^{-\tau-1}(f(v,\tau)-M(v))+(1-e^{-\tau-1})Q(f)(v,\tau)\bigg)(1+\log g(v,\tau)) dv\,.
\end{multline}
In particular taking $\vp(v)\equiv 1$ we see that $t\mapsto H(g(t))$
is absolutely continuous on every
finite intervals and using conservation of mass and energy we have
\begin{eqnarray}\label{(7.22)}
\fr{d}{d t}H(g(t)|M)
&=&e^{-t-1}\int_{{\bRN}}(f(v,t)-M(v)) \log g(v,t)dv\nonumber\\
 &+&(1-e^{-t-1})\int_{{\bRN}}Q(f)(v,t)\log g(v,t))\,dv
\quad {\rm a.e.}\quad t\ge 0 \,.\nonumber\\
\end{eqnarray}

Since $\sup_{t\ge 0}\|f(t)\log f(t)\|_{L^1}\le H(f_0)+C_N$, it follows that
\beas&& \int_{{\bRN}}(f(v,t)-M(v)) \log g(v,t))dv
=\int_{{\bRN}}f\log g dv+\int_{{\bRN}}M\log(1/g)dv
\\
&&\le \|f(t)\log ^+f(t)\|_{L^1}+ C(1+t)\int_{{\bRN}}\langle{v}\rangle^2 M(v)dv\le C(1+t)\,.\eeas
To estimate the second term in the right hand side of (\ref{(7.22)}), we let
\begin{equation}\label{(7.23)}
G(v,t)=\fr{1}{1-e^{-t-1}}g(v,t)=
f(v,t)+\zeta(t)M(v)\,,\quad \zeta(t)=\fr{e^{-t-1}}{1-e^{-t-1}}\,.
\end{equation}
Then ( recalling  $M'M_*'=MM_*$)
\begin{equation}\label{(7.24)}
f'f_*'-ff_*-(G'G_*'-GG_*)=\zeta(t)\bigg( M f_* +f M_*-M'f_*'-M_*'f'\bigg)\,.
\end{equation}
and
\begin{equation}\label{(7.25)}
G'G_*'-GG_*=(1-e^{-t-1})^{-2}(g'g_*'-gg_*)\,,\quad (1-e^{-t-1})\zeta(t)=e^{-t-1}
\end{equation}
and so we compute
\begin{eqnarray}\label{(7.26)}
&& (1-e^{-t-1})\int_{{\bRN}}Q(f)(v,t)\log g(v,t))\,dv
\nonumber\\
&=&e^{-t-1}\int_{{\bRRSN}}B
(M f_* +f M_*-M'f_*'-M_*'f')\log g \,d\sg dv_* dv
\nonumber\\
&-& (1-e^{-t-1})^{-1}D(g(t))\ .
\end{eqnarray}
Next using (\ref{(7.16)}), (\ref{(7.15)}) (for $\log^{+}(1/g)$
and $\log ^{+}g$) we compute
\beas&&
\int_{{\bRRSN}}B
(M f_* +f M_*-M'f_*'-M_*'f')\log g\, d\sg dv_* dv\\
&&=\int_{{\bRRSN}}B fM_*
\left(\log g +\log g_*+\log(1/g')+\log(1/g_*')\right) d\sg dv_*dv \\
&&\le A_0 \int_{{\bRRN}}\fr{1}{|v-v_*|^{|\gm|}}
f M_*\left(\log^{+}f+\log^{+}f_*
+ C(1+t)(\langle{v}\rangle^2+\langle{v_*}\rangle^2)\right)
dv_*dv\\
&&=A_0\int_{{\bRN}}\left(\int_{{\bRN}}\fr{M_*}{|v-v_*|^{|\gm|}}dv_*\right) f\log^{+}f dv
+A_0\int_{{\bRN}}\left(\int_{{\bRN}}\fr{1}{|v-v_*|^{|\gm|}}M_*\log^{+}f_* dv_*\right) f dv\\
&&+C(1+t)\int_{{\bRN}} \left(\int_{{\bRN}}\fr{M_* }
{|v-v_*|^{|\gm|}}dv_*\right) \langle{v}\rangle^2 f dv+C(1+t)\int_{{\bRN}}
\left(\int_{{\bRN}}\fr{\langle{v_*}\rangle^2 M_* }
{|v-v_*|^{|\gm|}}dv_*\right)f\, dv\\
&&\le C(1+t)+C\|f(t)\log^{+}f(t)\|_{L^1}\le C(1+t)\,.\eeas
Here we used the fact that $\langle{\cdot}\rangle^s M\in L^{\infty}\cap L^1({\bRN})$ and $\log ^{+}f\in L^p\cap L^1({\bRN)}$
for all $1<p<\infty$.

Summarizing the above estimates  we obtain (\ref{(7.18)}).
\smallskip

Proof of (\ref{(7.19)}): This is the main estimate in this step. To do this
we shall use the functions  $\Phi_R(v), S_R(v,v_*, v', v_*')$
defined in (\ref{(7.1)})-(\ref{(7.2)}) for $\Phi_R(v)=\min\{\langle{v}\rangle^k,\,R\}$ with
$k=s-2$.

Let
\beas&& H_{k,R}(g(t))=\int_{{\bRN}}\Phi_R(v) g(v,t)\log g(v,t) dv\,,\\
&&
H_{k,R}^{+}(g(t))=\int_{{\bRN}}\Phi_R(v) g(v,t)\log^{+}g(v,t) dv\\
&&
H_{k,R}^{-}(g(t))=\int_{{\bRN}}\Phi_R(v) g(v,t)\log^{+}(1/g(v,t)) dv\,.\eeas
 To prove (\ref{(7.19)}), it suffices to prove that
\begin{equation}\label{(7.27)}
H_{k,R}(g(t))\le C(1+t)^2+C \int_{0}^{t}e^{-\tau} H_{k,R}^{+}(g(\tau))d\tau\qquad
\,\forall\,t\ge 0
\end{equation}
where and below all the constants $C$ do not depend on $R$ and
$\|B\|_{L^{\infty}}$. In fact if (\ref{(7.27)}) holds true, then
using $H_{k,R}^{+}(g(t))=H_{k,R}(g(t))+H_{k,R}^{-}(g(t))$ and the
obvious estimate $H_{k,R}^{-}(g(t))\le
C(1+t)\max\{\|f(t)\|_{L^1_{k+2}},\,\|M\|_{L^1_{k+2}}\} \le C(1+t)^2$
we get
$$
H_{k,R}^{+}(g(t))\le C(1+t)^2+C\int_{0}^{t}e^{-\tau} H_{k,R}^{+}(g(\tau))d\tau\,,
\quad \forall\,t\ge 0\,.$$
So applying Gronwall's Lemma  gives
$$H_{k,R}^{+}(g(t))\le C(1+t)^2\exp\bigg(\int_{0}^{t}Ce^{-\tau}d\tau\bigg)
\le C(1+t)^2\,.$$
Then letting $R\to\infty$ leads to (\ref{(7.19)}) by Fatou's Lemma.

As usual, for notational convenience we also denote without confusion that
$$S_R=S_R(v,v_*,v',v_*')\,,\quad \Phi_R=\Phi_R(v)\,,\quad  \Phi_R'
=\Phi_R(v')\,,\quad
(\Phi_R)_*'=\Phi_R(v_*')\,,\quad {\rm etc.}$$

Now we begin to prove (\ref{(7.27)}). It has been shown in the above
that there are no problem of integrability and the function
$t\mapsto H_{k,R}(g(t))$ is absolutely continuous on any finite
intervals. We compute using (\ref{(7.21)}) with $\vp=\Phi_R$ that
\beas&& \fr{d}{dt}H_{k,R}(g(t))
=e^{-t-1}\int_{{\bRN}}\Phi_R(v)(f(v,t)-M(v))dv\qquad\quad  \\
&&\qquad \qquad \qquad +(1-e^{-t-1})\int_{{\bRN}}\Phi_R(v) Q(f)(v,t) dv
\qquad \quad  \\
&&
\qquad \qquad \qquad+e^{-t-1}\int_{{\bRN}}\Phi_R(v)(f(v,t)-M(v))\log g(v,t)dv
\qquad \quad \\
&&\qquad \qquad \qquad+(1-e^{-t-1})\int_{{\bRN}}\Phi_R(v)Q(f)(v,t)\log g(v,t)
dv \qquad \\
&&\qquad \qquad \qquad:=I_{k,R}^{(1)}(t)+
I_{k,R}^{(2)}+I_{k,R}^{(3)}(t) + I_{k,R}^{(4)}(t)\,.\qquad
\quad\eeas To estimate these terms we shall use the following
inequality:
\begin{equation}\label{(7.28)}
|\Phi_R(w)-S_{R}(v,v_*,v',v_*')|\le k
(\langle{v}\rangle^2+
\langle{v_*}\rangle^2)^{(k-|\gm|)/2} |v-v_*|^{|\gm|}
\,\quad \forall\,w\in\{v,v_*,v',v_*'\}\,.
\end{equation}
Now we are going to estimate $I_{k,R}^{(i)}\,(i=1,2,3,4$).
Let us emphasize again that  {\bf all constants $C$
are independent of $R$ and $\|B\|_{L^{\infty}({\bRSN})}$.}

The first one is easy: By moment estimate we have
$$I_{k,R}^{(1)}(t)\le
e^{-t-1}\|f(t)\|_{L^1_k} \le C(1+t)e^{-t}\,.
$$
For the second term
we use the vanishing property
\begin{equation}\label{(7.29)}
\int_{{\bRRSN}}B(v-v_*,\sg)S_R(v,v_*,v',v_*')
(f'f_*'-ff_*)d\sg dv_* dv=0
\end{equation}
which is due to the symmetry
$$S_R(v,v_*,v',v_*')=S_R(v_*,v,v_*',v')=S_R(v',v_*',v,v_*)\,,\quad {\rm etc.}$$
Then using (\ref{(7.28)}) we compute
\beas&&  I^{(2)}_{k,R}(t)
=(1-e^{-t-1})\int_{{\bRRSN}}B(\Phi_R-
S_R)
(f'f_*'-ff_*)d\sg dv_* dv
\\
&& \qquad \quad \,\le C\|f(t)\|_{L^1_{k-|\gm|}}\le C(1+t)\,. \eeas

\noindent For the third term $I_{k,R}^{(3)}(t)$ we use the control
$$f(v,t)\le \fr{1}{1-e^{-t-1}} g(v,t)$$ to see that
\beas&& I_{k,R}^{(3)}(t)\le e^{-t-1}\int_{{\bRN}}\Phi_R f \log ^{+}g dv
+e^{-t-1}\int_{{\bRN}}\Phi_R M\log^{+}(1/ g)
dvd\tau\\
&&\qquad \quad \,\,\le e^{-t} H_{k,R}^{+}(g(t)) +C(1+t)e^{-t}\,.\eeas

The estimate of the last term $I_{k,R}^{(4)}(t)$ is the key part of this section.
One will see that in this estimate the term $H_{k,R}^{+}(g(t))$ can not be
avoided and this is why we introduce the decay weight $e^{-t-1}$ rather than
the equal weight $1/2$.

Inserting the function
$G'G_*'-GG_*=(1-e^{-t-1})^{-2}(g'g_*'-gg_*)$
 (see (\ref{(7.23)}),(\ref{(7.25)})) we have
\beas&& I_{k,R}^{(4)}(t)
=(1-e^{-t-1})\int_{{\bRRSN}}B
\Phi_R(f'f_*'-ff_*)\log g
dv \\
&&\qquad \quad\, =(1-e^{-t-1})\int_{{\bRRSN}}B
\Phi_R\bigg(f'f_*'-ff_*-(G'G_*'-GG_*)\bigg)\log g
dv \\
&&\qquad \quad\,
+(1-e^{-t-1})^{-1}\int_{{\bRRSN}}B(\Phi_R-S_R)(g'g_*'-gg_*)\log g\,
dv \\
&&\qquad \quad\, +(1-e^{-\tau-1})^{-1}\int_{{\bRRSN}}B
S_R(g'g_*'-gg_*)\log g\,
dv \\
&&\qquad \quad \, :=I_{k,R}^{(4,1)}(t)+I_{k,R}^{(4,2)}(t)+I_{k,R}^{(4,3)}(t)\,.
\eeas

Further estimate using (\ref{(7.24)}) and $(1-e^{-t-1})\zeta(t)=e^{-t-1}$:
\beas&&I_{k,R}^{(4,1)}(t)\le e^{-t-1}
\int_{{\bRRSN}}B\Phi_R
M f_*\log ^{+}g \,d\sg dv_* dv\\
&&\qquad \quad \,\,\, +e^{-t-1} \int_{{\bRRSN}}B\Phi_R
fM_*\log ^{+}g \,d\sg dv_* dv\\
&&\qquad \quad \,\,\, +e^{-t-1}  \int_{{\bRRSN}}B\Phi_R
(M' f_*' +f' M_*')\log^{+}(1/g) d\sg dv_* dv
\\
&&\qquad \quad \,\,\, :=I_{k,R}^{(4,1,1)}(t)+I_{k,R}^{(4,1,2)}(t)
+I_{k,R}^{(4,1,3)}(t)\,,\eeas

\beas&&
I_{k,R}^{(4,1,1)}(t)
\le A_0e^{-t-1} \int_{{\bRN}}
\left(\int_{{\bRN}}\fr{\langle{v}\rangle^k M}{|v-v_*|^{|\gm|}}
\log ^{+}g\,dv\right)f_* dv_*
\\
&&\qquad \qquad \,\,\le Ce^{-t-1}\|f(t)\|_{L^1}\le C\,,\eeas
and (using $f(v,t)\le \fr{1}{1-e^{-t-1}} g(v,t)$)
\beas&& I_{k,R}^{(4,1,2)}(t)
\le A_0\fr{e^{-t-1}}{1-e^{-t-1}}\int_{{\bRN}}
\left(\int_{{\bRN}}\fr{M_* dv_*}{|v-v_*|^{|\gm|}}\right)
\Phi_R \,g\log ^{+}g\, dv\\
&&\qquad \qquad \,\, \le Ce^{-t}\int_{{\bRN}}\Phi_R\,g\log ^{+}g\, dv
=Ce^{-t} H_{k,R}^+(g(t))\,.\eeas

For $I_{k,R}^{(4,1,3)}(t)$ we change variables and use inequality
$$\Phi_R'\log^{+}(1/g')+(\Phi_R)_*'\log^{+}(1/g_*')
\le C(1+t)\langle{v}\rangle^{k+2}\langle{v_*}\rangle^{k+2}$$
recalling $k+2=s$ to get
\beas&&I_{k,R}^{(4,1,3)}(t)=e^{-t-1} \int_{{\bRRSN}}B\bigg( \Phi_R'
\log^{+}(1/g') +(\Phi_R)_*'
\log^{+}(1/g_*')\bigg)
M_*f d\sg dv_* dv
\\
&& \qquad \qquad \,\,\le CA_0(1+t)e^{-t-1} \int_{{\bRN}}\left(\int_{{\bRN}}
\fr{\langle{v_*}\rangle^s M_*}{|v-v_*|^{|\gm|}}dv_*\right)
\langle{v}\rangle^s
f dv\\
&&\qquad \qquad \,\,\le C
(1+t)e^{-t-1}\|f(\tau)\|_{L^1_{s}}
\le C (1+t)^2e^{-t}\,. \eeas
Thus
$$I_{k,R}^{(4,1)}(t)\le C e^{-t} H_{k,R}^+(g(t)) +C\,.$$

{\bf Estimate of $I_{k,R}^{(4,2)}(t)$}: Neglecting negative parts we have
\beas&& I_{k,R}^{(4,2)}(t)
=(1-e^{-t-1})^{-1}\int_{{\bRRSN}}B(\Phi_R'-S_R)gg_*\log g'\,d\sg dv_*dv\\
&&+(1-e^{-t-1})^{-1}\int_{{\bRRSN}}B(S_R-\Phi_R)gg_*\log g\,d\sg dv_*dv\\
&&\le 2\int_{{\bRRSN}}B(\Phi_R'-S_R)^{+}gg_*\log^{+}g'\,d\sg dv_*dv\\
&&+2\int_{{\bRRSN}}B(S_R-\Phi_R')^{+}gg_*\log^{+}(1/g')\,d\sg dv_*dv\\
&&+2\int_{{\bRRSN}}B(S_R-\Phi_R)^{+}gg_*\log^{+}g\,d\sg dv_*dv\\
&&+2\int_{{\bRRSN}}B(\Phi_R-S_R)^{+}gg_*\log^{+}(1/g)\,d\sg dv_*dv
\\
&&:=I_{k,R}^{(4,2,1)}(t)+I_{k,R}^{(4,2,2)}(t)+I_{k,R}^{(4,2,3)}(t)
+I_{k,R}^{(4,2,4)}(t)\,.\eeas
Using $B(v-v_*,\sg)\le b(\cos\theta)|v-v_*|^{-|\gm|}$ and Lemma 7.1
with $\beta=|\gm|$ we have
\beas&&
I_{k,R}^{(4,2,1)}(t)+I_{k,R}^{(4,2,3)}(t)\le C\left(\|g\log ^{+}g\|_{L^1}+
\|g\|_{L^1}\right)\|g\|_{L^1_{k-|\gm|}}\le C(1+t)\,,\\
\eeas
while using (\ref{(7.28)}) and $\log^{+}(1/g')\le C(1+t)(\langle{v}\rangle^2
+\langle{v_*}\rangle^2)$ gives
\beas&& I_{k,R}^{(4,2,2)}(t)+I_{k,R}^{(4,2,4)}(t)
\le C(1+t)\|g(t)\|_{L^1_{s-|\gm|}}\,.\eeas
Since $\|g(t)\|_{L^1_{s-|\gm|}}\le \|f(t)\|_{L^1_{s-|\gm|}}+C$, it follows
that
$$I_{k,R}^{(4,2)}(t)
\le C(1+t)(\|f(t)\|_{L^1_{s-|\gm|}}+1)\,.$$

The last term $I_{k,R}^{(4,3)}(t)$ is negative which is due to the
total symmetry of $S_{R}(v,v_*,v',v_*')$: Using classical derivation one has
$$ I_{k,R}^{(4,3)}(t)=-(1-e^{-t-1})^{-1}D_{S_R}(g(t))\le 0$$
where
$$D_{S_R}(g(t))=\fr{1}{4}\int_{{\bRRSN}}B
S_R(g'g_*'-gg_*)\log\left(\fr{g'g_*'}{gg_*}\right)\,
d\sg dv_*dv\ge 0\,.$$

Summarizing the above we obtain
$$I_{k,R}^{(4)}(t)\le  C(1+t)(\|f(t)\|_{L^1_{s-|\gm|}}+1)
+C e^{-t}H^{+}_{k,R}(g(t))
\,.$$
And therefore collecting all estimates of $I_{k,R}^{(1)}(t),
\,I_{k,R}^{(2)}(t),\,I_{k,R}^{(3)}(t),\,
 I_{k,R}^{(4)}(t)$ we obtain
$$\fr{d}{dt}H_{k,R}(g(t))\le  C(1+t)(\|f(t)\|_{L^1_{s-|\gm|}}+1)
+C e^{-t}H^{+}_{k,R}(g(t))\quad {\rm a.e.}
\quad t\ge 0\,.$$
Since $\int_{0}^{t}\|f(\tau)\|_{L^1_{s-|\gm|}}d\tau
\le C(1+t)$ and $H_{k,R}(g(0))\le \|f_0\log ^{+}f_0\|_{L^1_2}$,
it follows that
\beas&& H_{k,R}(g(t))\le  C
+ C\int_{0}^{t}(1+\tau)(\|f(\tau)\|_{L^1_{s-|\gm|}}+1)d\tau +C \int_{0}^{t}
e^{-\tau}H^{+}_{k,R}(g(\tau))d\tau\\
&&\qquad \qquad \quad \le C(1+t)^2+C \int_{0}^{t}
e^{-\tau}H^{+}_{k,R}(g(\tau))d\tau\,.\eeas
This proves (\ref{(7.27)}) and finishes the Step 1.

{\bf Step 2}. The method of proving (\ref{(7.13)}) is to establish an
inequality between
$H(g(t))$ and $D(g(t))$ as has been done in the case of hard
potentials, see e.g. \cite{3,4,16}. [ Note that $g(v,t)$ is not a solution
of Eq.(B),
but it is better than a solution in the present sense ...]
We shall prove that
\begin{equation}\label{(7.30)}
D(g(t))\ge c (1+t)^{-2\vep}[H(g(t)|M)]^{1+\vep}\qquad \forall\, t\ge 0
\end{equation}
 where
$$\vep=\fr{2+|\gm|}{k-2}<\fr{1}{2}\,,\quad k=s-2\,\,\,(\,>6+2|\gm|\,)\,.$$
If (\ref{(7.30)}) holds true, then by the differential inequality
(\ref{(7.18)}) we get
$$\fr{d}{dt}H(g(t)|M)\le
-c (1+t)^{-2\vep}[H(g(t)|M)]^{1+\vep}+C(1+t) e^{-t}\qquad {\rm a.e.}\,\quad
t\in[0,\infty)\,.  $$
Applying Lemma 7.2  we then obtain (with $\alpha=\fr{1-2\vep}{\vep}\ge 4\ld$)
$$H(g(t)|M)\le C(1+t)^{-\alpha}\le C(1+t)^{-4\ld}$$
and hence (\ref{(7.13)}) follows from (\ref{(7.17)}).

To prove (\ref{(7.30)}), we consider
$${\cal D}_k(g)=\fr{1}{4}\int_{{\bRRSN}}(1+|v-v_*|^2)^{k/2}
(g'g_*'-gg_*)\log\left(\fr{g'g_*'}{gg_*}\right) d\sg dvdv_*$$ and
make use of Villani's inequality (see Lemma 7.3 and note that
$H(g(t)|M)\le H(f(t)|M)\le H(f_0|M)$):
\begin{equation}\label{(7.31)}
C_{H_0}H(g(t)|M)\le
{\cal D}_2(g(t)) \end{equation}
 where
$$C_{H_0}=\fr{|{\bSN}|}{4(2N+1)}\inf_{f\in {\cal H}_{0}}(N-T^*_{f})>0\,,\quad
H_0=H(f_0|M)\,.$$

By assumption  $K_*(1+|z|^2)^{-|\gm|/2}\le B(z,\sg)$ and writing
$\fr{k}{2}=(1+\fr{|\gm|}{2(1+\vep)})\cdot \fr{1+\vep}{\vep}$ we have
\beas&& K_*^{\fr{1}{1+\vep}}(1+|z|^2)\le [B(z,\sg)]^{\fr{1}{1+\vep}}
\bigg((1+|z|^2)^{k/2}\bigg)^{\fr{\vep}{1+\vep}}\,.\eeas So by
H\"{o}lder's inequality
\begin{equation}\label{(7.32)}
K_*[{\cal D}_2(g(t))]^{1+\vep}\le
D(g(t))[{\cal D}_k(g(t))]^{\vep}\,,\quad t\ge 0\,.
\end{equation}
Now we prove that
\begin{equation}\label{(7.33)}
{\cal D}_k(g(t))\le C (1+t)^2\,.
\end{equation}
By symmetry we have
$${\cal D}_k(g)=\fr{1}{2}\int_{{\bRRSN}}(1+|v-v_*|^2)^{k/2}
1_{\{gg_*\ge g'g_*'\}}
(gg_*-g'g_*')\log\left(\fr{gg_*}{g'g_*'}\right) d\sg dvdv_*\,.$$
Note that using (\ref{(7.16)}) gives
$$\log(\fr{1}{g'g_*'})\le C(1+t)(\langle{v}\rangle^2+\langle{v_*}\rangle^2)\,.$$
So
if $gg_*\ge g'g_*'$ then \beas&&
(gg_*-g'g_*')\log\left(\fr{gg_*}{g'g_*'}\right) \le
gg_*\left(\log(gg_*)
+\log\left(\fr{1}{g'g_*'}\right)\right)\\
&&\le gg_*\log ^{+}g+gg_*\log^{+}g_*+C
gg_*(1+t)(\langle{v}\rangle^2+\langle{v_*}\rangle^2) \,.\eeas Since
$(1+|v-v_*|^2)^{k/2}\le 2^{k-1}(\langle{v}\rangle^k
+\langle{v_*}\rangle^k)$, it follows from the main estimate
(\ref{(7.19)}),   $\|g(t)\log^{+} g(t)\|_{L^1}\le \|f(t)\log^{+}
f(t)\|_{L^1} \le C$,  $\|g(t)\|_{L^1_s}\le C(1+t)$, and $k+2=s$ that
\beas&& {\cal D}_k(g(t))\le C\int_{{\bRRN}}(\langle{v}\rangle^k
+\langle{v_*}\rangle^k)
g\log^{+}(g)g_*dvdv_*\\
&&+C(1+t)\int_{{\bRRN}}(\langle{v}\rangle^{k+2} +\langle{v_*}\rangle^{k+2})
gg_*dvdv_*\\
&&\le C\bigg(\|g(t)\log^{+} g(t)\|_{L^1_k}+
\|g(t)\log^{+} g(t)\|_{L^1}\|g(t)\|_{L^1_k}+(1+t)\|g(t)\|_{L^1_s}\bigg)\\
&&\le C(1+t)^2\,.
\eeas
This proves (\ref{(7.33)}).  Combining (\ref{(7.31)})-(\ref{(7.33)}) we
obtain
$$K_* [C_{H_0} H(g(t)|M)]^{1+\vep}\le C(1+t)^{2\vep}D(g(t)) $$
which gives (\ref{(7.30)}).

\noindent{\bf Step 3}. Let $B(z,\sg)$ be given in the theorem. We
shall use approximate solutions. Let
$$B_n(z,\sg)=\min\{ B(z,\sg)\,,\, n\}\,,\quad n\ge K_*\,.$$
It is obvious that for all $n\ge K_*$, $B_n(z,\sg)\ge
K_*(1+|z|^2)^{-|\gm|/2}\,.$ For each $n\ge K_*$, let $f^n\in C^1([0,
\infty); L^1({\bRN}))
 \cap L^{\infty}([0,\infty); L^1_2\cap L^1{\rm log}L({\bRN}))$ be the
 unique mild solution of
Eq.(B) with the kernel $B_n$ and $f^n|_{t=0}=f_0$ and $f^n$ has all
properties as listed in Step 1 and Step 2. In particular $f^n$
satisfies for all $n\ge K_*$ and all $t\ge 0$
$$ H(f^n(t)|M)\le C(1+t)^{-4\ld}$$
with the same constants $\ld>0$ and $0<C<\infty$ (which is of course
independent of $n$). As argued in the proof of existence of weak
solutions, there exists a subsequence $\{f^{n_k}\}_{k=1}^{\infty}$
of $\{f^n\}$ and a weak solution $f$ of Eq.(B) satisfying
$f|_{t=0}=f_0$, such that
$$\forall\,t\ge 0\,,\quad
f^{n_k}(\cdot,t)\rightharpoonup f(\cdot,t) \,\,\,(k\to\infty) \quad
{\rm weakly\,\,\,in}\quad L^1({\bRN})\,.$$  Then by convexity and
weak convergence, we obtain
$$H(f(t)|M)\le \liminf_{k\to\infty}H(f^{n_k}(t)|M)\le C(1+t)^{-4\ld}
\quad \forall\,
t\ge 0\,.$$
 This completes the proof. $\quad \Box$


\medskip


\begin{thebibliography}{99}

\bibitem{ADVW} Alexandre,~R.; Desvillettes,~L; Villani,~C. and Wennberg, W.,
Entropy dissipation and long-range interaction, {\it  Arch. Rational
Mech. Anal.}, {\bf 152} (2000)), no.1, 327-355.

\bibitem{1} Arkeryd,~L.,Intermolecular forces of infinite range and the
Boltzmann equation, Entropy dissipation and long-range interaction,
{\it  Arch. Rational Mech. Anal.}, {\bf 77} (1981)), no.1, 11--21.

\bibitem{2} Bouchut,~F; Desvillettes,~L.,
A proof of the smoothing properties of the positive part of Boltzmann's kernel.
{\it  Rev. Mat. Iberoamericana}, {\bf 14} (1998), no.1, 47--61.

\bibitem{3} Carlen,~E.A.; Carvalho,~M.C.,
Entropy production estimates for Boltzmann equations with physically
realistic collision kernels,
{\it J. Statist. Phys.} {\bf 74} (1994), no.3-4, 743--782.

\bibitem{4} Carlen,~E.A.; Gabetta,~E.; Toscani,~G.,
Propagation of smoothness and the rate of exponential convergence to
equilibrium for a
spatially homogeneous Maxwellian gas,
{\it Comm. Math. Phys.} {\bf 199} (1999), no.3, 521--546.


\bibitem{5} Carlen,~E.A.; Lu,~X.G.,
Fast and slow convergence to equilibrium for Maxwellian molecules
via Wild sums, {\it J. Statist. Phys.} {\bf 112} (2003), no.1-2,
59--134.

\bibitem{6} Cercignani,~C.,
{\it The Boltzmann equation and its applications}
(Springer-Verlag, New York, 1988).

\bibitem{7} Cercignani,~C; Illner,~R; Pulvirenti,~M.,
{\it The mathematical theory of dilute gases } (Springer-Verlag, New York, 1994).

\bibitem{8} Chapman,~S.; Cowling,~T.G.,
{\it The Mathematical Theory of Non-Uniform Gases}, Third Edition
(Cambridge University
Press, 1970).

\bibitem{9} Desvillettes,~L.,
Some applications of the method of moments for the homogeneous Boltzmann
and Kac equations,
{\it Arch. Rational Mech. Anal.} {\bf 123} (1993), no.4, 387--404.

\bibitem{10} Desvillettes,~L.; Mouhot,~C.,
Large time behavior of the a priori bounds for the solutions to the
spatially homogeneous Boltzmann equations with soft potentials, {\it
Asymptot. Anal.} {\bf 54} (2007), no.3-4, 235--245.

\bibitem{DL}   DiPerna,~R. J.; Lions,~P.-L. Global solutions of Boltzmann's
equation and the entropy inequality. {\it Arch. Rational Mech. Anal.}
{\bf 114} (1991),
no. 1, 47--55.

\bibitem{11} Goudon,~T.,
On Boltzmann equations and Fokker-Planck asymptotics: influence of grazing
collisions,
{\it J. Statist. Phys.} {\bf 89} (1997), no.3-4, 751--776.

\bibitem{12} Lions,~P.L.,
Compactness in Boltzmann's equation via Fourier integral operators and
applications, I,
{\it J. Math. Kyoto Univ.} {\bf 34} (1994), no.2, 391--427.

\bibitem{Lions} Lions,~P.L.,
Regularity and compactness for Boltzmann collision operators without angular
cut-off,
{\it C. R. Acad. Sci. Paris}, {\bf 326} S\'erie I, 1(1998), 37Ð41.

\bibitem{13} Lu,~X.G.,
A direct method for the regularity of the gain term in the Boltzmann equation,
 {\it J. Math. Anal. Appl.} {\bf 228} (1998), no. 2, 409--435.

\bibitem{14} Lu,~X.G.,
Conservation of energy, entropy identity, and local stability for
the spatially homogeneous Boltzmann equation, {\it J. Statist.
Phys.} {\bf 96} (1999), no. 3-4, 765--796.

\bibitem{15} Mouhot,~C.,
Rate of convergence to equilibrium for the spatially homogeneous Boltzmann
equation with
hard potentials,{\it Comm. Math. Phys.} {\bf 261} (2006), no. 3, 629--672.

\bibitem{16} Toscani,~G.; Villani,~C.,
Sharp entropy dissipation bounds and explicit rate of trend to equilibrium
for the spatially
homogeneous Boltzmann equation,
{\it Comm. Math. Phys.} {\bf 203} (1999), no. 3, 667--706.

\bibitem{17} Toscani,~G.; Villani,~C.,
On the trend to equilibrium for some dissipative systems with slowly
increasing a priori bounds,
{\it J. Statist. Phys.} {\bf 98} (2000), nos. 5-6, 1279--1309.

\bibitem{18} Villani,~C.,
On a new class of weak solutions to the spatially homogeneous Boltzmann
and Landau equations,
{\it Arch. Rational Mech. Anal.} {\bf 143} (1998), no. 3, 273--307.

\bibitem{V} Villani,~C., Regularity estimates via the entropy dissipation
for the
spatially homogeneous Boltzmann equation, {\it  Rev. Mat. Iberoam.}
{\bf 15}, no. 2 (1999), 335--352.

\bibitem{19} Villani,~C.,
A review of mathematical topics in collisional kinetic theory,
{\it Handbook of mathematical fluid dynamics}, Vol.I, 71--305, North-Holland,
Amsterdam, 2002.

\bibitem{20} Villani,~C.,
Cercignani's conjecture is sometimes true and always almost true,
{\it Comm. Math. Phys.} {\bf 234} (2003), no. 3, 455--490.

\bibitem{21} Wennberg,~B.,
Regularity in the Boltzmann equation and the Radon transform,
{\it Comm. Partial Differential Equations}, {\bf 19} (1994), no.11-12,
2057--2074.

\bibitem{22} Wennberg,~B., The geometry of binary collisions and generalized
Radon transforms,
 {\it Arch. Rational Mech. Anal.} {\bf 139} (1997), no. 3, 291--302.

\bibitem{23} Wennberg,~B.,  Entropy dissipation and moment production for
the Boltzmann equation,
{\it Jour. Stat. Phys.} {\bf 86}, nos. 5-6, (1997) 1053-1066.



\end{thebibliography}
\end{document}